\documentclass{aa}  
\usepackage{graphicx}
\usepackage{natbib} \bibpunct{(}{)}{;}{a}{}{,} 
\usepackage[varg]{txfonts}
\usepackage{hyperref}
\def\deg{\ifmmode{^\circ}\else$^{\circ}$\fi}			       
\def\kms{\ifmmode {{\rm \;km\;s^{-1}}}		    	      
       \else {\hbox{$\,${\rm km$\;$s$^{\rm -1}$}}}\fi}
\def\solar{\ifmmode_{\mathord\odot} \else $_{\mathord\odot}$\fi} 
\begin{document}
\title{The foot points of the Giant Molecular Loops in the Galactic center region}

    \author{D. Riquelme \inst{1} 
    \and M.A. Amo-Baladr\'on \inst{2} 
    \and J. Mart\'{i}n-Pintado \inst{2}
    \and R. Mauersberger \inst{1} 
    \and S. Mart\'{i}n \inst{3} \inst{4}
    \and M. Burton \inst{5} \inst{6}
    \and M. Cunningham \inst{5}
    \and P. A. Jones \inst{5}
    \and K. M. Menten \inst{1}
    \and L. Bronfman \inst{7}
    \and R. G\"usten \inst{1}}
    \institute{Max-Planck-Institut f\"ur Radioastronomie, Auf dem H\"ugel 69, 53121 Bonn, Germany\\
    \email{riquelme@mpifr-bonn.mpg.de} 
    \and Centro de Astrobiolog\'ia (CSIC/INTA), Ctra. de Torrej\'on a Ajalvir km 4, E-28850, Torrej\'on de Ardoz, Madrid, Spain
    \and European Southern Observatory, Alonso de C\'ordova 3107, Vitacura, Santiago, Chile
    \and Joint ALMA Observatory, Alonso de C\'ordova 3107, Vitacura, Santiago, Chile
    \and School of Physics, University of New South Wales, NSW 2052, Australia
    \and Armagh Observatory and Planetarium, College Hill, Armagh BT61 9DG, Northern Ireland, UK
    \and Departamento de Astronom\'{i}a, Universidad de Chile, Casilla 36-D, Santiago, Chile}
   \date{Received ; accepted }

 
  \abstract
   {}
   {To reveal the morphology, chemical composition, kinematics and to establish the main processes prevalent in the gas at the foot points of the giant molecular loops (GMLs) in the Galactic center region}
   {Using the 22-m Mopra telescope, we mapped the M$-3.8+0.9$ molecular cloud, placed at the foot points of a giant molecular loop, in 3-mm  range molecular lines. To derive the molecular hydrogen column density, we also observed the $^{13}$CO $(2-1)$ line at 1 mm using the 12-m APEX telescope. From the 3 mm observations 12 molecular species were detected, namely HCO$^+$, HCN, H$^{13}$CN, HNC, SiO, CS, CH$_3$OH, N$_2$H$^+$, SO, HNCO, OCS, and HC$_3$N. } 
   {Maps revealing the morphology and kinematics of the M$-3.8+0.9$ molecular cloud in different molecules are presented. We identified six main molecular complexes. We derive fractional abundances in 11 selected positions of the different molecules assuming local thermodynamical equilibrium.}
   {Most of the fractional abundances derived for the M$-3.8+0.9$ molecular cloud are very similar over the whole cloud. However, the fractional abundances of some molecules show significant difference with respect to those measured in the central molecular zone (CMZ). The abundances of the shock tracer SiO are very similar between the GMLs and the CMZ. The methanol emission is the most abundant specie in the GMLs. This indicates that the gas is likely affected by moderate $\sim $ 30 km s$^{-1}$ or even high velocity (50 km s$^{-1}$) shocks, consistent with the line profile observed toward one of the studied position. The origin of the shocks is likely related to the flow of the gas throughout the GMLs towards the foot points.}

   \keywords{ISM: molecules -- ISM: clouds-- Galaxy: center}
   \titlerunning{the Giant Molecular Loops in the Galactic center region}
   \maketitle
%
\section{Introduction \label{intro}}
\begin{figure*}
\centering
\includegraphics[width= 0.9 \textwidth]{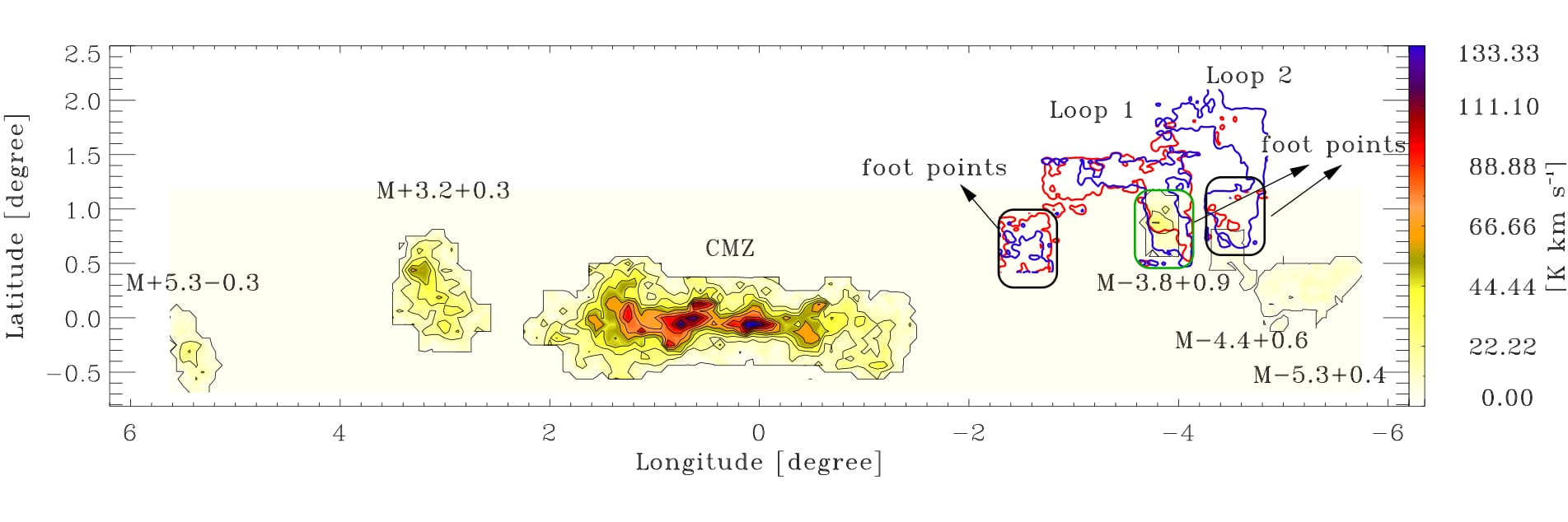}
\caption{Overview of the large scale GC region showing the features discussed in this work. The CMZ and five molecular clouds are indicated by the HCO$^+$ $(1-0)$ emission from \citet{Riquelme_et_al_2010b}.  Loops 1 and 2 are plotted in the CO $(2-1)$ emission from \citet{Kudo_et_al_2011}. Loop 1 is integrated in the velocity range $-180$ to $-90$ \kms (blue contour), and loop 2 from $-90$ to $-40$ \kms (red contour). The foot points are indicated with the black ellipse. The green ellipse shows the M$-3.8+0.9$ molecular cloud studied in this work.}
\label{overview}
\end{figure*}

The central regions of galaxies interact and exchange matter and radiation with their surroundings. 
This interaction strongly affects and modifies the physical properties and the chemistry in their nuclear regions. 
Due to its proximity, the central region of the Milky Way  allows detailed high resolution studies of the role of 
a variety of phenomena, namely  magnetic loops \citep{Fukui_et_al_2006}, galactic winds \citep{Bland-Hawthorn_Cohen_2003}, gas accretion by e.g., a barred Galactic model \citep{Binney_et_al_1991}.\\
The molecular component of the Galactic center (GC) region (i.e., the inner $\sim 1$ kpc of the Galaxy in the context of this work) is composed by a large 
molecular complex in the central $\sim 500$ pc  known as the ``Central Molecular Zone''
\citep[CMZ,][]{Morris_Serabyn_1996}, and several molecular clouds with high CO luminosity and large velocity dispersion 
\citep{Bitran_et_al_1997} outside the CMZ from $l\sim-6$ \deg to $l\sim 6$ \deg and $b\sim -1$\deg to $b\sim 2$\deg (Fig. \ref{overview}) that are also placed in the GC region. 
There are many molecular line emission surveys of the CMZ \citep[e.g., ][]{Jones_et_al_2012}, but few of 
them cover the molecular clouds beyond the CMZ.
\citet{Bitran_et_al_1997} observed a region of \mbox{$-12$\deg $<l<$ $12$\deg, $-2$\deg $<b<$ $2$\deg} in \mbox{CO$(1-0)$}, 
identifying 5 large velocity width and high CO luminosity clumps 
outside the CMZ located in the GC. \citet{Fukui_et_al_2006} extended the coverage of maps in this line to a larger latitude range  with better spatial resolution and identified huge loop structures in the negative velocity range from $-180$ \kms to $-40$ \kms with filamentary structure with a width of $\sim 30$ pc and a length of $\sim 400$ pc and heights of
$\sim 2$\deg\, from the Galactic plane. These features are coherent in velocity, with velocity gradients of $0.2-0.35 $\kms pc$^{-1}$.
\citet{Fukui_et_al_2006} proposed that these ``giant molecular loops''  (GMLs) placed in the GC are formed by a magnetic buoyancy
caused by a Parker instability. According to the model presented in \citet{Fukui_et_al_2006}, the gas
of the loops would flow down their sides, along the magnetic field
lines, and join with the gas layer of  the Galactic plane, generating
shock fronts at the ``foot points'' of the loops.  The presence of shocked gas is
supported by the broad velocity features of \mbox{$\sim$ 40} to  \mbox{80
\kms} width observed by \citet{Bitran_et_al_1997} and \citet{Fukui_et_al_2006}.\\
\indent Additional evidence of shocked gas at the ``foot points'',
comes from a survey of the GC region in  \mbox{HCO$^+(1-0)$}, \mbox{H$^{13}$CO$^+(1-0)$}, and
\mbox{SiO$(2-1)$} lines  \citep{Riquelme_et_al_2010b}.  The mapped area ($-5\deg.75 <l< 5\deg.63$ and
$-0\deg.7< b <1\deg.35$)  includes the CMZ and the 5 clumps
observed by \citet{Bitran_et_al_1997} mentioned above.  They found an enhancement of the SiO
emission (an archetypical tracer of shocked gas) at the ``foot points'' zones with  respect to HCO$^+$. 
This strongly suggests the presence of shocks \citep{Martin-Pintado_et_al_1992,Martin-Pintado_et_al_1997}.

Despite there are still no confirmatory magnetic field measurements in these features, multi-transitional CO observations towards both the foot points
and the complete loops, and magnetohydrodynamical simulations support the GMLs scenario. 
Table \ref{loopscharacteristics} summarizes the characteristics of the GMLs.
\citet{Torii_et_al_2010b} studied in detail the foot point of the GMLs towards $l\sim 356^{\circ}$ (hereafter ``M$-3.8+0.9$ molecular cloud''). They identified and analyzed several features including two ``U-shapes'', which they propose to be formed by the merging of two downward, i.e., to lower latitude flows between two loops as predicted by magneto-hydrodynamics numerical simulations \citep{Takahashi_et_al_2009, Machida_et_al_2009}. \citet{Fujishita_et_al_2009} discovered the ``loop 3'', placed in the positive-velocity range in $l\simeq354\deg-359\deg$, and \citet{Fukui_et_al_2006} proposed the existence of other GMLs connecting molecular clouds with large velocity widths at $l\sim 1\deg.3, 3\deg.2$ \citep{Bania_et_al_1986}, and $5\deg.3$.
\citet{Kaneda_et_al_2012} found that the polycyclic aromatic hydrocarbon (PAH) infrared emission at 9 $\mu$m is spatially correlated with the loops; however, it is suppressed in the foot points as compared with the IRAS 100  $\mu$m emission.  This suggested the destruction of PAHs relative to sub-micron dust grains by shocks.
Isotope studies \citep{Riquelme_et_al_2010a} suggest that gas has been accreted
towards the foot point of the loops, and metastable inversion transitions of the ammonia \citep{Riquelme_et_al_2013} revealed
high kinetic temperatures ($> 90$ K) in the ``foot points'' of the loop at $l\sim 5\deg.2$. 
Fig. \ref{overview} presents an overview of the large scale GC region, showing the CMZ, the five molecular clouds outside the CMZ, and the location of the giant molecular loops discussed in this paper.\\
\indent This paper presents high angular resolution mapping of 3-mm molecular lines
toward the  M$-3.8+0.9$ molecular cloud, placed at the foot points of the molecular loops
discovered by \citet{Fukui_et_al_2006} in the GC.  These observations allow us to derive
the morphology, chemistry and the kinematics of both the quiescent and the shocked gas at small spatial scales of order 1 pc.
The projected distance from Sgr A* of the M$-3.8+0.9$ molecular cloud is 564 pc, assuming a distance of 8.5 kpc (the IAU recommended value).

\begin{table*}
\caption{Summary of the characteristics of the giant molecular loops}
\label{loopscharacteristics}
\centering 
\begin{tabular}{lcccccl}
\hline\hline
Loop     & Longitude range & Latitude range &  Velocity range & Mass & Foot point location\\
         & [Degree]         &  [Degree]      &  [km\,$s^{-1}$] &M\solar &  \\\hline
Loop 1   & $355\deg.8$ -- $358\deg.0$ &$0^{\circ}.5$ -- $1^{\circ}.6$ &$-180$ -- $-90$ &$7-12\times 10^5$ & $(l,b)\sim (356^{\circ}, 1^{\circ})$ and $(l,b)\sim (357^{\circ}.4, 0.^{\circ}8)$ \\
Loop 2   & $355\deg.2$ -- $356^{\circ}.6$ &$0^{\circ}.6$ -- $2^{\circ}.2$ & $-90$ -- $-40$ & $8-16\times 10^5$ & $(l,b)\sim (355^{\circ}.4, 0^{\circ}.8)$ and $(l,b)\sim (356^{\circ}.1, 0^{\circ}.8)$ \\
Loop 3   & $354\deg-359\deg$ &$0^{\circ}$ -- $2^{\circ}$ & $30$ -- $160$ & $3\times 10^3$ & $(l,b)\sim 355\deg.5, 0\deg.7)$,$(l,b)\sim 359\deg, 0\deg.6)$ \\
\end{tabular}
\end{table*}

\section{Observations and data reduction}
\subsection{Mopra observations and data reduction}\label{observationdetails}
The observations were carried out using the 22-m Mopra telescope
during September 2008 and August 2009. Located in the Southern 
hemisphere, and due to its high angular resolution and wide-bandwidth
spectrometer, the Mopra telescope offered excellent capabilities to establish the chemical abundances of the molecular gas at the foot
points of the GC loops. We used the digital mode filter bank MOPS in broad-band mode, covering 8
GHz of bandwidth simultaneously in four $2.2$ GHz sub-bands, each of them with
$8192$ channel spaced by $0.27$ MHz. Two polarizations were measured
simultaneously.

We mapped a selected region of the M$-3.8+0.9$ molecular cloud
\citep{Riquelme_et_al_2010b} using the on-the-fly (OTF) mapping mode \citep{Ladd_et_al_2005, Mangum_2007}.
Observations of tiles of $5'\times5'$ size with overlaps of $30''$ were used to cover the complete region.
The complete maps covered the region of $356\deg.06<l<356\deg.29$ and
$0\deg.645<b<1\deg.1$. We used position switching mode with the off
position placed at $(l,b)=(356\deg.375,1\deg.5)$, which was checked to be free
of emission, and observed in symmetric mode (one off per OTF scan). The
spacing between scan rows was $12''$, and each tile of $5\times5$ ${\rm arcmin}^2$ 
took 55 min to complete. To establish pointing parameter corrections we observed before each map the SiO maser source
AH$\,$Sco. The spectra were read out with 2 seconds of integration
time.  The system temperature was calibrated with a noise diode and
a hot/cold load (paddle) every 30 min. We observed two frequency setups, one centered at
$89.41$ GHz and the other at $99.72$ GHz covering the ranges between 
$85.275$ and $93.555$ GHz and from $95.585$ to $103.866$ GHz.

\indent The data were reduced using the LIVEDATA and GRIDZILLA packages.
LIVEDATA is the processing software used to apply system temperature
calibration, bandpass calibration, heliocentric correction, spectral
smoothing and to write out the data in sdfits \citep{Garwood_2000}
format. GRIDZILLA is a regridding software package to convert the
sdfits files to the FITS data cube \citep{Jones_et_al_2008}. A first order polynomial baseline was subtracted with LIVEDATA, 
and the data were regridded with GRIDZILLA into data cubes, using a Gaussian smoothing
interpolation.\\
\indent The final spatial resolution of the data cubes is between $49''$ and $51''$ at
115 and 86 GHz respectively, which is obtained after convolution of the
Mopra beam width of $33\pm 3''$ at 115 GHz and $36\pm 3''$ from 86 to 100 GHz, as measured using Jupiter in 2004  \citep{Ladd_et_al_2005}, with a Gaussian of $36''$ full width half maximum (FWHM).
This FWHM for the Gaussian size improves the signal-to-noise of the data, albeit with a modest loss in spatial resolution.
We calibrated the data in the main beam brightness temperature ($T_{\rm mb}$) scale. The main beam efficiency of Mopra varies between 0.49 at 86 GHz 
and 0.42 at 115 GHz. However, to convert to $T_{\rm mb}$ we used the values for the extended 
beam efficiency which are more appropriate for the extended emission of the Galactic center region 
(0.65 at 86 GHz and 0.55 at 115 GHz) \citep{Ladd_et_al_2005}. The spectral resolution of the data 
is 269.5 kHz (0.94 -- 0.78 \kms). One data cube per molecular line was made. The size of the pixel 
is $15$ arc sec in the final cube. We produces 13 data cubes for the detected molecules in the Mopra survey (see Table \ref{table:1}). In 5 of those data cubes (CS, SiO, HC3N-10-9, H13CN, CH3OH), it was necessary to subtract 3$^{rd}$ order baselines for $\sim 20$ \% of the data using MADCUBAIJ\footnote{\url{http://www.cab.inta-csic.es/madcuba}} software.

\subsection{APEX observations and data reduction} 
The $J=2-1$ rotational transition of $^{13}$CO was mapped using the 12-m Atacama Pathfinder EXperiment (APEX) 
telescope \citep{Gusten_et_al_2006}, covering a similar region as the Mopra observations (see Fig. \ref{clumpC_total13CO}). 
The observations were carried out on 24 Jun, 1, 2 and 3 July 2014 under the APEX project code M-093.F-008-2014 
using the APEX-1 (SHIFI) receiver \citep{Vassilev_et_al_2008} and the eXtended bandwidth Fast Fourier
Transform Spectrometer (XFFTS) backend \citep{Klein_et_al_2012}.  OTF position-switching observing mode was used, with a close off position 
slightly contaminated with $^{13}$CO emission ($\alpha$(J2000): $17{\rm h}33{\rm m}13.0{\rm s}$, $\delta$(J2000): $-31\deg30'13''.9$) mainly at the velocity range from $-10$ to 30 km s$^{-1}$ with an intensity peak $T_{\rm MB}=0.6$ K, 100 to 110 km s$^{-1}$ with an intensity peak of $T_{\rm MB}=0.2$ K, and in much less amount from $-80$ to $-70$ km s$^{-1}$ and from $-50$ to $-20$ km s$^{-1}$ with $T_{\rm MB}<0.1$ K. This contamination  
was corrected later with observations against a clean off position  (RA(J2000): $17{\rm h}30{\rm m}48.0{\rm s}$, DEC(J2000): $-31\deg11'48''.1$). 
This observing strategy ensure flat baselines across the map. The pointing was checked every 1.5-2 hours on IRAS17150-3224; corrections smaller than $2''$ were determined. The system temperature ($T_{{\rm sys}}$) ranged from 125 to 204 K, with an average value of 155 K. 
The calibration was done using the standard APEX calibration procedure, with an estimated error of $\sim 10\%$. 

The data was reduced using the CLASS package from the GILDAS software\footnote{\url{http://www.iram.fr/IRAMFR/GILDAS}}. The antenna temperature 
($T_{\rm A}^*$) was converted to $T_{\rm MB}$  using the Ruze formula\footnote{$B_{\rm eff}(\lambda) = B_0\times exp(-(4\times\pi \times \sigma/\lambda)^2)$} with B$_0=0.69$ and $\sigma=0.19$ for an extended source (see Appendix \ref{cal}). All spectra were taken into account since the observed rms noise was in all cases lower than 1.5$\times$ the theoretical noise \citep{Ao_et_al_2013, Ginsburg_et_al_2016} \footnote{where the theoretical noise was estimated from $T_{\rm sys}/\sqrt{\Delta \nu \times T_{\rm exp}}$ for the OTF observations and $2\times T_{\rm sys}/\sqrt{\Delta \nu \times T_{\rm exp}}$ for the position switching observations in the reference position \citep{Mangum_2007}}.
Third order polynomial baselines were subtracted for the OTF mapping observations of the M$-3.8+0.9$ molecular cloud, and a fifth order polynomial baselines for the position switching observations in the reference position. Then, each individual spectrum for the mapping was combined with the reference position spectrum using the accumulate function in CLASS with an equal weighting. In this way, the baseline time-dependence is removed. Since line widths $>10$ \kms were expected, the final spectra were smoothed with the box function in CLASS to reach a final velocity resolution of $1.04$ \kms which is more than enough to resolve all the kinematic structures of this molecular cloud.  The data were regridded in equatorial coordinates and then converted to Galactic coordinates for comparison with the Mopra data using standard CLASS routines. The average root-mean-square (rms) noise for the spectra in the data cube is 127 mK at a velocity resolution of $\approx 1$~km~s$^{-1}$. The final data cube was corrected for additional baseline subtraction using MADCUBAIJ software in small regions when needed (2-3 order polynomial base line subtraction in the 30\% of the data).

\section{Results}

Following the list of most prominent 3 mm wavelength molecular lines 
observed in Sgr\,B2 by \citet[see their Table 2 in ][]{Jones_et_al_2008}, we made one
data cube per molecule\footnote{all data cubes publicly available}. 
Tables \ref{table:1} and \ref{table:2} show the molecules and the rms noise values reached 
in each molecular line cube. The results are presented both in the main text and in the Appendices A to D. 

\begin{table}
\caption{List of imaged molecular transitions, their rms noise level ($T_{\rm A}^*$) and the main beam efficiency.}
\label{table:1}
\centering 
\begin{tabular}{lcccl}
\hline\hline
Molecule     & Transition & Rest. Freq. &  rms$^3$  & $\eta_{\rm MB}$\\
             &            &  [GHz]      & [mK]  & \\\hline

H$^{13}$CN   & $1-0^1$        & 86.340  &  54 & 0.65\\
SiO          & $2-1$ $v=0$   & 86.847   &  51 & 0.65\\
HNCO         &$4(0,4)-3(0,3)$& 87.925   &  53 & 0.64 \\
HCN          & $1-0^1$        & 88.632  &  58 & 0.64 \\
HCO$^+$      & $1-0$         & 89.188   &  34 & 0.64\\
HNC          & $1-0^1$        & 90.664  &  38 & 0.63 \\
HC$_3$N      & $10-9$        & 90.980   &  40 & 0.63 \\
N$_2$H$^+$   & $1-0^1$        & 93.174  &  38 & 0.62 \\
CH$_3$OH     & $2_K-1_K^1$    & 96.74   &  50 & 0.61 \\  
OCS          & $8-7$         & 97.300   &  57 & 0.61\\%
CS           & $2-1$         & 97.980   &  50 & 0.61 \\
SO           & $3(2)-2(1)$   & 99.300   &  58 & 0.60  \\ %
HC$_3$N      & $11-10$       & 100.08   &  53 & 0.60\\
$^{13}$CO    & $2-1^2$       & 220.398  & 127$^4$ & 0.67\\
\end{tabular}
\tablefoot{$^1$ have fine or hyperfine structure transitions. $^2$ observed with the APEX telescope. $^3$ the spectral resolution of the data is 0.27 MHz which corresponds to a velocity resolution from 0.78-0.94 \kms depending of the frequency of the species. $^4$ velocity resolution of 1.04 km/s.}
\end{table}    

\begin{table}
\caption{List of non detected molecular transitions and their rms noise level in a 0.27 MHz wide channel.}
\label{table:2}
\centering 
\begin{tabular}{lccc}
\hline\hline
Molecule     & Transition & Rest. Freq. &  rms   \\
             &            &  [GHz]      & [mK]  \\\hline
CH$_3$CCH    & $5-4^1$     & 85.457      &  48 \\             
HOCO$^+$     & $4(0,4)-3(0,3)$& 85.530   &  56 \\
SO           & $2(2)-1(1)$   & 86.093   &  43 \\ 
H$^{13}$CO$^+$& $1-0$     & 86.754      &  53 \\
HN$^{13}$C   &  $1-0^1$    & 87.091      &  55 \\
CCH          & $1-0^1$     & 87.328      &  56 \\
CH$_3$CN     & $5-4^1$     & 91.979      &  50 \\
$^{13}$CS    & $2-1$      & 92.494      &  50 \\
C$^{34}$S    & $2-1$      & 96.410       &  46 \\\
CH$_3$OH     & $2(1,1)-1(1,0)A-$&97.582 &  77 \\
NH$_2$CN     & $5(1,4)-4(1,3)$& 100.63  &  63 \\
H$_2$CS      & $3(1,3)-2(1,2)$&101.48   &  79 \\
CH$_3$CCH    & $6-5^1$     & 102.530      &  77 \\
H$_2$CS      & $3(0,3)-2(0,2)$&103.04   &  54 \\
\end{tabular}
\tablefoot{$^1$ have fine or hyperfine structure transitions.}
\end{table}

\subsection{Morphology \label{morphologykinematics}}
\indent The morphology and velocity structure are illustrated by the HCN and $^{13}$CO molecular lines, which show the most intense emission among all detected molecules.

\begin{figure*}
\centering
\includegraphics[angle=0,width=1.0\textwidth]{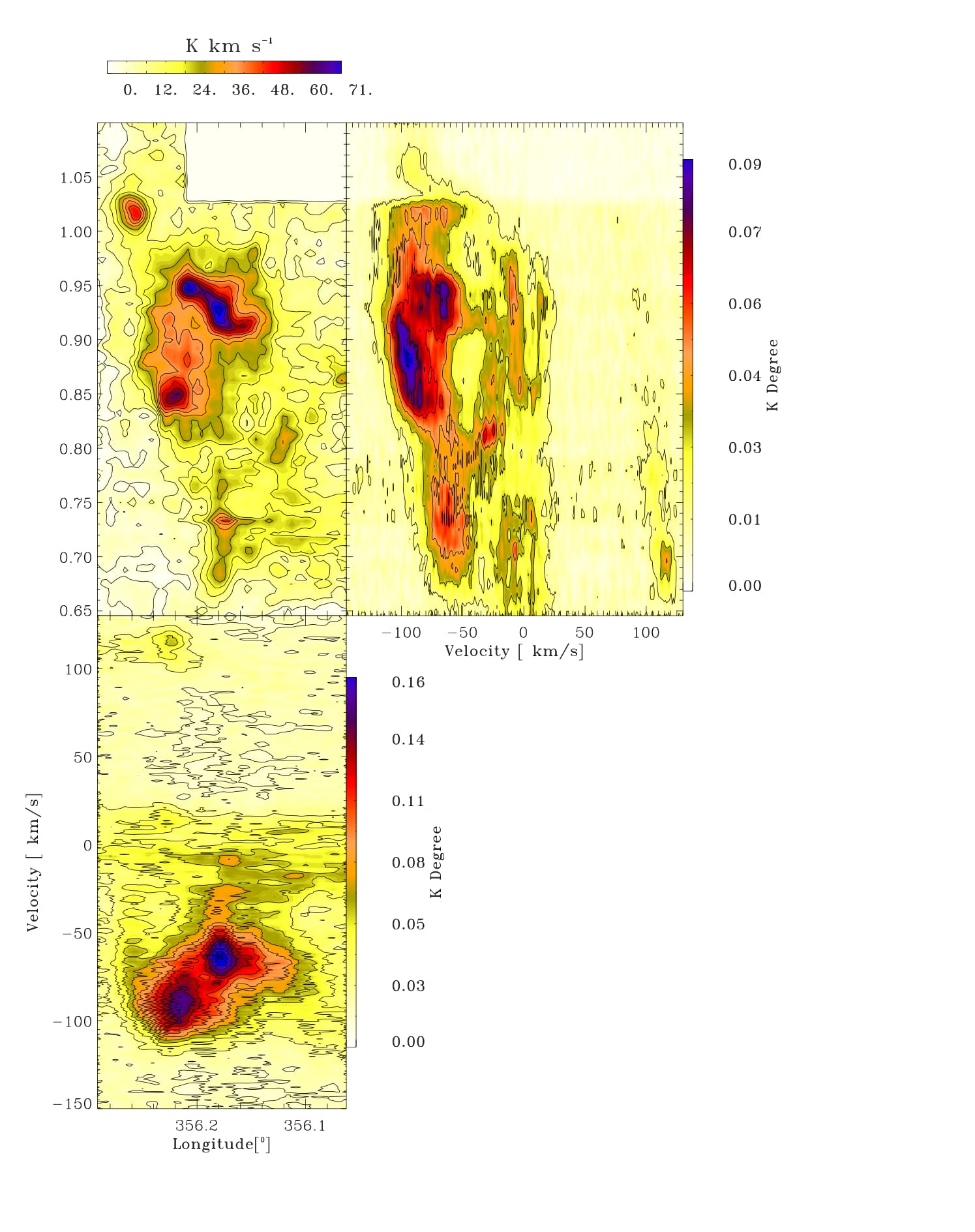}
\caption{HCN $(1-0)$ emission of the M$-3.8+0.9$ cloud: Left. Integrated brightness temperature map
in the velocity range from $-140$ to $-20$ km\,s$^{-1}$.  Right. Latitude-velocity
plot integrated over the whole longitude range (from $l=356.29\deg$ to
$356.06\deg$). Bottom. Longitude-velocity plot integrated over the whole
latitude range (from $b=0.64\deg$ to $b=1.10$\deg).}
\label{clumpC_totalHCN}
\end{figure*}

\begin{figure*}
\centering
\includegraphics[angle=0,width=1.0\textwidth]{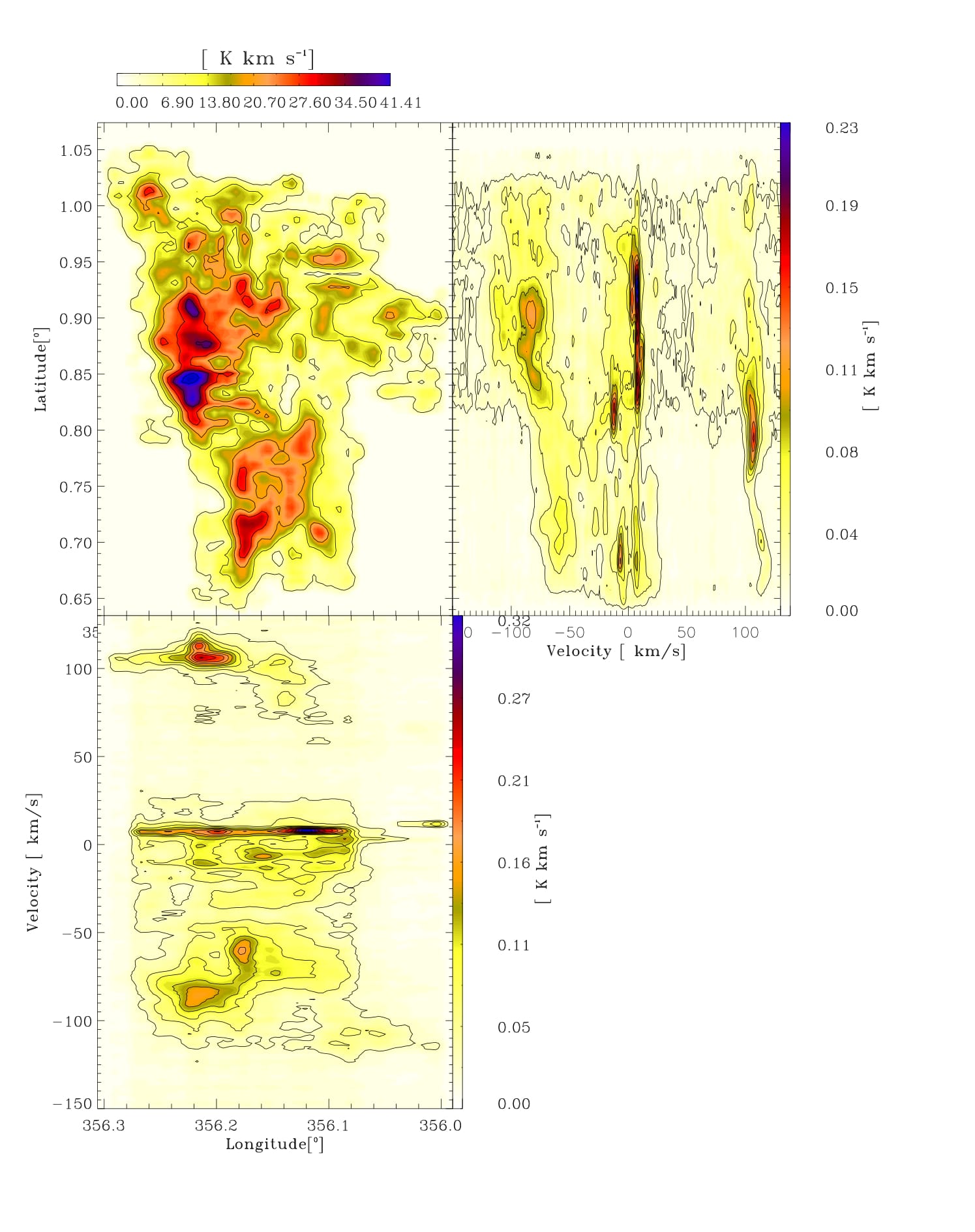}
\caption{$^{13}$CO (2-1) emission of the M$-3.8+0.9$ cloud: Left. Integrated brightness temperature map
in the velocity range from $-140$ to $-20$ km\,s$^{-1}$. The dashed lines show
the mapped region. Right. Latitude-velocity plot integrated over the whole longitude range 
(from $l=356\deg$ to $356\deg.294$). Bottom. Longitude-velocity plot integrated over the whole
latitude range (from $b=0\deg.6423$ to $b=1\deg.051$)}
\label{clumpC_total13CO}
\end{figure*}

\begin{figure*}
\centering
\includegraphics[width= 0.9 \textwidth]{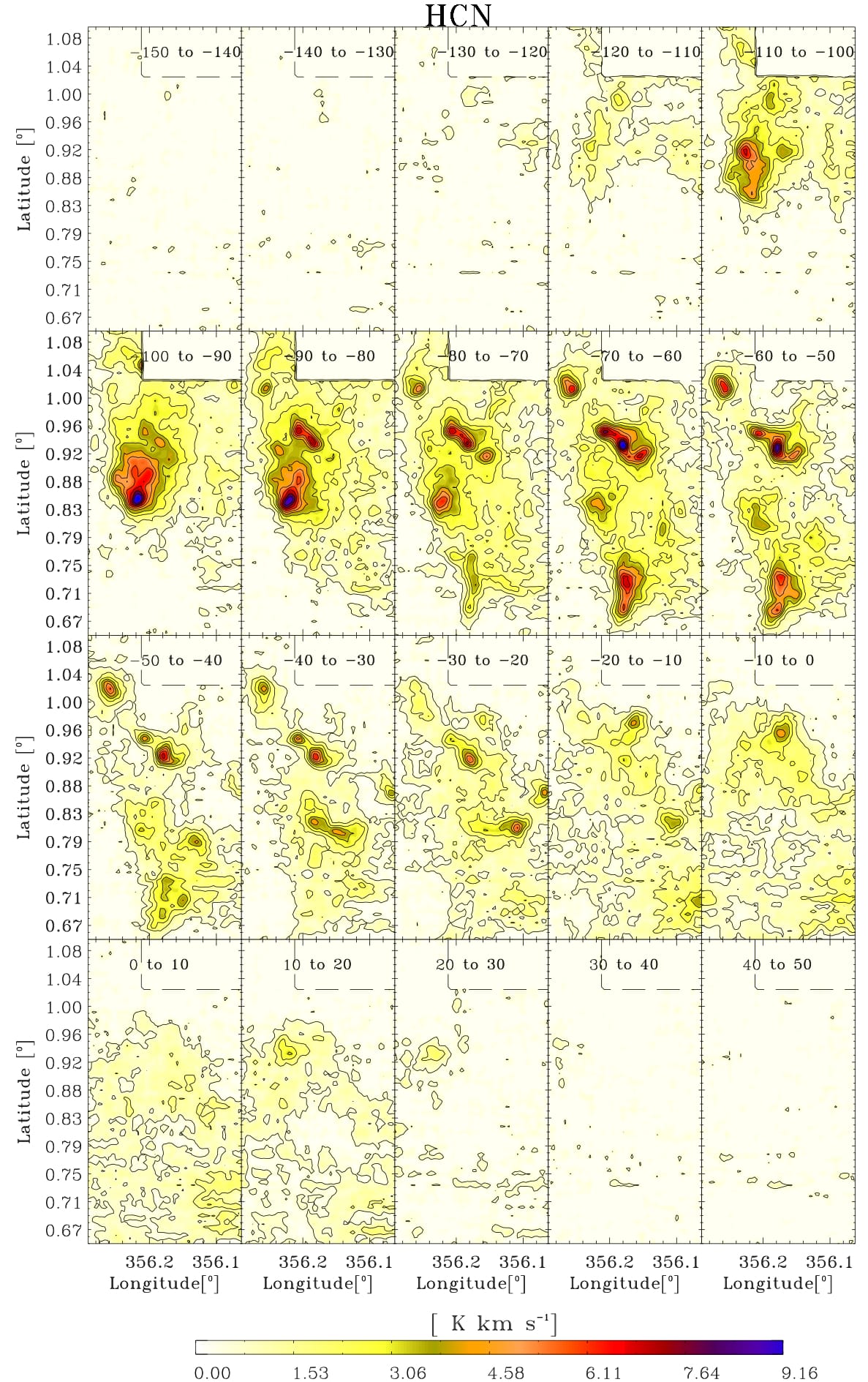}
\caption{Integrated brightness temperature of the M$-3.8+0.9$ molecular cloud in HCN $(1-0)$
in velocity intervals of $10$ \kms.}
\label{clumpC_vel}
\end{figure*}

\begin{figure*}
\centering
\includegraphics[width= 0.9 \textwidth]{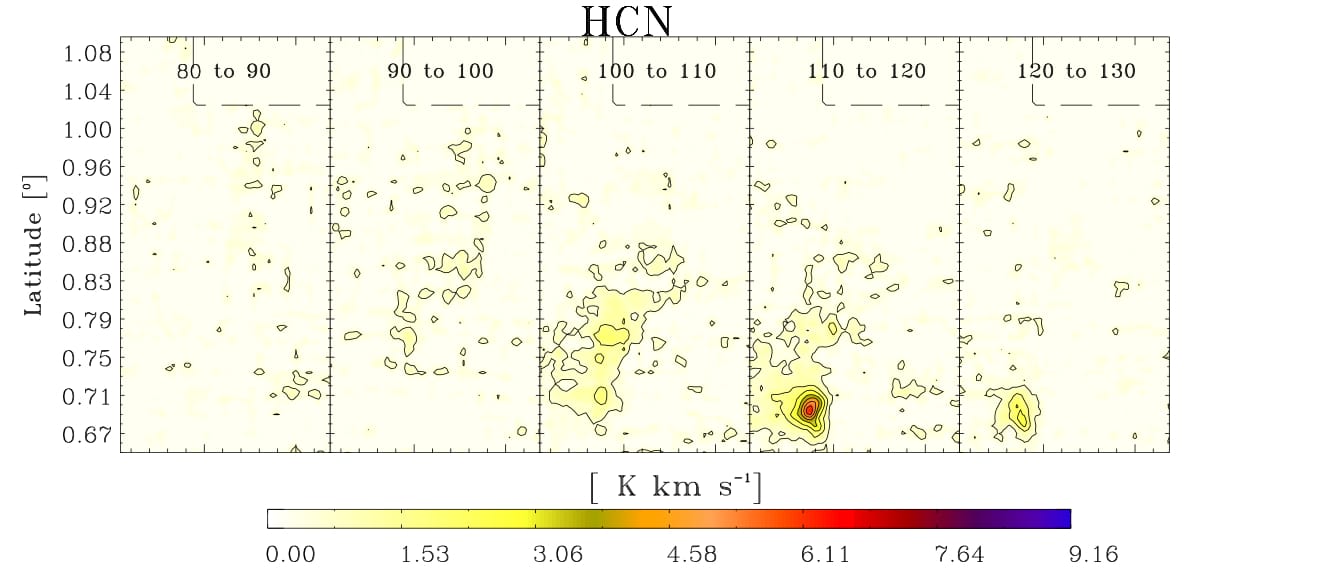}
\caption{Integrated brightness temperature of the M$-3.8+0.9$ molecular cloud in HCN $(1-0)$
in velocity intervals of $10$ \kms.}
\label{clumpC_vel2}
\end{figure*}

Figs. \ref{clumpC_totalHCN} and \ref{clumpC_total13CO} show: I) the integrated brightness temperature maps of HCN and $^{13}$CO $(2-1)$, for the
observed region of M$-3.8+0.9$, in the velocity range from $v_{\rm LSR}=-140$ to$v_{\rm LSR}=-20$
\kms, II) the longitude-velocity map integrated over the whole observed latitude range, and III) the latitude-velocity map integrated in the whole observed longitude range. Fig. \ref{clumpC_vel} and \ref{clumpC_vel2} show the integrated brightness temperature of the molecular
emission in velocity intervals of 10 km\,s$^{-1}$ in HCN, and in
Fig. \ref{clumpC_vel13CO} and \ref{clumpC_vel213CO} we show the $^{13}$CO $(2-1)$ emission. The M$-3.8+0.9$ cloud has a narrow line width emission at positive velocities ($\sim $120 \kms) at $(l,b)\sim (356\deg.25,0\deg.70)$, which can be seen in the HCN emission (Fig. \ref{clumpC_vel2}). Its position coincides with that of an ultra compact H\,II region \citep{Caswell_Haynes_1987}, but this source has a radial velocity of $113$ \kms, and therefore is not associated to the foot point as noted by \citet{Torii_et_al_2010b}.
Therefore this source will not be further discussed in this work. This source was only partially covered by the $^{13}$CO observation. 
There are also several molecular clouds showing positive velocities in the $^{13}$CO emission (Fig. \ref{clumpC_vel213CO}). The emission at $\sim 70-90$ \kms could be associated to the far-3 kpc arm \citep{Dame_Thaddeus_2008} and the emission at $90-130$ \kms could be associated to the 135 \kms arm \citep{Bania_1980}), as noted by \citet{Riquelme_et_al_2010b}, which are not associated with the loops 1 and 2, and they will not be discussed in this work.

Figs. \ref{clumpC_vel} and \ref{clumpC_vel13CO} show the presence of a velocity gradient from higher to lower velocities in the north-western to the south-eastern direction. Four main velocity components were identified: from
$-140$ to $-70$ \kms, from $-70$ to $-40$ \kms, from $-40$ to $-20$ \kms, and
from $-20$ to $20$ \kms. In Fig. \ref{inicio} to \ref{final} we show the
integrated intensities maps between $-140$ \kms and $-20$ \kms as well as the integrated intensities maps in the four velocity ranges for all the detected transitions listed in Table \ref{table:1}. In these velocity ranges, six main molecular complexes are identified, indicated by green boxes in Fig. \ref{rois} on the HCN maps. These complexes were identified by visual inspection and may be spatially correlated if the M$-3.8+0.9$ cloud has a velocity gradient of 2 km $^{-1}$ pc$^{-1}$. For an easy comparison, the $^{13}$CO emission is also plotted in this figure.
The velocity structure is also shown in the latitude-velocity domain 
integrated in longitude steps of 108'' (from Fig. \ref{bvini} to Fig. \ref{bvfinal}). 
In those figures we can see the two ``U-shapes'' identified by \citet{Torii_et_al_2010a} (see Section \ref{discusioncomparison}).

\subsection{Complex 3}
In the velocity range from $-140$ to $-70$ \kms, the most prominent feature is the Complex 3, which shows an elongated structure perpendicular to the Galactic plane with an abrupt sharp intensity decrease towards the eastern edge. This complex also contains two intensity peak which are detected in most of the molecules (see the Figs. in Appendix \ref{velocityintegratedmolecules}) at $(l,b)=(356\deg.22, 0\deg.91)$ 
and $(l,b)=(356\deg.21, 0\deg.84)$.
The $^{13}$CO, HC$_3$N, N$_2$H$^+$, and SO molecular emissions have the intensity peak in the north, 
in contrast with the HCN , HCO$^+$, and CS emissions which have the intensity peak in the south. Some molecules only appear in the north (e.g., HNCO, N$_2$H$^+$, SO, and HC$_3$N). 

\subsection{Complexes 1, 2, 4, 5 and 6}
\begin{figure*}
\centering
\vbox{
\includegraphics[width= 0.9 \textwidth]{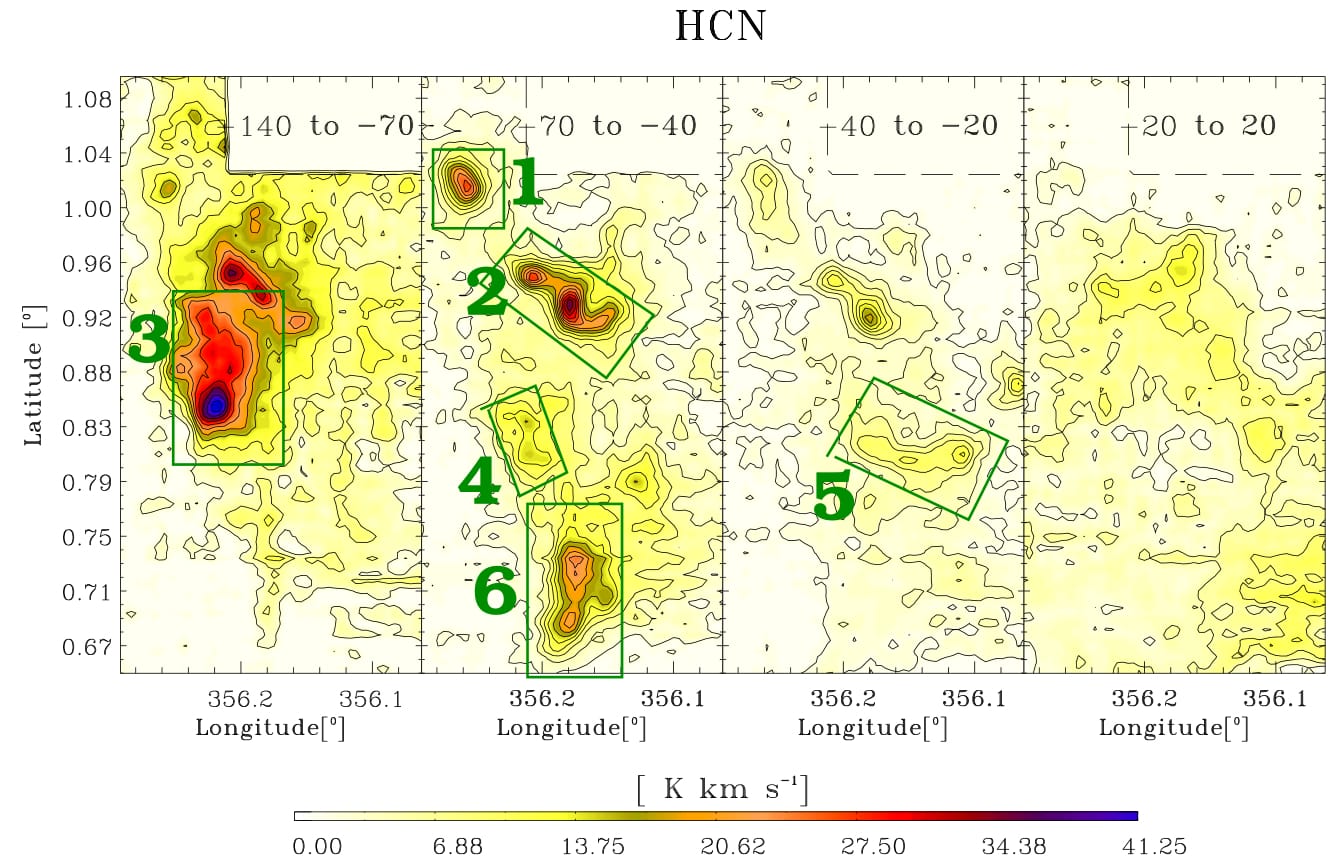}
\includegraphics[width= 0.9 \textwidth]{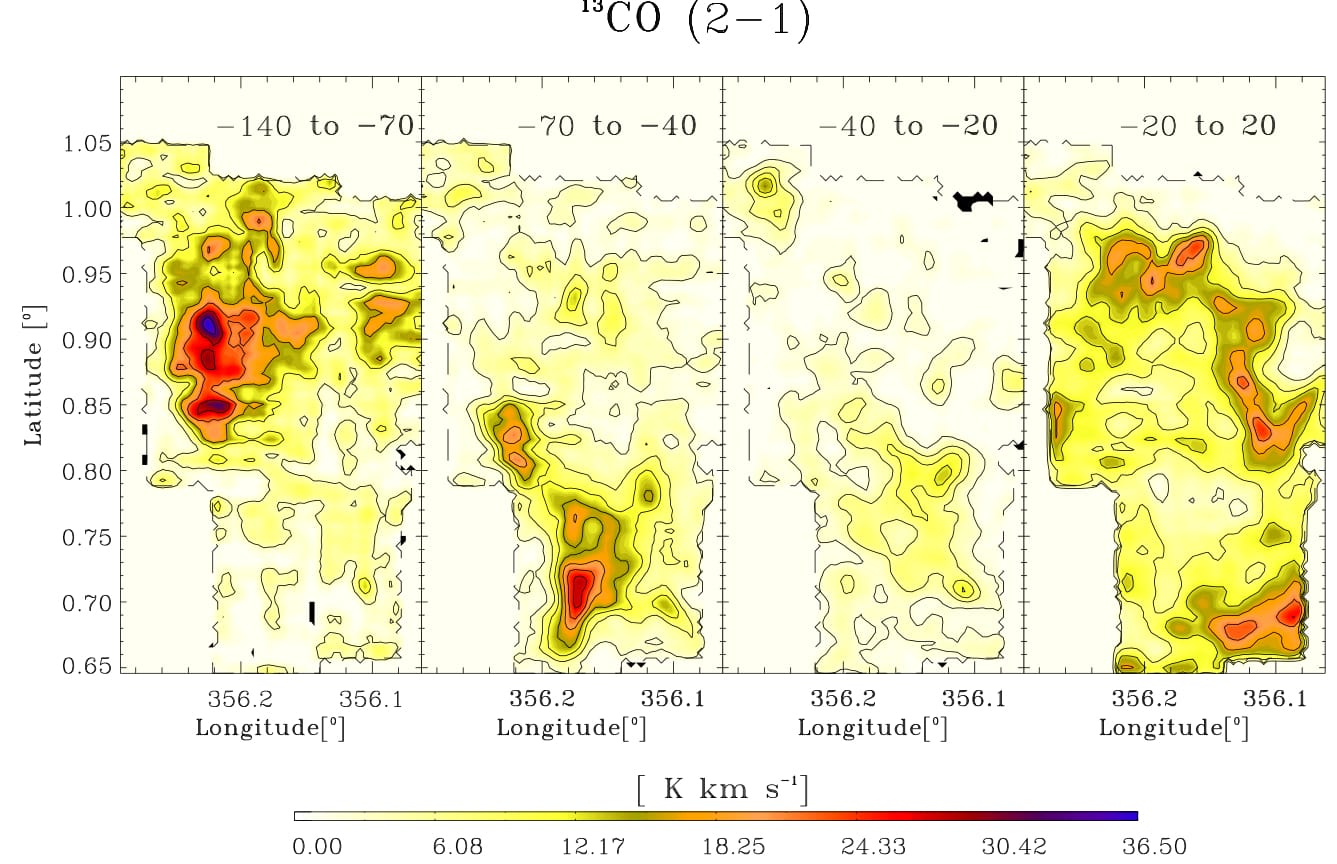}
}
\caption{Position of the six molecular complexes discussed in the text. The
green boxes show the complexes discussed in the text. }
\label{rois}
\end{figure*}

In the velocity range from $-70$ to $-40$ \kms, the Complexes 1, 2, 4 and 6 can be identified. Complex 1 
has a very large line-width ($-110$ to 0 \kms). It is also strong in HCN emission, while in HCO$^+$ emission 
the intensity is weaker but still cover the same velocity range. In the other molecules, 
the emission shows two kinematic components with the intensity peak at $\sim -40$ \kms.
This complex is very prominent in the HCN, HCO$^+$, CS, HNC, and CH$_3$OH emission (see from Fig. \ref{inicio} to \ref{final}). 
Although the Complex 2 is not visible in all the detected molecules, it is the 
region which presents the largest line width in the M$-3.8+0.9$ molecular cloud, as can be seen in the Appendix \ref{velocityintegratedmolecules} 
which is consistent with previous works \citep[see, ][]{Torii_et_al_2010a,Torii_et_al_2010b}.  
The Complex 4 looks like connecting the Complex 3 and 6. This complex is very prominent in $^{13}$CO, CS, HCO$^+$, HCN, and HNC.
It is remarkable that Complex 6 also shows an elongated structure perpendicular 
to the Galactic plane similar to that observed for Complex 3. This complex also shows 
an abrupt sharp intensity decrease towards the eastern edge.  In the
velocity range from $-40$ to $-20$ \kms, we can see the Complex 5 which is an
elongated feature, parallel to the Galactic plane. In the last velocity
range, from $-20$ to $20$ \kms, the emission shows a shell-like structure which
appears over the complete molecular cloud and is clearly seen in the $^{13}$CO, HCO$^+$ and HNC
maps (Figs. \ref{rois}, \ref{HCO+_intervals} and \ref{HNC_intervals}). As can be seen in the channel maps of $^{13}$CO plots (Fig. \ref{clumpC_vel13CO}, this feature is indeed narrower in velocity, from -10 to 10 \kms.  This emission could be associated to local gas along the line of sight and since it is probably not associated to the GMLs, this feature will not be discussed in this work.

\section{Analysis}

\subsection{$^{13}$CO (2-1) emission. H$_2$ Column density estimate}
\label{H2colum}

Since the $^{13}$CO emission is optically thin (see optical depth estimations for CO in \citep{Torii_et_al_2010b}) 
and the critical density is relatively low ($n_{\rm crit} \sim 10^4$ cm$^{-3}$), this molecule is a good tracer of the total H$_2$ column density using a proper conversion factor 
($N({\rm H_2})=N({\rm ^{13}CO})\times [{\rm ^{13}CO/H_2}]$). Since we only have one transition, we assume local thermodynamical equilibrium (LTE) at a excitation temperature ($T_{\rm{ex}}$) of 10 K to derive the column density, N:
\begin{equation}
 \label{formula1} 
 N=\frac{8\pi  k  \nu^2  Q(T_{\rm ex})}{c^3 A_{ij} g_u h} \exp{\Bigg(\frac{E_u}{k T_{\rm ex}}\Bigg)} \frac{1}{\Big(1-\frac{J(T_{\rm bg})}{J(T_{{\rm ex}})}\Big)}\int T_b dv
\end{equation}
where $k$ is the Boltzmann constant, $\nu$ the frequency of the transition, $h$ the Planck constant, 
$Q(T)$ the partition function at the assumed excitation temperature, $g_u$ the upper state degeneracy, $A_{ij}$ the Einstein coefficient, $E_u$ the energy of the upper state, and 
$J(T)=\frac{h \nu}{k} \Big(\frac{1}{(\exp((h \nu)/(k T))-1}\Big)$ is the source function at a temperature T. The molecular parameters were taken from the Cologne Database 
for Molecular Spectroscopy (CDMS) catalog \citep{Muller_et_al_2005, Muller_et_al_2001}, and T$_{\rm{bg}}=2.73$ K is the cosmic background radiation temperature.
If we use the T$_{\rm{ex}}=40$ K as derived by the multi-J transition study of $^{12}$CO by \citet{Torii_et_al_2010b}, the N($^{13}$CO) increase by less than 10\% (see Fig. \ref{variationTex}). Thus, we decided to use 10 K which is consistent with our estimations in \citet{Riquelme_et_al_2013} for CS, and with the discussion in Section \ref{comparisonCMZ}.

For our calculations, we assume an abundance ratio CO/H$_2$ of 10$^{-4}$ \citep{Frerking_et_al_1982}. This abundance ratio was also used by \citet{Rodriquez-Fernandez_et_al_2001}, \citet{Dahmen_et_al_1998} and \citet{Huettemeister_et_al_1998} for their large scale studies of the GC, which cover the complete CMZ and the Bania's clump2. As noted by \citet{Huettemeister_et_al_1998}, \citet{Farquhar_et_al_1994} showed that this ratio is stable against a possibly enhanced cosmic ray flux to the GC. \citet{Riquelme_et_al_2010a} derived a high $^{12}$C/$^{13}$C isotopic value ($34-73$) in several positions in this molecular cloud which is higher than the typical values (20-25) found in the GC region \citep[see, e.g., ][]{Langer_Penzias_1990,Wilson_Matteucci_1992}. Therefore, the value of 53 was used, corresponding to the typical value found in the 4 kpc molecular ring \citep{Wilson_Rood_1994}, which was also used by \citet{Torii_et_al_2010b}, \citet{Kudo_et_al_2011} and \citet{Riquelme_et_al_2013} in the GMLs regions. 
This translates into a $\mathrm{[^{13}CO/H_2]}$ conversion factor of $1.9\times 10^{-6}$.

\subsection{Comparison of the emission between different molecules}
To compare and quantify the differences and similarities between the emission distribution throughout the M$-3.8+0.9$ molecular cloud in the different detected molecules, we perform a principal component analysis and we derive the fractional abundances in selected positions. 

\subsubsection{Principal component analysis \label{pcasection}}
A principal component analysis \citep[PCA, see, e.g, ][]
{Heyer_Schloerb_1997,Shlens_2014, Ungerechts_et_al_1997, Lo_et_al_2009, Jones_et_al_2012} was performed using the most intense molecular lines in Table \ref{table:1}, namely $^{13}$CO, HCN, HCO$^+$, HNC, CH$_3$OH, and CS. 
As mentioned in Section \ref{observationdetails}, the Mopra beam varies from $33\pm3$'' at 115 GHz and $36\pm 3$'' from $86-100$ GHz. Then, the beams sizes for the molecules used in the PCA analysis are identical within the uncertainties. The resolution of the pixel in all data cubes is 15''. The $^{13}$CO $(2-1)$ data was converted to Galactic coordinates using standards class routines using one of the Mopra datacube (HCN) as a pattern, imposing the pixel size of 15''. Therefore, all data cube used in the PCA have a uniform spatial resolution.
To implement the method, a python script with the PCA 
module\footnote{\url{http://folk.uio.no/henninri/pca_module}} is used in a similar way as \citet{Jones_et_al_2012}, using a covariance matrix  method. Because the PCA analysis works with normalized data, we only can compare data with a good signal-to-noise ratio.
Since the spectroscopic parameters from the different molecules and $^{13}$CO, as well as the critical densities are different, 
it is possible that the different molecules are not tracing the same gas. 
We used the integrated emission in the velocity range from $-140$ to $-20$ km\,s$^{-1}$, which covers the complete velocity 
range corresponding to the GC region since the emission from $-20$ to $20$ \kms  correspond to local gas in the 
line of sight. The PCA is also restricted to an area defined using a mask in the HCN emission at a 14-$\sigma$ level (13.29 K \kms). 
This threshold was chosen to include all the important features visible in the Figs. \ref{inicio} to \ref{final} 
and excluding the regions with low signal-to-noise ratio which can add spurious features in the PCA.

\begin{figure*}
\hspace{-0.5in}
\includegraphics[width=1.1\textwidth]{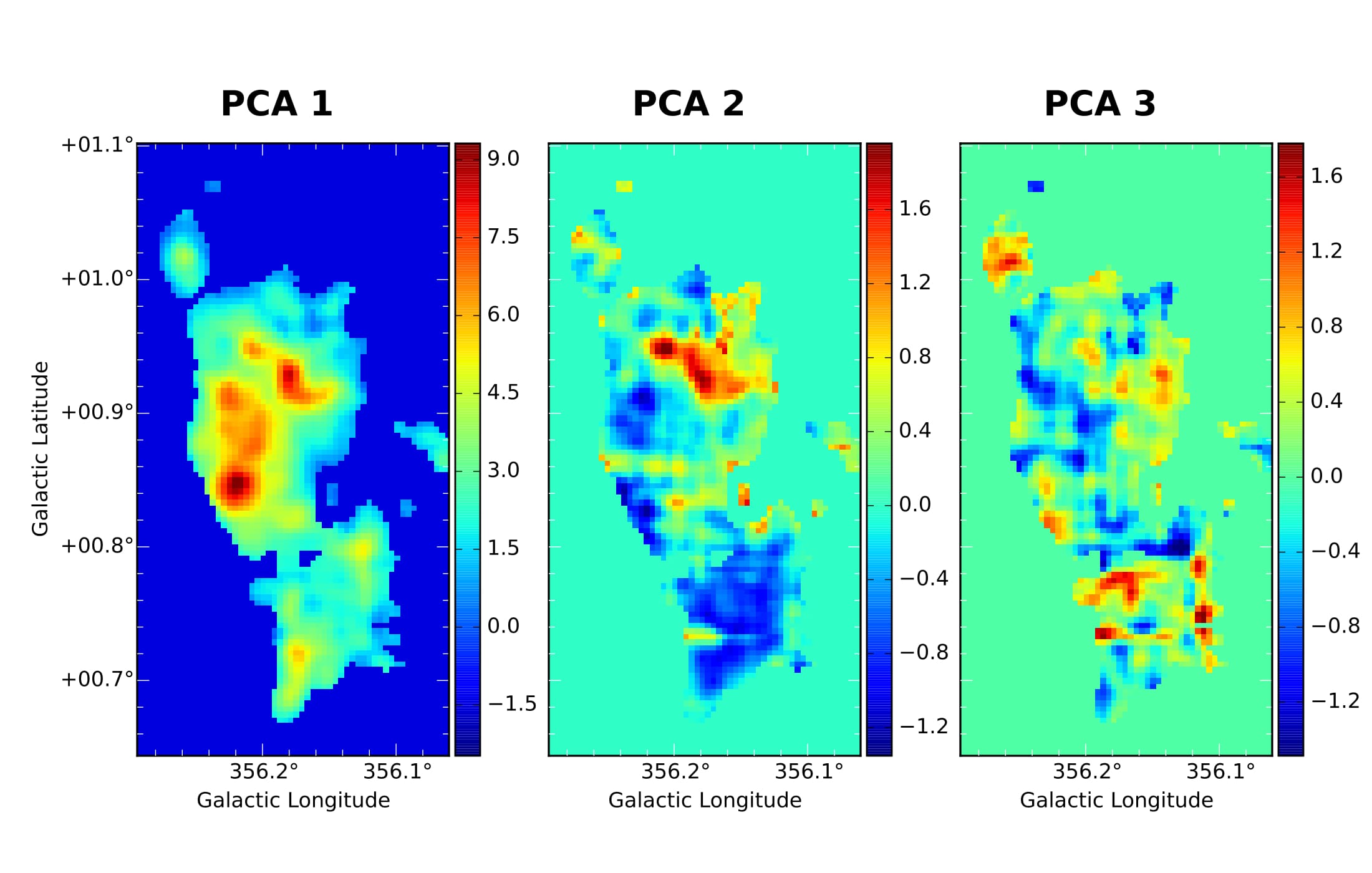}
\caption{Images of the first three PC of the M$-3.8+0.9$ molecular cloud (see text in Section \ref{pcasection}).}
\label{PCA7molecules}
\end{figure*}

\begin{figure*}
\hbox{
\includegraphics[angle=-90,width=0.5\textwidth]{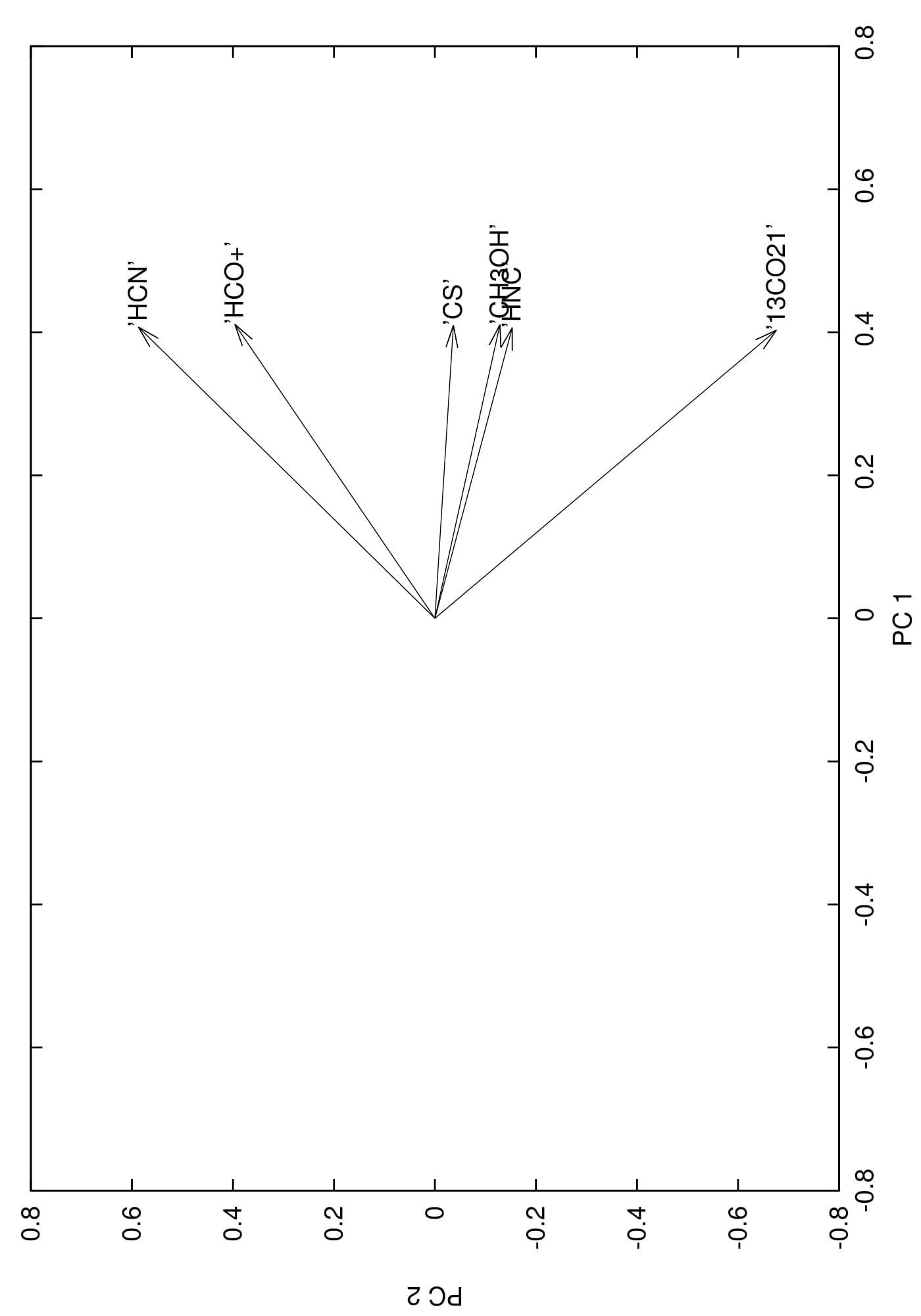}
\includegraphics[angle=-90,width=0.5\textwidth]{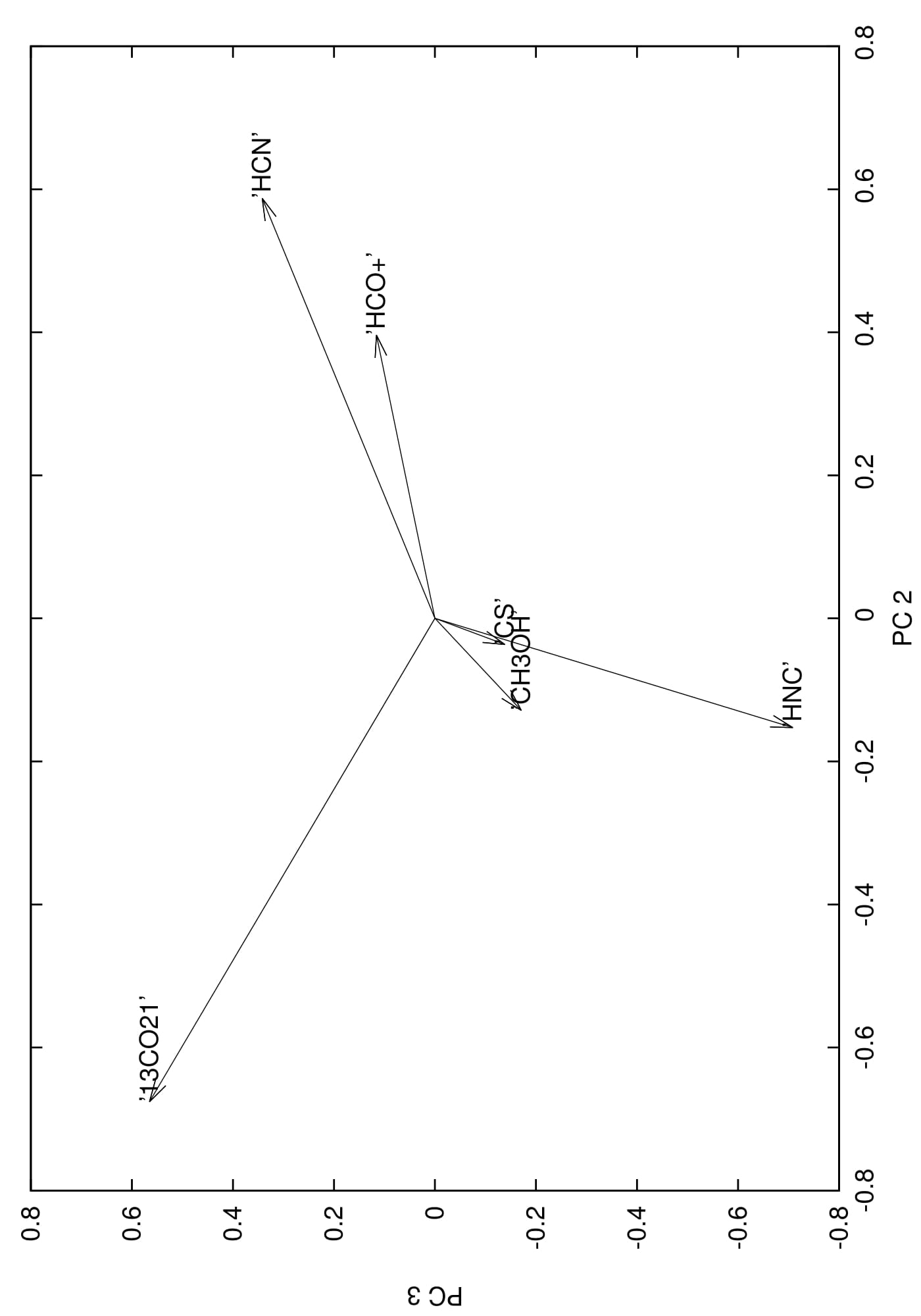}
}
\caption{Plots of eigenvectors of the first three PC. }
\label{plots7molecules}
\end{figure*}

The PCA shows that the emission distribution of all the molecular lines studied are closely related. The first three principal components
describe the 94, 2, and 1.5 per cent of the variance in the data. Fig. \ref{PCA7molecules} shows the results of the PCA.  
The color scale indicates the correlation between the molecules. 
A positive value indicates that the emission from these molecules are correlated and the negative values show an anti-correlation. The first principal 
component axis in Fig. \ref{plots7molecules} shows that all 6 molecules are positively correlated. The PCA1 image in Fig. \ref{PCA7molecules} shows emission common to the six molecular lines with a large fraction of the variance (94 per cent) indicating that all molecular lines are similar distributed throughout the molecular cloud.
The major features are the complex 2 and 3 which show the largest correlations in the cloud. 
The PCA2 image shows the major difference among the six remaining molecules. Despite the value of the variance is small (2\%), 
these differences are still physically significant because the positive features follow very well the Complex 2 which is dominant 
in HCN and HCO$^+$, and the negative features correspond to the north part of the complexes 3, 4, 5, and 6 where $^{13}$CO, HNC, CH$_3$OH, 
and, to to less extent the CS emission are enhanced.  The PCA3 has only a minimum value of the variance (1.5 \%)
and shows smaller variations mainly between the $^{13}$CO and HNC emissions.

\subsubsection{Spectra towards selected positions}
\begin{figure*}
\hspace{-0.5in}
\includegraphics[width=1.0\textwidth]{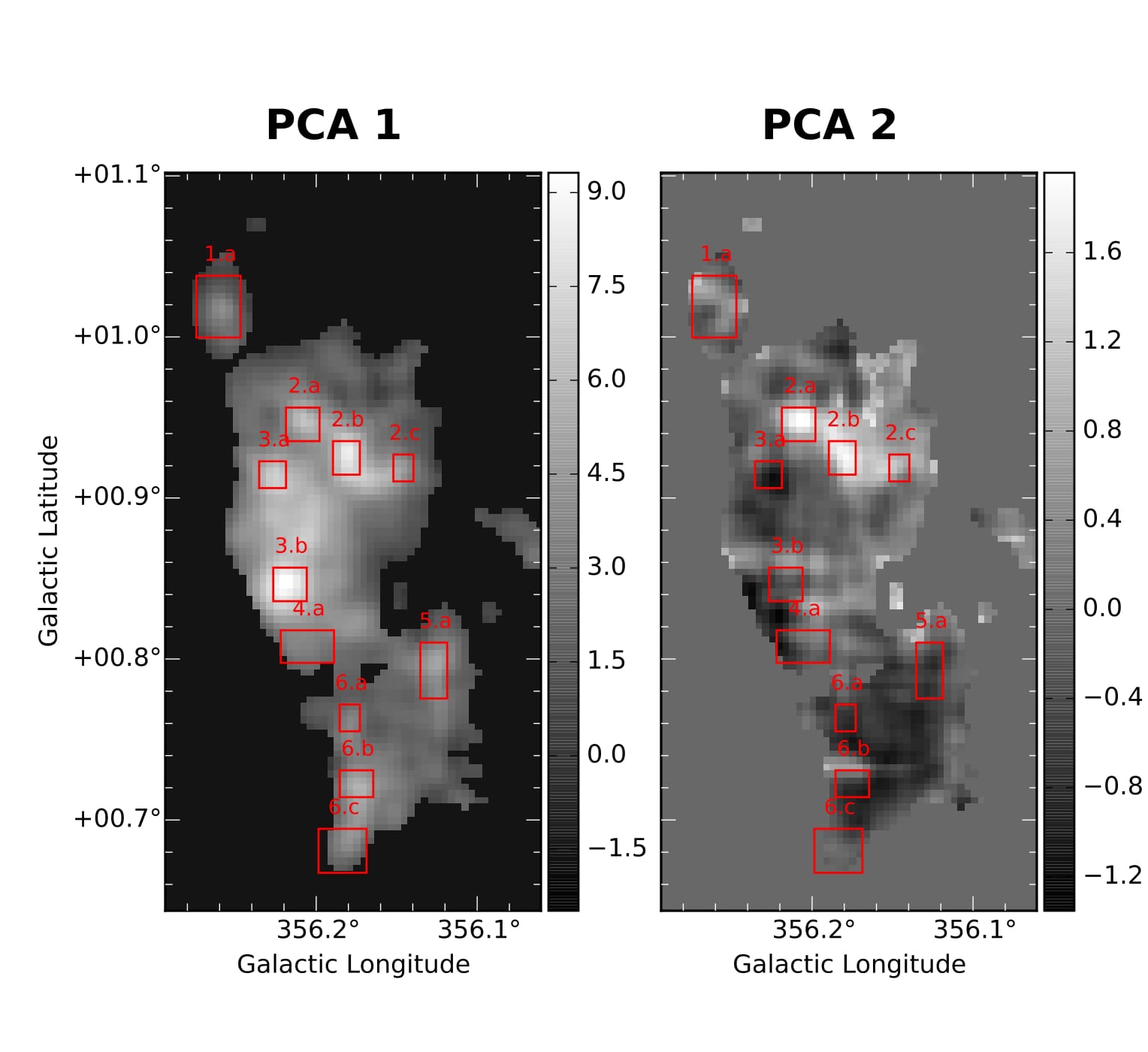}
\caption{Selected regions to extract average spectra and to derive fractional abundances of the detected molecules.}
\label{PCAmoleculesregiones}
\end{figure*}

\begin{figure*}
\vbox{
\hbox{
\includegraphics[angle=0,width=0.3\textwidth]{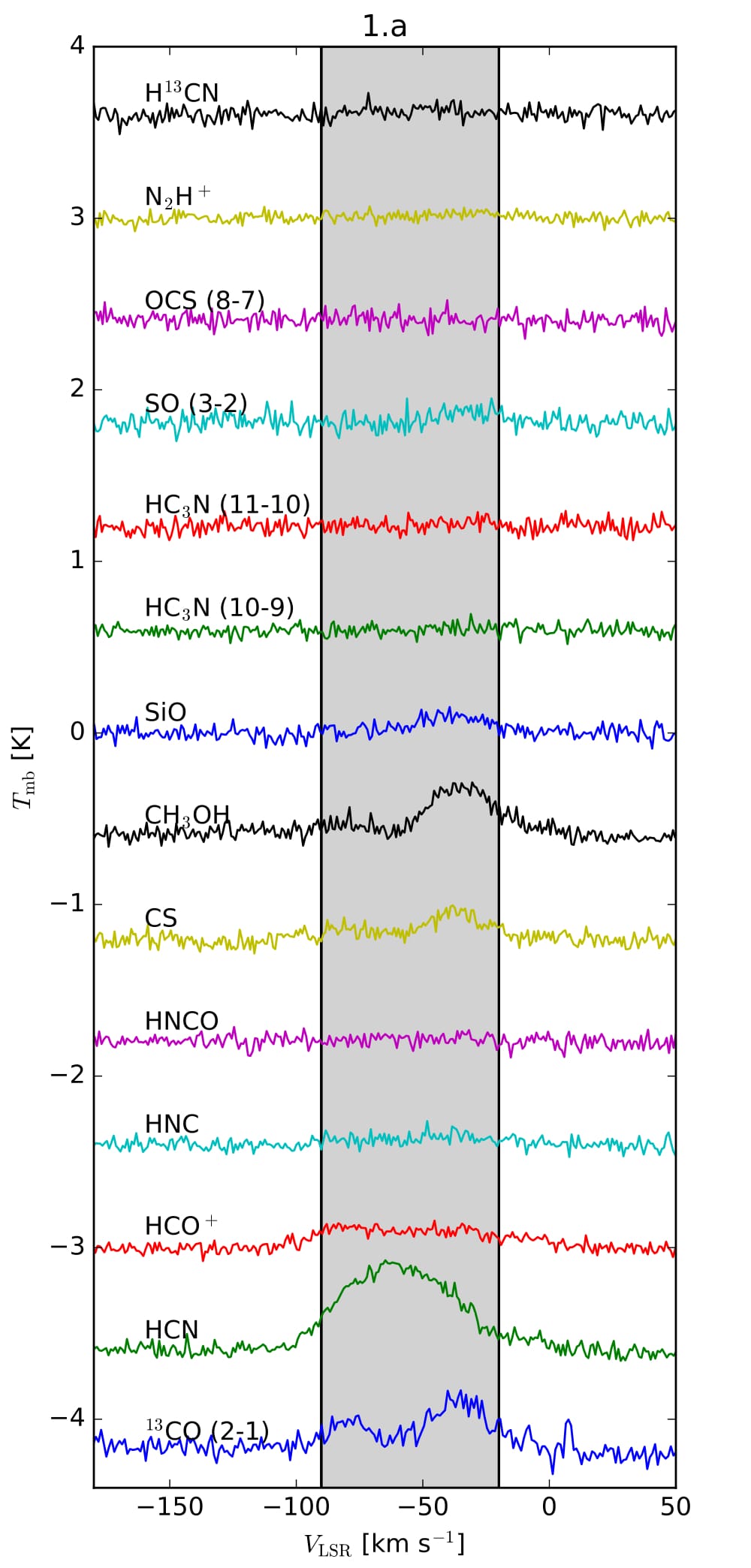}
\includegraphics[angle=0,width=0.3\textwidth]{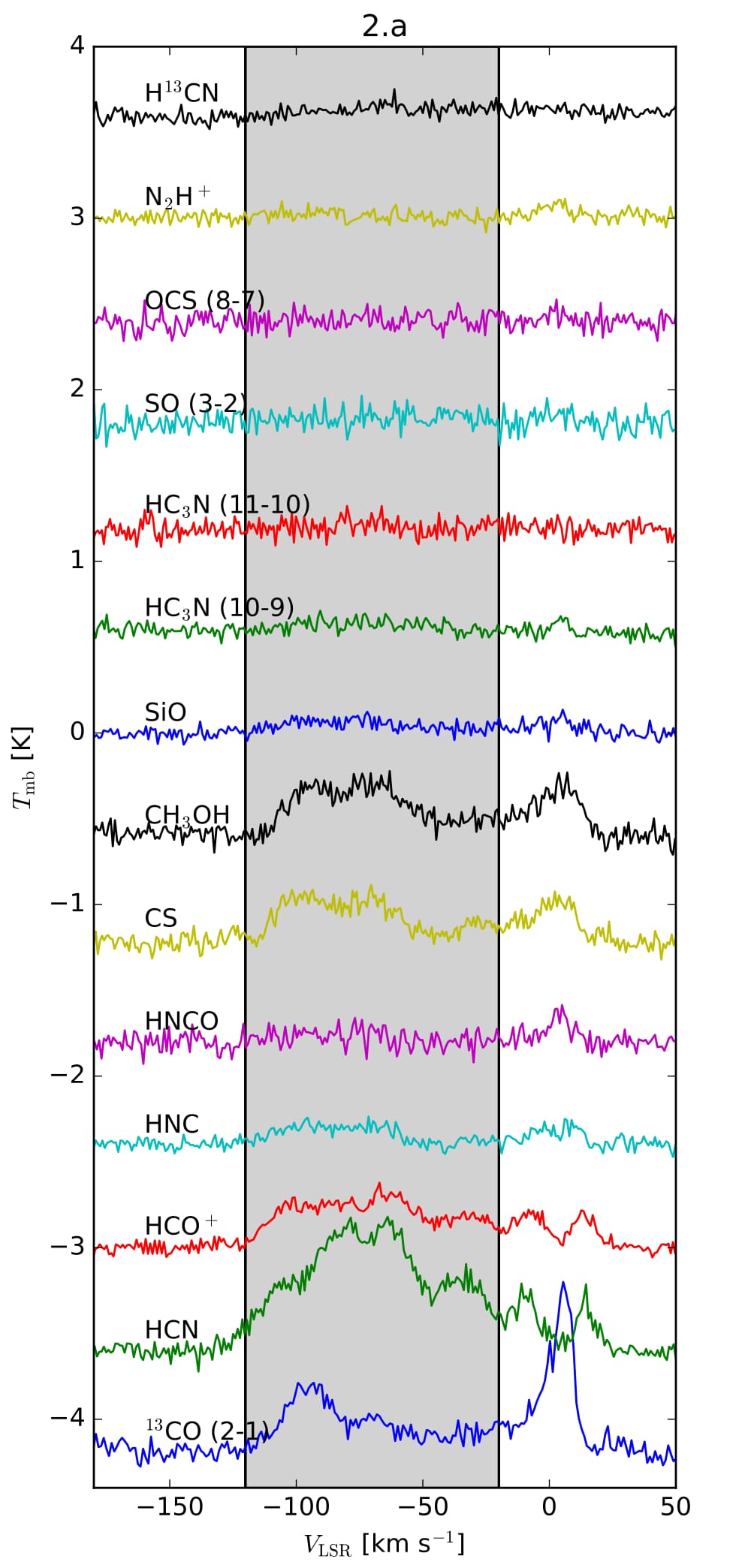}
\includegraphics[angle=0,width=0.3\textwidth]{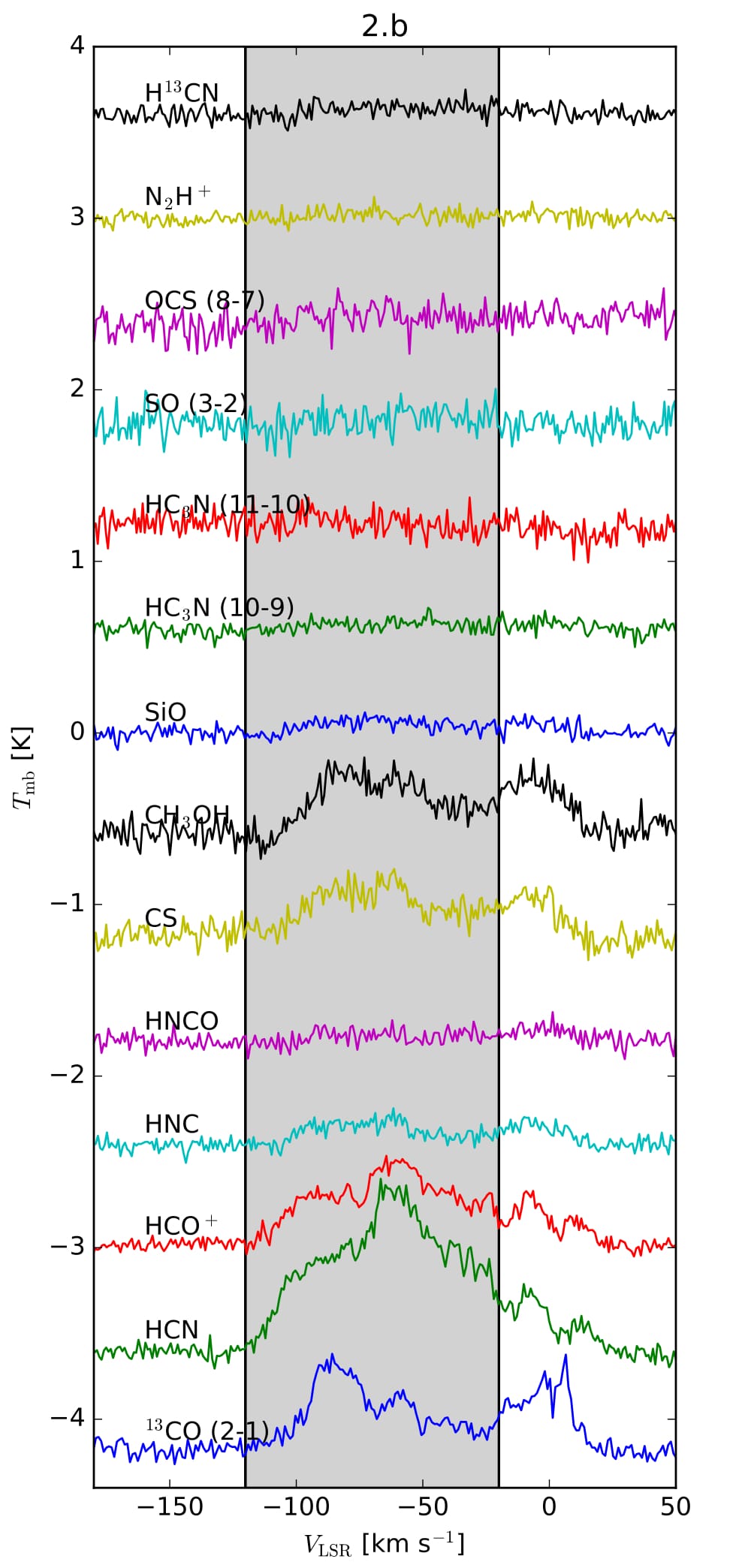}
}
\vspace{0.5in}
\hbox{
\includegraphics[angle=0,width=0.3\textwidth]{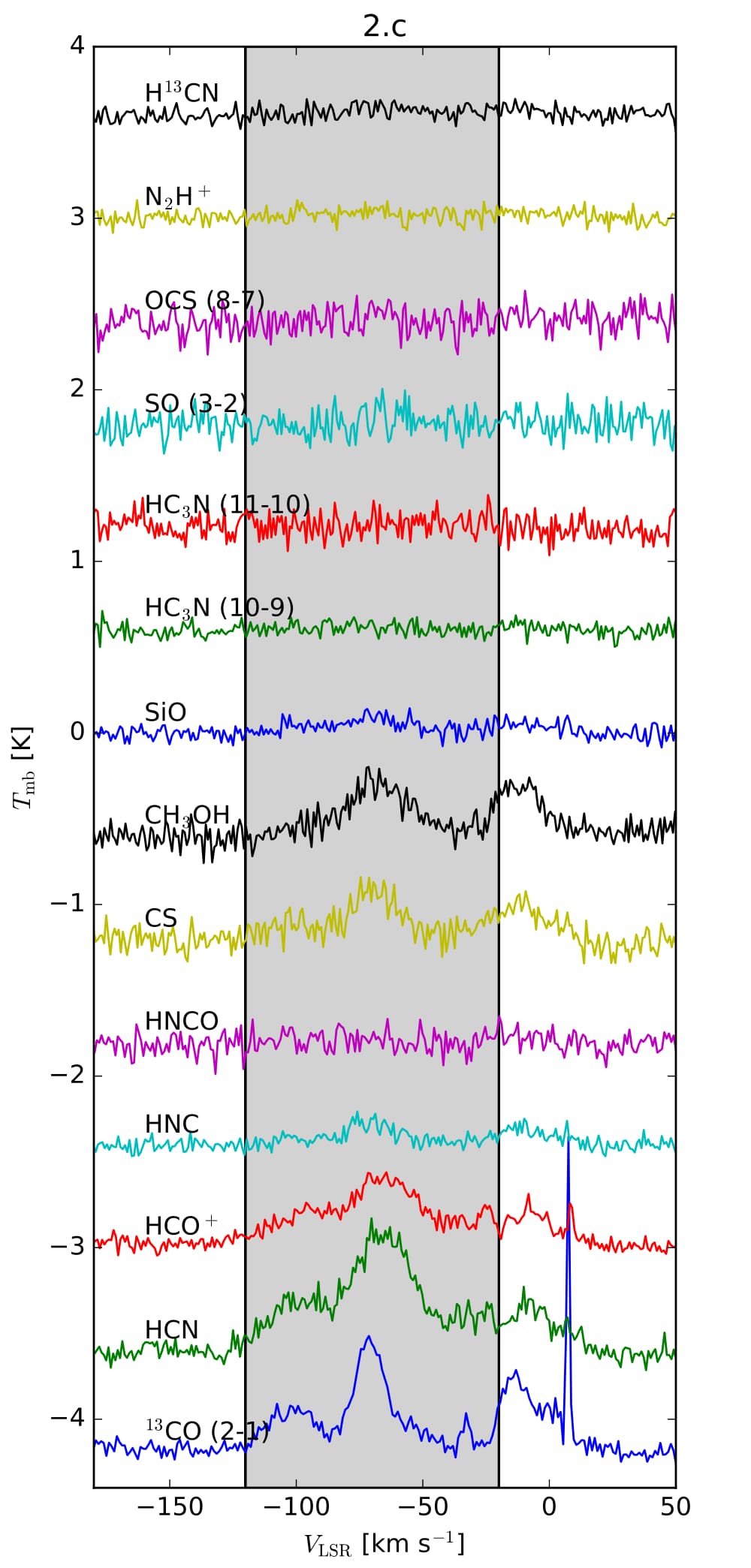}
\includegraphics[angle=0,width=0.3\textwidth]{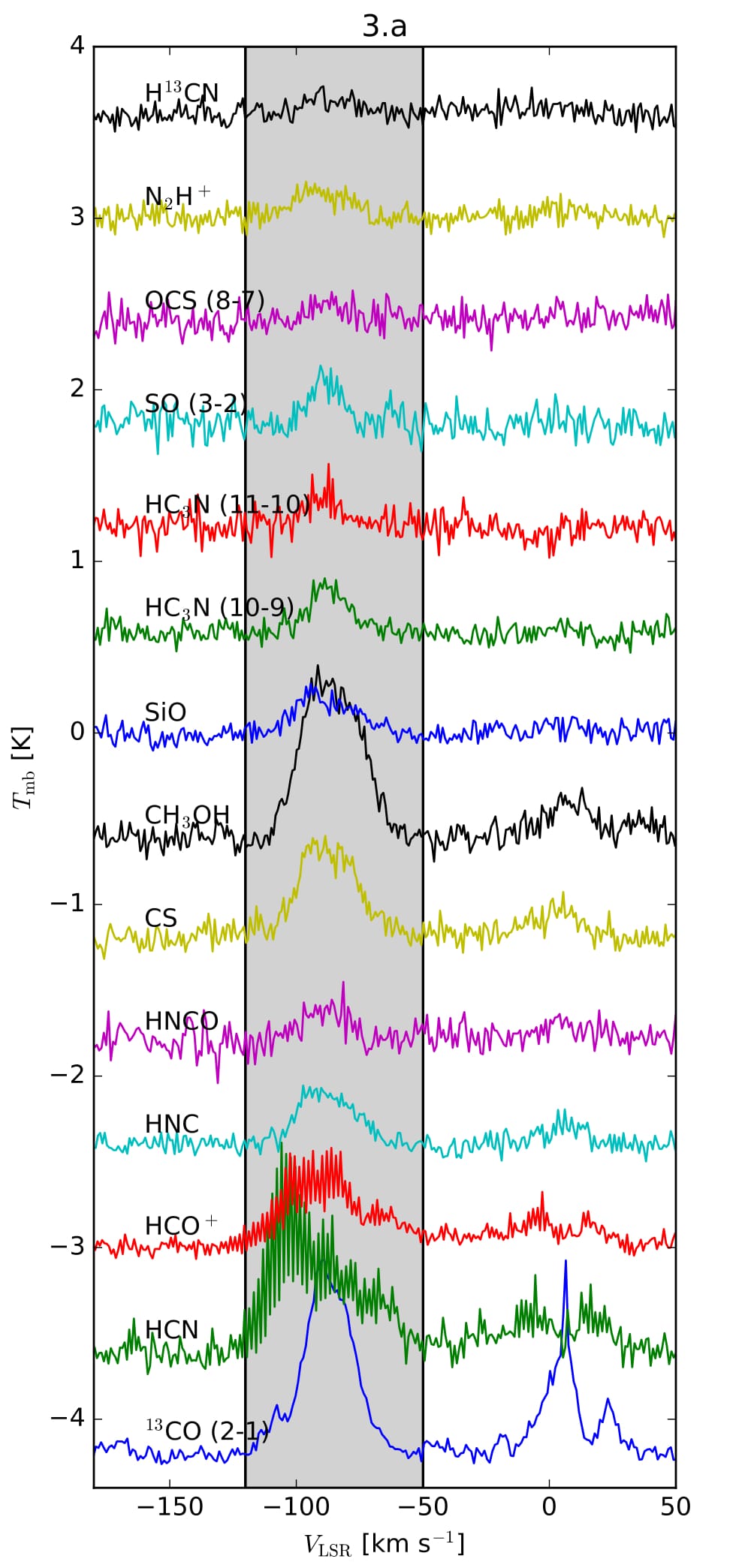}
\includegraphics[angle=0,width=0.3\textwidth]{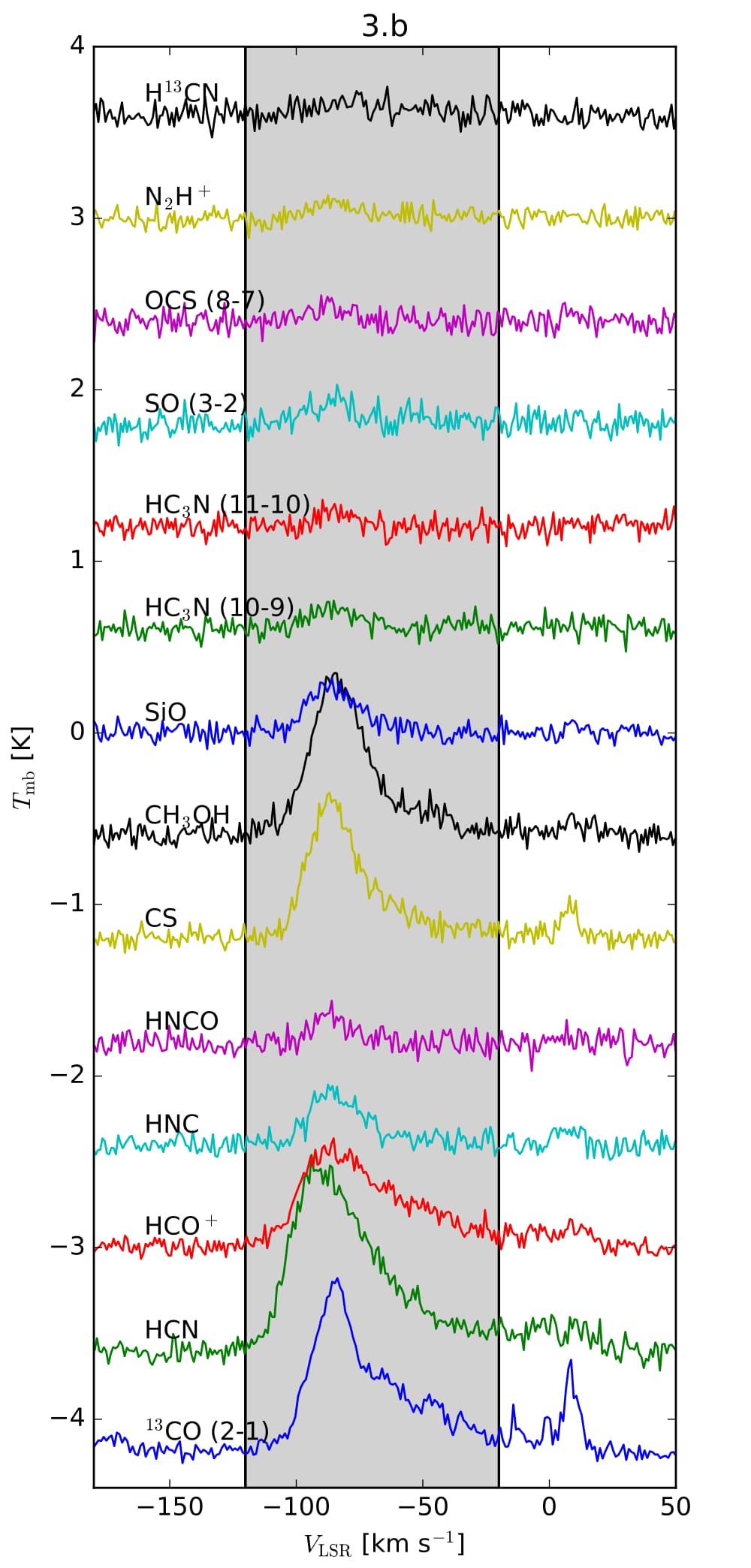}
}
}
\caption{Average spectra for each box shown in Fig. \ref{PCAmoleculesregiones}. $^{13}$CO (2-1), HCO$^+$, HCN, CH$_3$OH, CS spectra are scaled by a factor 0.5. }
\label{espectraregiones1}
\end{figure*}

\begin{figure*}
\vbox{
\hbox{
\includegraphics[angle=0,width=0.3\textwidth]{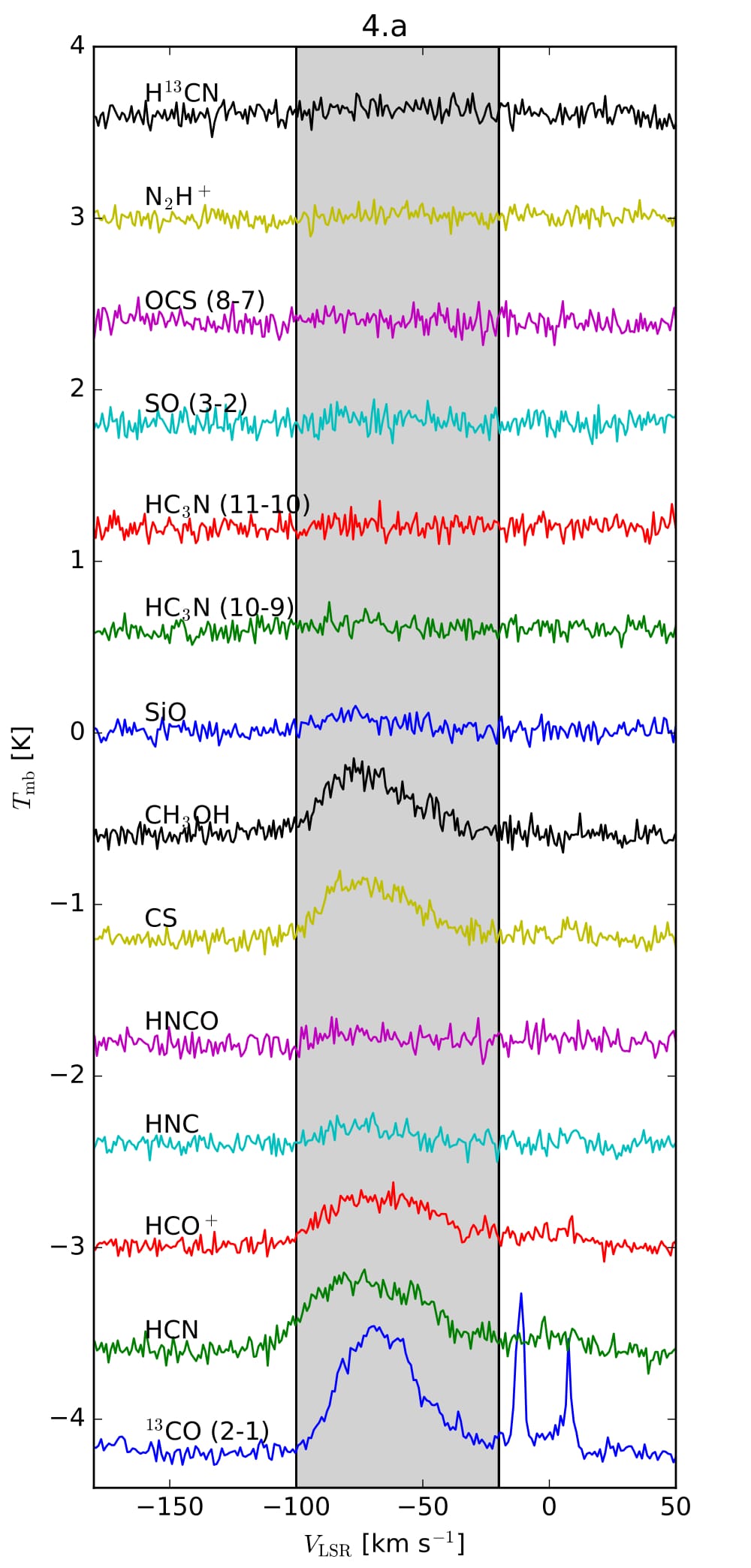}
\includegraphics[angle=0,width=0.3\textwidth]{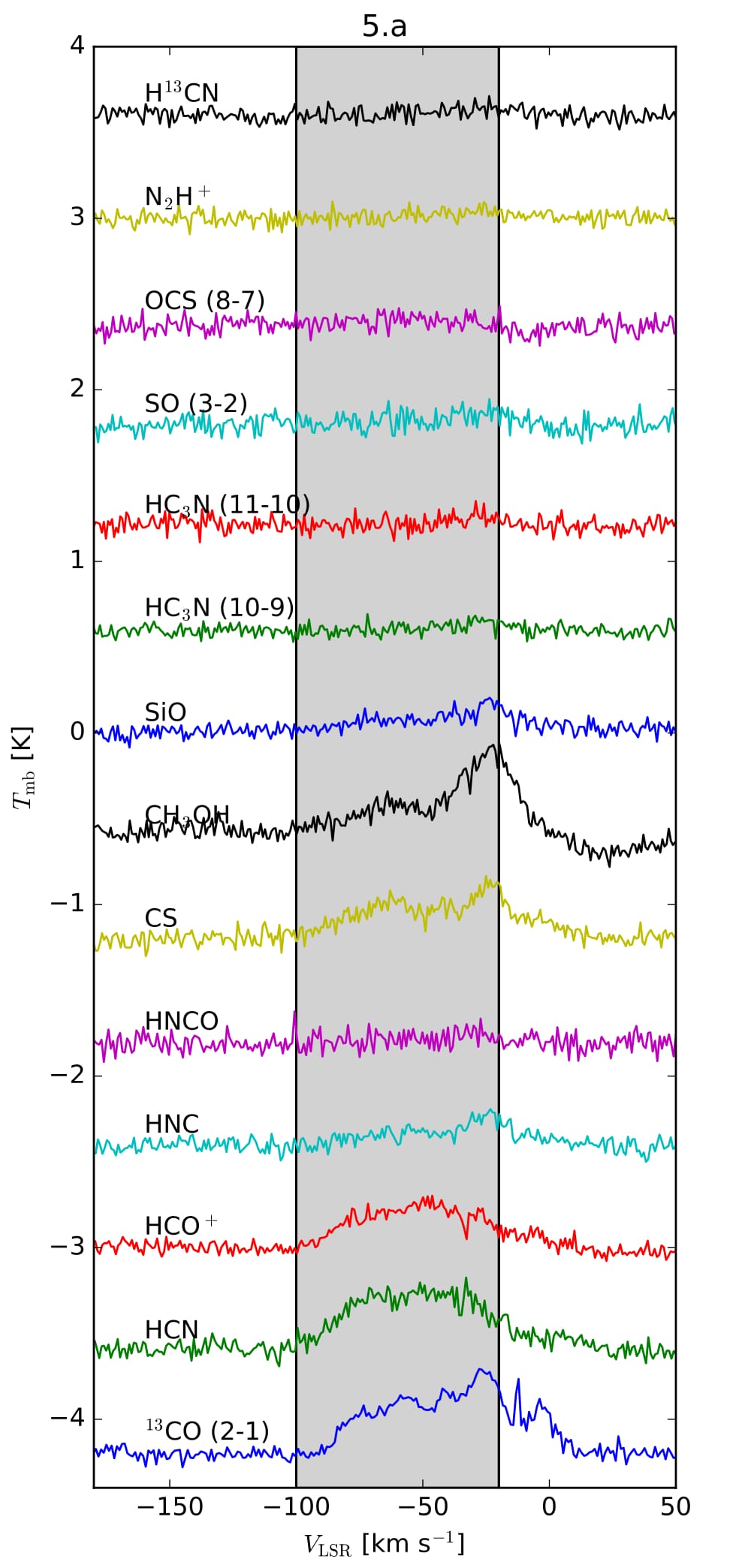}
\includegraphics[angle=0,width=0.3\textwidth]{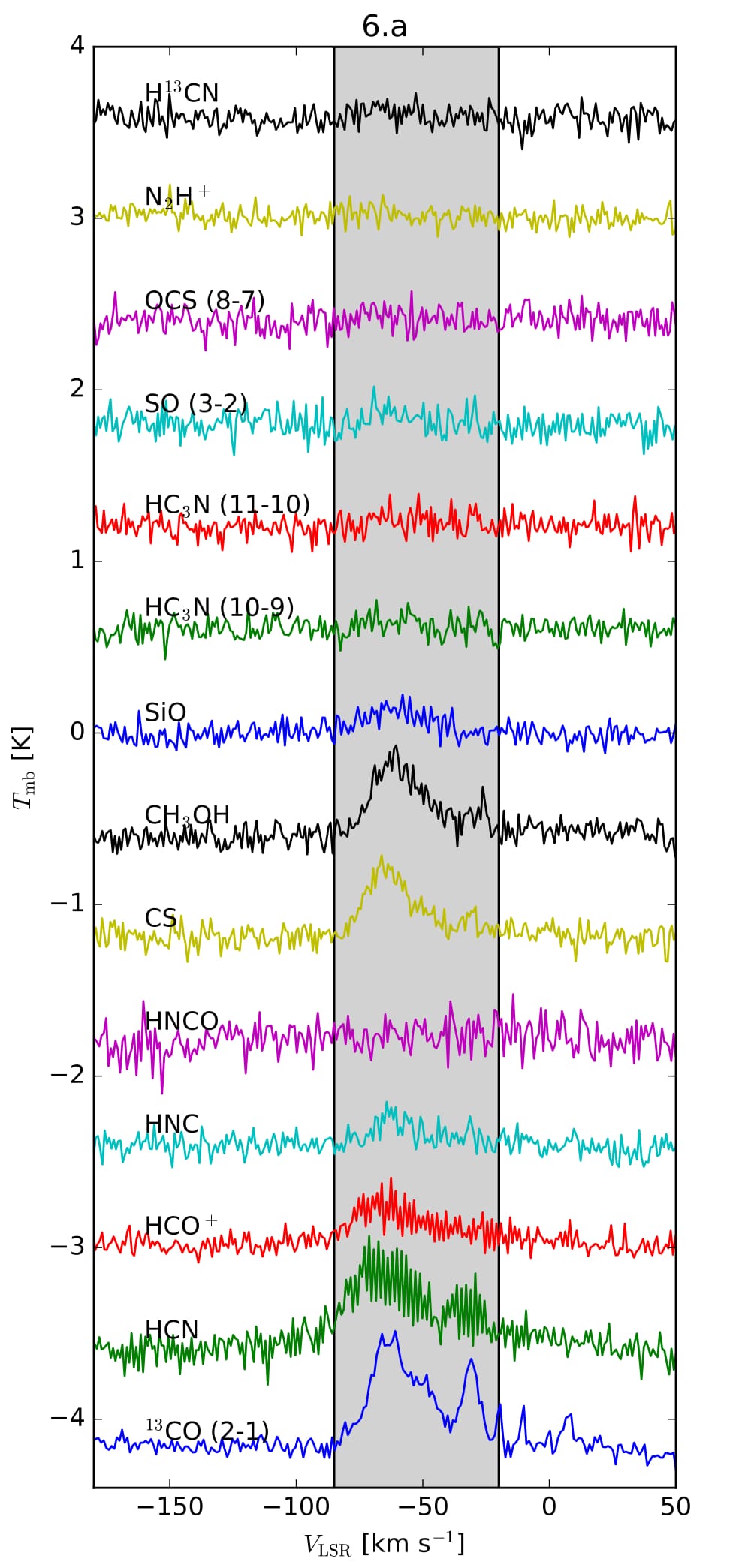}
}
\vspace{0.5in}
\hbox{
\includegraphics[angle=0,width=0.3\textwidth]{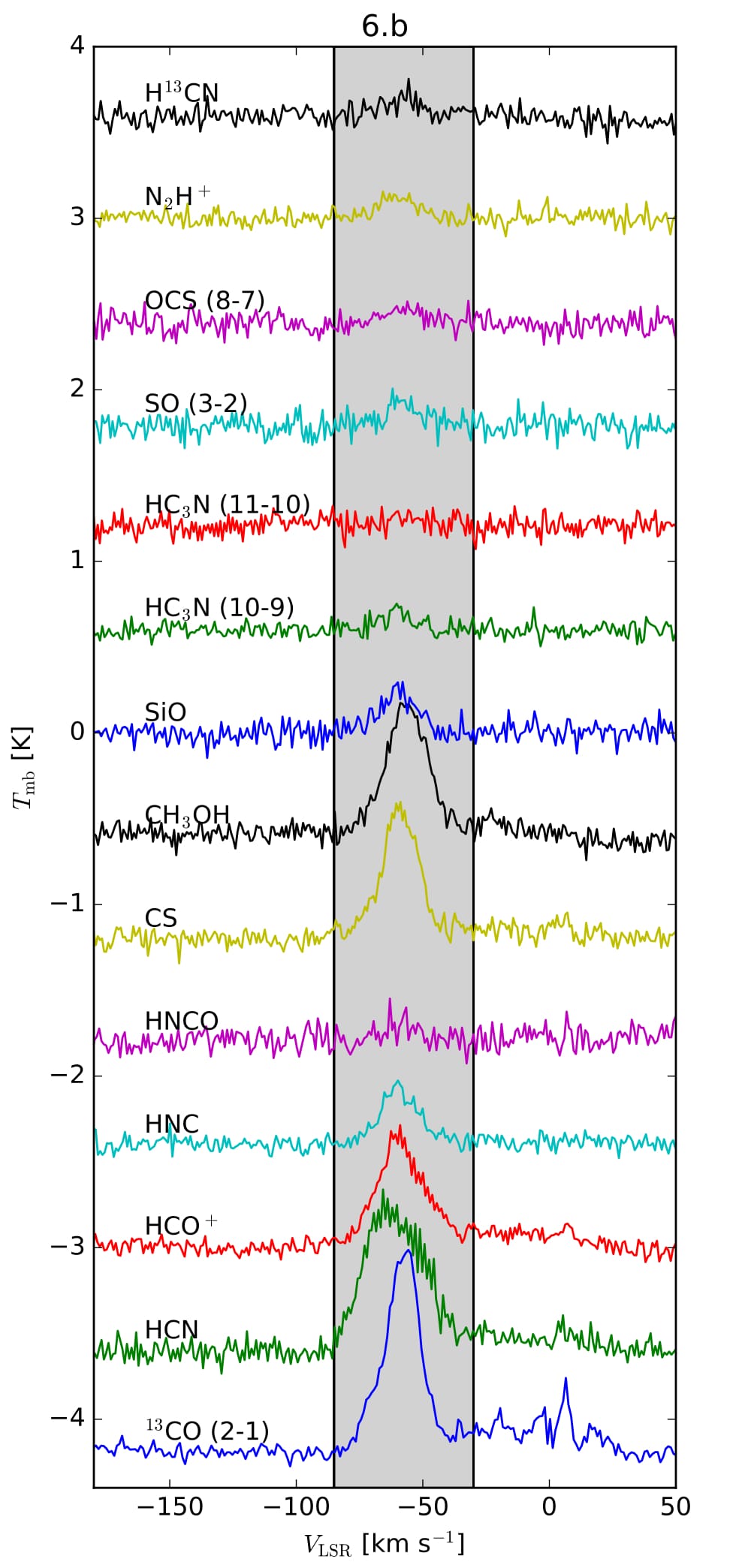}
\includegraphics[angle=0,width=0.3\textwidth]{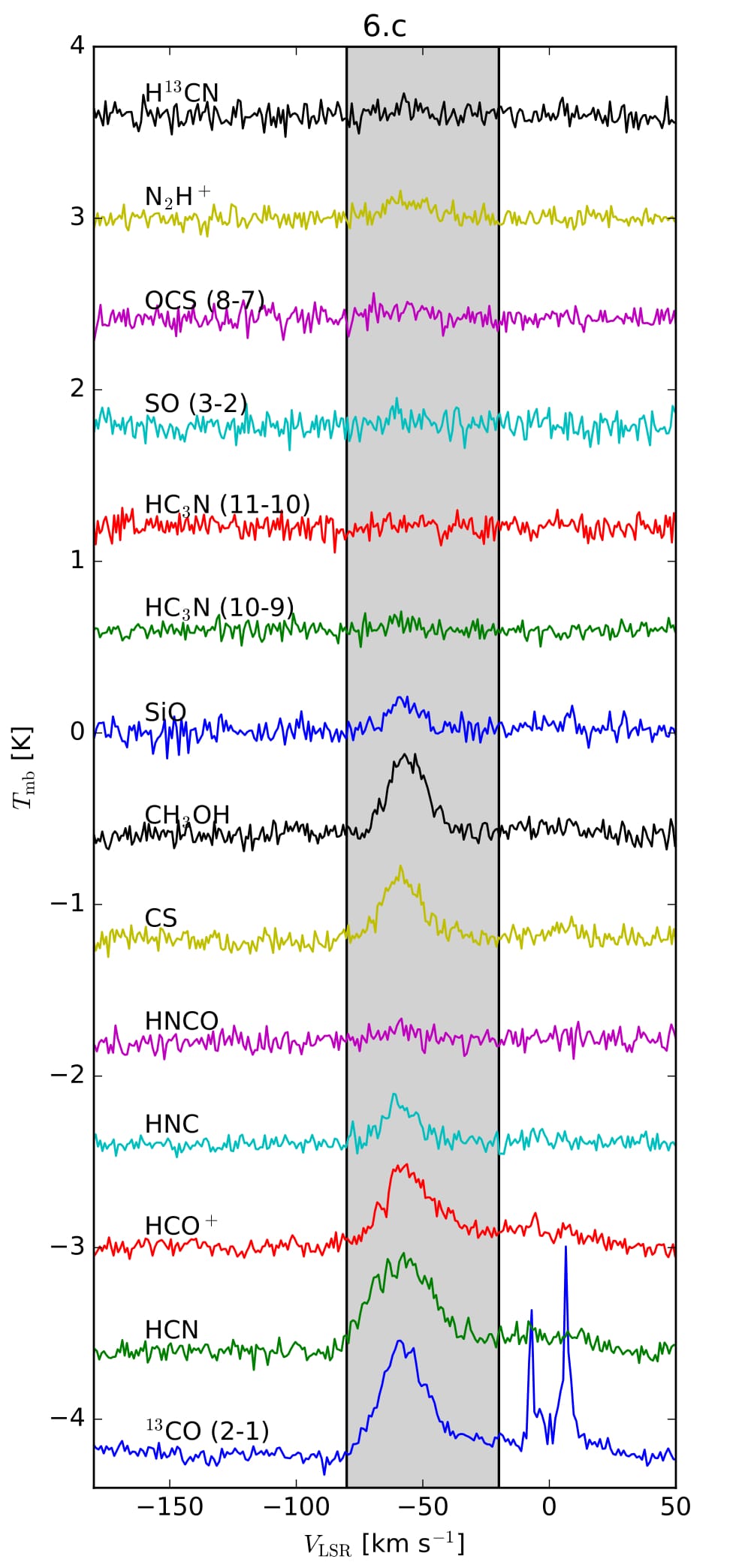}
}
}
\caption{Average spectra for each box shown in Fig. \ref{PCAmoleculesregiones}. }
\label{espectraregiones2}
\end{figure*}

Using the images of the different PCA previously discussed, we select 11 positions which have an intensity peak in the PCA1, which corresponds to the regions where the most intense molecules are correlated, and peaks in the PCA2 which corresponds to regions with the largest anti correlation between them. We selected at least one position per complex (complex 1 to 6 as defined in Section \ref{morphologykinematics}), then for example, the position chosen towards the complex 4 is not spotlighted in any of the PCA results. Because of the high correlation between the molecular transitions, this selection is equivalent to select the peaks in  any of the 6 molecules considered in the PCA analysis. We obtain the integrated spectra in the boxes shown in Fig. \ref{PCAmoleculesregiones} in all the detected molecules 
(Figs. \ref{espectraregiones1} and \ref{espectraregiones2}). From those figures, we can see the complex velocity structure in the M$-3.8+0.9$ molecular cloud, with large velocity widths in all the selected regions.
Most of the molecules show a similar line profile within each region, with the clear exception of HCN, which in some regions has the intensity peak at a very different velocity. A clear example is shown in positions 1.a, 2.a, 2.b, and 5.a, with the most extreme case in position 1.a  where the profile of HCN differs completely from the one shown by the other molecules. The differences in the intensity peak of HCN could be explained by opacity effects since the N(HCN)/N(H$^{13}$CN) ratio range from 6 to 21 which indicates that the HCN is optically thick. 
Position 3.b shows a characteristic shock profile \citep[see e.g., ][]{Jimenez-Serra_et_al_2008}, with a prominent wing at the redshifted part of the spectra. The strong shock observed in this position produces a gas acceleration up to 50 km s$^{-1}$.
As a good tracer of the column density, the $^{13}$CO spectra also clearly show  the local gas in the line of sight as narrow emission at $\sim 0$ km s$^{-1}$, which is not seen in other molecules. This narrow emission is superposed to a broader one which corresponds to gas in the GC.   

\subsubsection{Molecular abundances across the M$-3.8+0.9$ molecular cloud} 
The column density for all the detected molecules was derived as indicated in Section 3.2. 
Because we only have one transition for each molecule, we assume LTE at a $T_{\rm{ex}}=10$ K. If we assume a $T_{\rm{ex}}$ of, e.g., 24 K \citep{Jones_et_al_2012} column density varies a factor $\sim 1.5$ in most of the molecules considered here, as can be seen in Fig. \ref{variationTex} with the exception of CH$_3$OH which varies a factor of 2.5, and HC$_3$N (with a factor of 0.4).  
The molecular parameters are taken from the CDMS catalog for all the molecules with the exception of CH$_3$OH which are taken from the Jet Propulsion Laboratory (JPL) catalog \citep{Picket_et_al_1998}. 
The column density is derived for the 11 regions defined in Fig. \ref{PCAmoleculesregiones} for the velocity range shown in the shadow rectangle in Fig. \ref{espectraregiones1} and \ref{espectraregiones2}, and in the Table \ref{columndensity_completerange}. The velocity range for each position was chosen to include all the emission for the M$-3.8+0.9$ cloud avoiding the possible contamination from local gas between $-20$ to $20$ \kms.
Table \ref{columndensity_completerange} shows the results. HC$_3$N has two detected rotational transitions. This allows to better estimate the column density and rotational temperature using rotational diagrams \citep{Goldsmith_Langer_1999}. However, because the energy involved of both transitions are very close, the rotational temperature is poorly constrained, and only positions 3.a and 3.b have adequate signal-to-noise ratio in both transitions. Therefore, to derive the column density of HC$_3$N, we used only the (10-9) transition (which is the one with the better signal-to-noise ratio) assuming a temperature of 10 K like with the others species in this work. 

\begin{figure*}
\hbox{
\includegraphics[angle=90,width=0.27\textwidth]{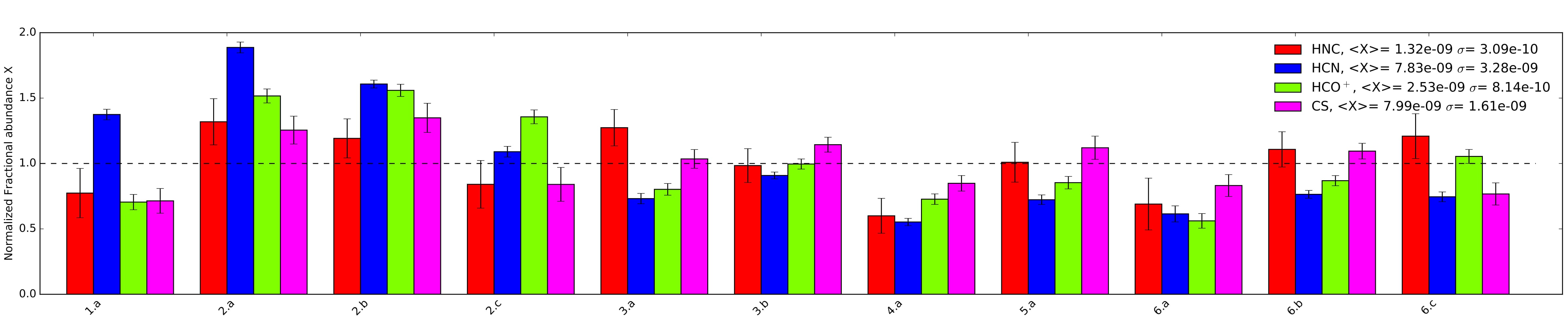}
\includegraphics[angle=90,width=0.27\textwidth]{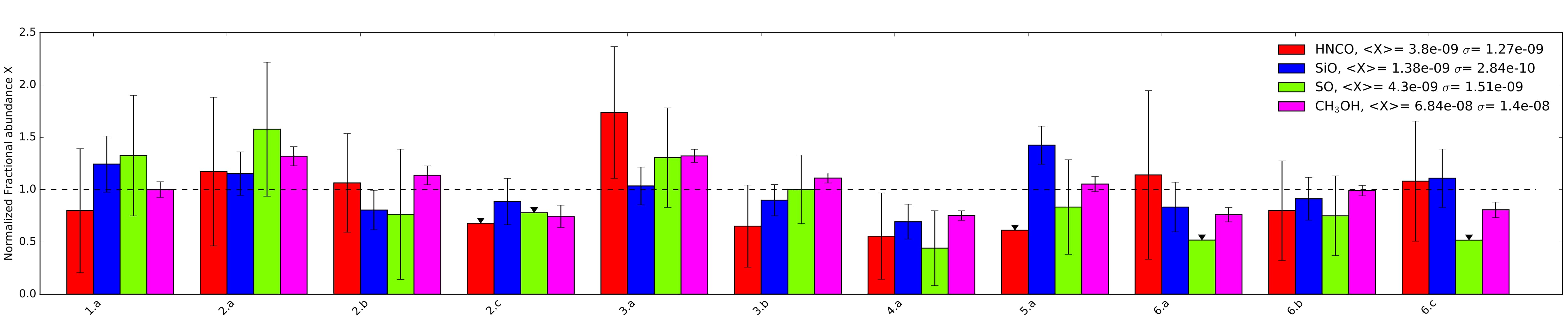}
\includegraphics[angle=90,width=0.27\textwidth]{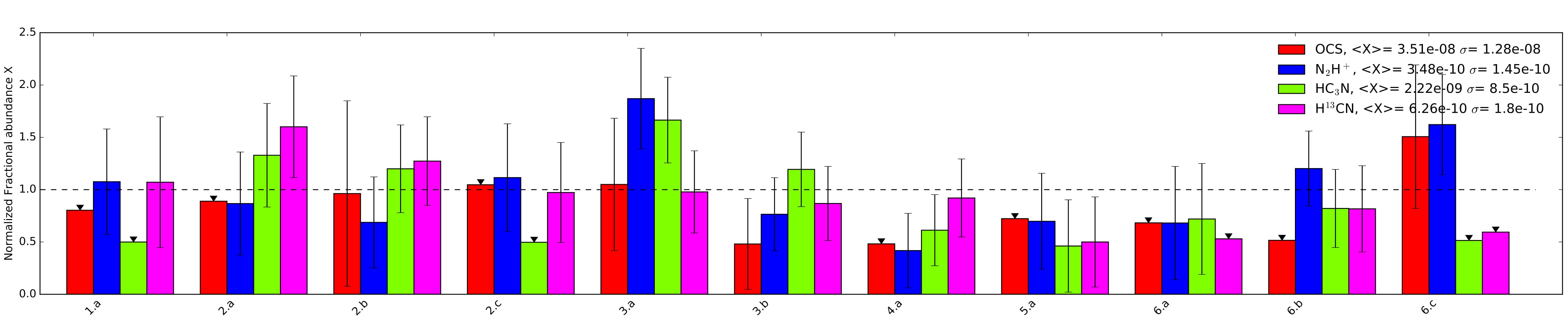}
}
\caption{Molecular fractional abundance ($X({\rm mol})=N({\rm mol})/N({\rm H_2})$ normalized by the average value for each molecule. The label indicates the average value and the standard deviation for each molecule. We include all detected molecules shown in Table \ref{table:1}. }
\label{fractionalabundancetotal}
\end{figure*}

To compare the emission from the different positions, we plot the fractional abundances ($X({\rm mol})=N({\rm mol})/N({\rm H_2})$, where \emph{mol} corresponds to every detected molecule) normalized by the average value for each molecule. The fractional abundance is computed for the velocity range shown in Table \ref{columndensity_completerange}, and Figs. \ref{espectraregiones1} and \ref{espectraregiones2}. The plots show in the upper right corner, the average value and the standard deviation of the fractional abundance for each molecule, which is computed using only detected emission and not the upper limits. The error bars corresponds to the estimated 3-sigma uncertainties. Despite that the PCA analysis showed the large correlation between the most intense molecules across the M$-3.8+0.9$ molecular cloud, Fig. \ref{fractionalabundancetotal} shows that there are significant differences ($3-\sigma$) between the abundance of different molecules in the selected regions. The fractional abundances are shown in Table \ref{table:Xcompletevelocity}.

\section{Discussion \label{discussion}}

\subsection{Comparison between the molecular fractional abundances in the foot points and in the CMZ \label{comparisonCMZ}}
The fractional abundances derived toward the foot points of the giant molecular loops are compared with those derived in the CMZ (Table \ref{table:XCMZ}). In order to obtain a set of fractional abundances values for the CMZ as homogeneous as the one presented in this work, we used the Mopra 3-mm data cubes from \citet{Jones_et_al_2012} to derive the molecular column densities, and for estimating the H$_2$ column density, we used the ``APEX CMZ SHFI-1 survey'' $^{13}$CO (2-1) data cube from \citet[][available at \url{http://doi.org/10.7910/DVN/27601}]{Ginsburg_et_al_2016}. Since the $^{12}$C/$^{13}$C isotopic value in the CMZ is 24 \citep{Langer_Penzias_1990}, we use a conversion factor [$^{13}$CO/H$_2$] of $5\times 10^{-6}$]. 
We select representative clouds in the CMZ shown in  Fig. \ref{coverturaCMZ}. The regions were chosen to be of similar size in Fig. \ref{PCAmoleculesregiones} $(75''x75'')$. The column densities are derived assuming LTE with an excitation temperature of 10 K, using Eq. \ref{formula1} for the optically thin emission; and for the optically thick emission (HCN, HCO$^+$ and HNC), we use the equations in \citet{Jones_et_al_2012}, computing the optical depth along the velocity axis in the same way than in \citet{Jones_et_al_2012}. To assume that the $^{13}$CO emission is optically thin is a reasonable approximation as shown by \citet{Rodriguez-Fernandez_et_al_2001} from $^{13}$CO and C$^{18}$O  (J=1-0, 2-1) observations in many sources widespread in the CMZ and also in the clump 2 \citep{Bania_et_al_1986}.  We did not use the $T_{\rm ex}=24$ K because \citet{Jones_et_al_2013} found that the excitation temperature for the lower rotational transitions of the molecules that they have 3-mm and 7-mm transitions observations available (SiO, HNCO, HC$_3$N, $^{13}$CS, HOCO$^+$) was between 2 to 9 K, concluding that the excitation temperature have to be much lower than the kinetic temperature of 30 K. To derive the column density of CS, we use their optically thin isotopomer $^{13}$CS and a assumed $^{12}$C/$^{13}$C isotopic value of 24. The SO emission from \citet{Jones_et_al_2012} is not included because they observed a transition not detected by us, and because they only detected it clearly in Sgr\,B2. Using upper limit information is not meaningful because of the poor baselines of the spectra in the data cube. 
Table \ref{table:XCMZ} also includes the fractional abundances derived in selected line-of-sights toward the CMZ from recent publications covering most of the species observed in this work.  
The ratio of the fractional abundances between the CMZ (\citet{Jones_et_al_2012} data) and the GMLs range from a factor 1.1 up to 5.1 for the different molecules. The largest differences are found for HNCO, N$_2$H$^+$, HNC. The smallest differences are found for SiO, with a factor of only $1.1$, which indicate similarities in the chemistry of this molecule in the CMZ and in the foot points. If we consider the same conversion factor between the CMZ and the M-3.8+0.9 cloud (see Section \ref{H2colum}) the ratio of the fractional abundances between the CMZ and the GMLs is 0.5 up to 2 for the different molecules.

\begin{sidewaystable}
\caption{Fractional abundances of all detected molecules in the selected regions defined in Fig. \ref{PCAmoleculesregiones} integrated over the complete velocity range showed in Table \ref{columndensity_completerange}. }
\label{table:Xcompletevelocity}
\centering 
\begin{tabular}{lccccccccccl}
\hline\hline
Molecule     & 1.a& 2.a & 2.b & 2.c & 3.a & 3.b & 4.a & 5.a & 6.a & 6.b & 6.c \\  
             & [x 10$^{-9}$] & [x 10$^{-9}$]& [x 10$^{-9}$]& [x 10$^{-9}$]&[x 10$^{-9}$] &[x 10$^{-9}$] &[x 10$^{-9}$] &[x 10$^{-9}$] & [x 10$^{-9}$]& [x 10$^{-9}$] & [x 10$^{-9}$] \\\hline
H$^{13}$CN &  0.67 (0.13) &  1.00 (0.10) &  0.80 (0.09) &  0.61 (0.10) &  0.61 (0.08) &  0.54 (0.07) &  0.58 (0.08) &  0.31 (0.09) &  0.33        &  0.51 (0.09) &  0.37        \\
SiO        &  1.72 (0.12) &  1.59 (0.10) &  1.11 (0.09) &  1.22 (0.10) &  1.43 (0.08) &  1.24 (0.07) &  0.96 (0.08) &  1.97 (0.08) &  1.15 (0.11) &  1.26 (0.09) &  1.53 (0.13) \\
HNCO       &  3.04 (0.75) &  4.46 (0.90) &  4.04 (0.60) &  2.58        &  6.60 (0.80) &  2.48 (0.50) &  2.11 (0.52) &  2.33        &  4.33 (1.02) &  3.03 (0.60) &  4.11 (0.73) \\
HCN        & 10.77 (0.10) & 14.79 (0.11) & 12.59 (0.08) &  8.54 (0.11) &  5.73 (0.10) &  7.12 (0.07) &  4.33 (0.07) &  5.66 (0.10) &  4.82 (0.16) &  5.99 (0.08) &  5.84 (0.10) \\
HCO$^+$    &  1.78 (0.05) &  3.83 (0.04) &  3.94 (0.04) &  3.43 (0.04) &  2.03 (0.04) &  2.52 (0.03) &  1.84 (0.03) &  2.16 (0.04) &  1.42 (0.05) &  2.19 (0.03) &  2.66 (0.04) \\
HNC        &  1.02 (0.08) &  1.74 (0.08) &  1.57 (0.07) &  1.11 (0.08) &  1.68 (0.06) &  1.30 (0.06) &  0.79 (0.06) &  1.33 (0.07) &  0.91 (0.09) &  1.46 (0.06) &  1.60 (0.08) \\
HC$_3$N    &  1.11        &  2.94 (0.37) &  2.66 (0.31) &  1.10        &  3.69 (0.30) &  2.64 (0.26) &  1.36 (0.25) &  1.02 (0.33) &  1.59 (0.39) &  1.82 (0.28) &  1.14        \\
N$_2$H$^+$ &  0.37 (0.06) &  0.30 (0.06) &  0.24 (0.05) &  0.39 (0.06) &  0.65 (0.06) &  0.27 (0.04) &  0.15 (0.04) &  0.24 (0.05) &  0.24 (0.06) &  0.42 (0.04) &  0.57 (0.06) \\
CH$_3$OH   & 68.44 (1.70) & 90.26 (2.09) & 77.77 (2.05) & 50.9 (2.42) & 90.4 (1.44) & 75.9 (1.08) & 51.4 (1.05) & 72.0 (1.63) & 52.0 (1.55) & 67.79 (1.13) & 55.2 (1.68) \\
OCS        & 28.16        & 31.20        & 33.76 (10.3) & 36.7        & 36.8 (7.39) & 16.8 (5.07) & 16.9        & 25.3        & 23.9        & 18.08        & 52.8 (8.02) \\
CS         &  5.71 (0.25) & 10.04 (0.28) & 10.79 (0.30) &  6.72 (0.35) &  8.27 (0.19) &  9.15 (0.15) &  6.78 (0.16) &  8.96 (0.24) &  6.65 (0.22) &  8.75 (0.16) &  6.13 (0.23) \\
SO         &  5.69 (0.83) &  6.78 (0.92) &  3.28 (0.89) &  3.35        &  5.61 (0.68) &  4.31 (0.47) &  1.89 (0.51) &  3.58 (0.65) &  2.23        &  3.23 (0.55) &  2.23        \\
\\ \hline \hline %
\end{tabular}
\tablefoot{}
\end{sidewaystable}

\begin{table*}
\caption{Fractional abundances for the CMZ from previous works}
\label{table:XCMZ}
\centering 
\begin{tabular}{lcccccccl}
\hline\hline
Molecule     &1.6 complex$^a$&1.3 complex$^a$ & Sgr\,B2$^a$& G0.4$^a$ & G0.25$^a$ & 50km/s$^a$ & 20km/s$^a$ & Sgr\,C$^a$\\  
             & (20,210)$^e$ &(40, 230)$^e$ &(20, 140)$^e$ &(-20,60)$^e$ & (5,100)$^e$&(0,100)$^e$ & (-20, 80)$^e$&(-80,-5)$^e$ \\\hline
             &[x 10$^{-9}$] &[x 10$^{-9}$] &[x 10$^{-9}$] & [x 10$^{-9}$]& [x 10$^{-9}$]& [x 10$^{-9}$]&[x 10$^{-9}$] &[x 10$^{-9}$] \\ \hline
H$^{13}$CN   &  1.11 (0.06) &  1.34 (0.03) &  0.66 (0.02) &  1.02 (0.03) &  1.17 (0.02) &  1.63 (0.01) &  1.53 (0.02) & 0.61 (0.03) \\
SiO          &  2.47 (0.08) &  2.60 (0.04) &  1.12 (0.02) &  1.40 (0.03) &  1.05 (0.02) &  1.40 (0.01) &  1.60 (0.03) & 0.50 (0.05) \\
HNCO         & 41.13 (0.40) & 10.39 (0.23) & 26.32 (0.17) & 21.12 (0.19) & 14.32 (0.17) & 10.52 (0.09) & 17.62 (0.23) & 3.62 (0.26) \\
HCN          & 14.92 (0.05) & 17.18 (0.08) & >10.34  & 13.27 (0.03) & 15.95 (0.02)  & >23.19         & 22.30 (0.03)  & 8.22 (0.05)  \\
HCO$^+$      & 4.17 (0.04)  & 5.17 (0.03)  & >3.96   & 2.25 (0.01)  & 2.74 (0.02)   & 3.50 (0.01)    & 4.60 (0.01)   & 1.87 (0.02)  \\
HNC          & 5.14 (0.05)  & 2.43 (0.02)  & >3.48   & 8.37 (0.02)  & 6.93 (0.02)   & 3.95 (0.01)    & 8.20 (0.02)   & 2.02 (0.02)  \\
HC$_3$N      &  7.84 (0.22) &  4.71 (0.13) & 12.41 (0.09) &  5.16 (0.08) &  7.00 (0.07) &  7.46 (0.04) &  8.45 (0.06) & 1.63 (0.11) \\
N$_2$H$^+$   &  1.85 (0.04) &  0.88 (0.02) &  1.39 (0.01) &  2.30 (0.02) &  2.28 (0.01) &  1.48 (0.01) &  2.58 (0.01) & 1.03 (0.01) \\
CH$_3$OH     &              & 		   & 		  & 		 & 		& 	       & 	      &             \\
OCS          &  	    & 		   & 		  & 		 &		&              & 	      & 	    \\
CS           &  9.34 (0.00) & 18.60 (1.77) & 32.40 (1.60) & 12.43 (1.04) & 10.69 (0.92) & 23.37 (0.58) & 23.27 (1.06) &15.39 (1.49) \\
SO           &		    & 		   & 		  & 		 & 		& 	       & 	      &             \\
\hline \hline %
Molecule     & +0.693$^c$&-0.11$^c$ & Sgr\,B2N$^{d,f}$& Sgr\,B2M $^{d,f}$ & CND$^{g}$ & & & \\  
             &  & & & &(-30,-30) & & & \\\hline
             &[x 10$^{-9}$] &[x 10$^{-9}$] &[x 10$^{-9}$] &[x 10$^{-9}$]& [x 10$^{-9}$]& & \\ \hline
H$^{13}$CN   &  >6        & >9        &  8.72 & 1.30  & 3.48& & &\\
SiO          &1.4 (0.2) &7.2 (0.2)  &  0.49 & 0.65  & 2.85& & &\\
HNCO         & 24.7 (1.4)&28.3 (3.0) &  2.78 & 85.07 &  & & &\\
HCN          &  >12  &>180       & 173.70 & 25.94 & 121.25& & &\\
HCO$^+$      &  9.6 (2.2) &15.5 (3.0) & 8.23 & 10.80  & 18.87& & &\\
HNC          & >42 &>60        & 7.78 & 8.67   & 6.82& & &\\
HC$_3$N      & 9.4 (1.0)& 11.2 (1.0)& 19.48 & 0.65  & 1.03& & &\\
N$_2$H$^+$   &        	   & 0.81 & 1.42  & 0.53& & &\\
CH$_3$OH     &  2160 (257)& <1670     & 28571.43 & 23.62 &  & & &\\  
OCS          &  31.8 (1.5)& 36.4 (6.5)& 292.21 & 4.06  & & & & \\%
CS           &  12.43 (1.04) &    & 420.45 & 21.59  & 38.38& & &\\
SO           & 10.4 (8.8)&7.8 (2.0)  & 1316.56 & 13.01 & 16.0 & & &\\ 
\hline \hline %
\end{tabular}
\tablefoot{$^a$ \citet{Jones_et_al_2012}, $^b$ we use the $^{13}$CS emission  and a $^{12}$C/$^{13}$C ratio of 24 to derive the fractional abundance of $^{12}$CS, The column densities of HCN, HCO$^+$ and HNC were corrected by opacity following the formulation in \citet{Jones_et_al_2012}. In SgrB, the $^{13}$C isotopic substitution could be affected by opacity as shown in Fig. \ref{espectraregionesCMZ_jones}, therefore the fractional abundances are a lower limits. The same is true in the 50 km/s cloud for H$^{13}$CN.  
$^c$\citet{Armijos-Abedano_et_al_2015}, $^d$ \citet{Belloche_et_al_2013}, $^e$ integrate velocity range [\kms]. $^f$ fractional abundance of the velocity component closest to 
the nominal velocity of Sgr\,B2N and Sgr\,B2M. The N(H$_2$) was derived from N($^{13}$CO) with the conversion factor used for the CMZ in this work. $^g$ \citet{Harada_et_al_2015}. 
The H$_2$ column density was estimated from multi-transition LVG analysis of CO using an abundance ratio CO/H$_2$=8x10$^{-5}$. }
\end{table*}

\subsection{High velocity shocks}

The species observed in this work provide a set of key molecules to derive the physical and chemical properties of the clouds. Emission from medium and high density tracers is intense and widespread in the M$-3.8+0.9$ molecular cloud. For example, CS can trace densities of $n>10^4$ cm$^{-3}$ \citep{Mauersberger_Henkel_1989}. It is only marginally enhanced in UV \citep{Martin_et_al_2008} and
shock-dominated environments \citep{Requena-Torres_et_al_2006}. HCN, HNC, HCO$^+$ with higher critical densities ($n\sim10^5- 10^6$ cm$^{-3}$) are expected to trace higher density gas rather than diffuse emission from the surrounding lower density cloud. Other molecules such as, e.g. HNCO, can trace gas even denser \citep[$n>10^6$ cm$^{-3}$][]{Jackson_et_al_1984}, but the abundance of this molecule is also driven by shock chemistry (see below). %

We have three shock tracers in our samples: SiO, HNCO and CH$_3$OH. The enhancement of the SiO abundance can be explained by the sputtering of grain cores after the passage of magnetohydrodynamics shocks \citep[see ][for details of the shocks properties]{Jimenez-Serra_et_al_2008}. The Si or directly the SiO is released into the gas phase.  
However, the SiO emission could also be enhancement by X-rays, as suggested by the correlation between the 6.4 keV Fe line with the SiO emission \citep{Martin-Pintado_et_al_2000, Amo-Baladron_et_al_2009}, and by cosmic-rays \citep{Yusef-Zadeh_et_al_2013} as suggested by the correlation with the 74 MHz nonthermal emission. \citet{Brogan_et_al_2003} present large scale observations of the GC region at 74 MHz, and we can see that there is no enhancement of this emission at the position of M$-3.8+0.9$ cloud. The X-ray scenario requires a population of very small grains to produce the SiO abundance enhancement, together with a past episode of bright X-ray emission from some source in the GC \citep{Amo-Baladron_et_al_2009}. Only few X-ray sources are detected close to  M$-3.8+0.9$ cloud \citep{Roberts_et_al_2001}, therefore we do not expect that neither X-ray nor cosmic rays produce a significant enhancement of SiO in this cloud.
SiO has been extensively studied in the GC region, finding high abundance which is associated with large scale shocks \citep[see e.g., ][among others]{Martin-Pintado_et_al_1997, Huettemeister_et_al_1998, Menten_et_al_2009, Salii_et_al_2002, Riquelme_et_al_2010b, Mihn_et_al_2015,Tsuboi_et_al_2015}. The fractional abundances of SiO derived in this work are similar to those derived in the CMZ, which is a clear indication that, like in the CMZ, the chemistry in the M$-3.8+0.9$ molecular cloud is driven by shocks. The highest SiO abundances are found in positions 1.a, 2.a, 3.a, 3.b and 5.a. 

Large abundances of HNCO and CH$_3$OH can also be explained by shocks that release these species from the icy mantles of dust grains. 
HNCO is formed efficiently in the solid phase \citep{Hasegawa_Herbst_1993}, and also can be formed by gas-phase reactions \citep{Iglesias_1977}. Its abundance is enhanced by grain erosion and disruption by low-velocity shocks ($< 26$ km s$^{-1}$) \citep{Zinchenko_et_al_2000} and decreases in the presence of high-velocity shocks (40-50 km s$^{-1}$). Therefore, the shocks that desorb the molecule should be slow enough ($<26$ km s$^{-1}$) in order do not to dissociate it. HNCO is also easily photodissociated in the presence of UV radiation, and also by the UV radiation field induced by shocks \citep{Viti_et_al_2002}. \citet{Martin_et_al_2008} performed a systematic study of 13 sources throughout the CMZ and found that this molecule is an excellent discriminator between chemistry driven by shocks and photodissociation. They determined differences up to a factor of 30 in the intensity ratio of HCNO/$^{13}$CS between shielded molecular clouds mostly affected by shocks and those pervaded by intense UV radiation. Using 
this intensity ratio, they clustered the sources into three groups; typical Galactic center clouds, hot cores, and Photon Dominated Regions (PDRs) and high-velocity shocks, from higher to lower values of this ratio. The fractional abundances of HNCO derived in this work are lower than the ones  derived for the typical Galactic center clouds which are affected by shocks, but similar to the hot core sources. It is clear that in the position-position-velocity space occupied by the foot point of the loops, the UV radiation field is not strong, as indicated by the enhanced HNCO abundances (Fig. \ref{HNCO}). The highest abundances are found in positions 2.a, 2.b, 3.a, and 3.b. As discussed earlier, the HNCO line observed toward position 3.b has a typical shock profile, with a prominent wing, as also seen in the others shock tracer (SiO, CH$_3$OH). However, here HNCO is intense only in the peak of the profile, vanishing in the wing, consistent with the idea that this molecule in enhanced in low velocity shocks and is destroyed in higher velocity shocks, as shown by \citet{Zinchenko_et_al_2000}.

Methanol, also ejected from grain mantles by shocks, is another well known shock tracer. \citet{Requena-Torres_et_al_2006} studied several complex organic molecules, in particular, CH$_3$OH, in 40 GC molecular clouds. They found high fractional abundances of CH$_3$OH ($2.4\times 10^{-8}$ to $1.1\times 10^{-6}$) similar to those in Table \ref{table:XCMZ}, and estimated that frequent ($\sim 10^5$ years) shocks with velocities $>6$ \kms are required to explain the high abundances in the gas phase of complex organic molecules in the GC molecular clouds. \citet{Jimenez-Serra_et_al_2008} show that CH$_3$OH show high fractional abundances for low velocity shock (20 \kms) but the abundances are increased for moderate (30 \kms) and high velocity shock (40 \kms). The methanol line detected in this work corresponds to the $2_k-1_k$ quartet centered in 96.74 GHz, a blend of the $2_{-1}- 1{-1}E, 2_0-1_0A^+,2_0-1_0E, 2_1-1_1E$ lines, which is considered as one line with the spectroscopic parameters of the most intense one ($2_0-1_0A^+$). \citet{Menten_et_al_2009} observed this line and also SiO and CS towards a molecular cloud affected by shocks at the edge of the CMZ and found that this methanol line was the most intense.
 
In our work, the CH$_3$OH emission is one of the most intense emission line (after those of HCN), with the largest fractional abundances in positions 2.a, 2.b, 3.a, 5.a. This, the large terminal velocity found in position 3.b (which indicate high velocity shock), and the similar abundances of SiO in the M$-3.8+0.9$ molecular cloud and in the CMZ, indicate  that the chemistry in the M$-3.8+0.9$ molecular cloud may be driven by moderate or high velocity shocks.

\subsection{High temperature gas}
\citet{Amo-baladron_et_al_2011} suggest a correlation between the abundance ratio HNC/HCN  and temperature, with values close to 1 in quiescent cool dark clouds, a decrease by 1-2 orders of magnitudes in the warmer giant molecular clouds near sites of massive star formation, and with   values ranging between $0.013-0.2$ near the PDR in the giant molecular cloud OMC-1, and in the immediate vicinity of the hot core Orion-KL \citep{Schilke_et_al_1992}. The decrease of this ratio is due to destruction processes of HNC produced by neutral-neutral reactions with an activation barrier \citep{Pineau-des-Forets_et_al_1990} higher than 190 K \citep{Hirota_et_al_1998} or 300 K \citep{Talbi_et_al_1996}. From our PCA analysis, Fig. \ref{plots7molecules} shows that the HCN and HNC molecular emission appears to be anti-correlated but only at 2\% variance, in contrast with higher values found in other places in the Galaxy \citep{Lo_et_al_2009}. Fig. \ref{PCA7molecules} shows the spatial distribution of this anti-correlation, which indicates that in positions 1.a, 2.a, 2.b,and 2.c the HNC abundance decreases, suggesting that those regions should be the warmer regions in the M$-3.8+0.9$ molecular cloud. This is consistent with the kinetic temperatures derived by \citet{Torii_et_al_2010b} who find their highest value in our position 2.b. From Table \ref{columndensity_completerange} we can see that the HNC/HCN abundance ratio ranges from 0.1 to 0.15 in those regions (1.a to 2.c).

\subsection{Comparison with previous work on the Foot Points \label{discusioncomparison}}
Our results indicate that the chemistry of the M$-3.8+0.9$ molecular cloud is mainly driven by shocks. As discussed in Sect. \ref{intro}, \citet{Fukui_et_al_2006} proposed that the GMLs are formed by a  magnetic buoyancy caused by a Parker instability, and they argue that the foot point of these loops, which are two bright spots at both ends, are formed by the accumulation of gas that flows down along the loops.
If the GMLs scenario applies, two foot points should coexist in the M$-3.8+0.9$ cloud, the western side of the loop 1 and the eastern side of the loop 2. 
These foot points were studied by \citet{Torii_et_al_2010b} and \citet{Kudo_et_al_2011} using multi-transition observations of CO and $^{13}$CO. They found several features in the M$-3.8+0.9$ molecular cloud, in particular two ``U-shapes'' in the longitude-latitude and latitude-velocity space, which they explain as a consequence of the magnetic loops as predicted by magnetohydrodynamical (MHD) simulations \citep{Takahashi_et_al_2009}. These features are also seen in the molecular emission presented in this work. \citet{Torii_et_al_2010b} also found an inverted-triangle feature that they called ``protrusion'', which corresponds to our Complex 6. Complexes 3 and 6 both show a sharp intensity gradient studied by \citet{Torii_et_al_2010b} and both mark the eastern side of the ``U-shape'', which is an indication that they share similar physical properties. They also identified three additional broad velocity features, which we identified as the positions 1.a, 2.b and 5.a in Fig. \ref{PCA7molecules}. They connect 
both sides of the ``U-shapes''  which are not  explained yet by the MHD models.

\section{Conclusions} 
We have mapped the M$-3.8+0.9$ molecular cloud in various 3-mm molecular lines using the 22-m Mopra telescope, and the J$=2-1$ rotational transition of $^{13}$CO using 12-m APEX telescope. Eleven molecular species were detected from the 3-mm survey and the $^{13}$C isotopic substitution of HCN. These molecules  encompass tracers of different physical processes, such as shock tracer (SiO, HNCO, CH$_3$OH), medium and high density tracers (CS, HCN, HNC, HCO$^+$, N$_2$H$^+$), and one with a high photodissociation rate (HNCO). 

The molecular cloud shows a velocity gradient from higher to lower velocities from the north-western to the south-eastern
direction. Both the longitude-latitude and the latitude-velocity plots show the ``U-shapes'' observed in the CO emission in previous work which supports the giant molecular loops phenomenon. We identified 6 molecular complexes throughout the molecular cloud. We performed a PCA analysis using the 6 most intense and widespread lines. Based on it, 11 positions were selected that show higher/lower correlations between species. We derived the fractional abundances for these positions for all detected molecules. The $^{13}$CO emission was used as a tracer of the molecular hydrogen and to derive its column density, N(H$_2$). The fractional abundances in the foot point of the GMLs were compared with the fractional abundances in the CMZ, and we found that SiO abundance in the CMZ and GMLs are similar. Based on the high abundance of shock tracers in the M$-3.8+0.9$ molecular cloud, in particular SiO and the moderate abundance of HNCO and CH$_3$OH, we conclude that moderate (30 km/s) or  even high velocity shocks (40-50 km/s) are the dominant physical process heating and driving the chemistry of the molecular gas in the  M$-3.8+0.9$ molecular cloud.

\begin{acknowledgements}
D.R. acknowledges fruitful discussions with colleagues from Nagoya University, in special with Yasuo Fukui, Kazufumi Torii and Rei Enokiya. We thank to Izaskun Jimenez-Serra and Esteban F.E. Morales for useful discussions. This work was partially carried out within the Collaborative Research Council 956, subproject A5, funded by the Deutsche Forschungsgemeinschaft (DFG). D.R. was supported by DGI grant AYA 2008-06181-C02-02 during the observations. LB acknowledges support by CONICYT grant PFB-06. JM-P acknowledges partial support by the MINECO under grants AYA2010-2169-C04-01, FIS2012-39162-C06-01, ESP2013-47809-C03-01 and ESP2015-65597-C4-1. We thank the anonymous referee for critical reading and constructive comments that helped to improve this manuscript.
The Mopra radio telescope is part of the Australia Telescope National Facility which is funded by the Australian Government for operation as a National Facility managed by CSIRO. The University of New South Wales Digital Filter Bank used for the observations with the Mopra Telescope was provided with support from the Australian Research Council. 
This publication is based on data acquired with the Atacama Pathfinder Experiment (APEX). APEX is a collaboration between the Max-Planck-Institut fur Radioastronomie, the European Southern Observatory, and the Onsala Space Observatory.
This research made use of Astropy, a community-developed core Python package for Astronomy \citep{2013A&A...558A..33A}.
            
\end{acknowledgements}

\bibliographystyle{aa} 
\bibliography{referencias} 
\clearpage
\Online

\begin{appendix}
\section{Complementary tables and figures}

\begin{figure*}
\centering
\includegraphics[width= 1.0 \textwidth]{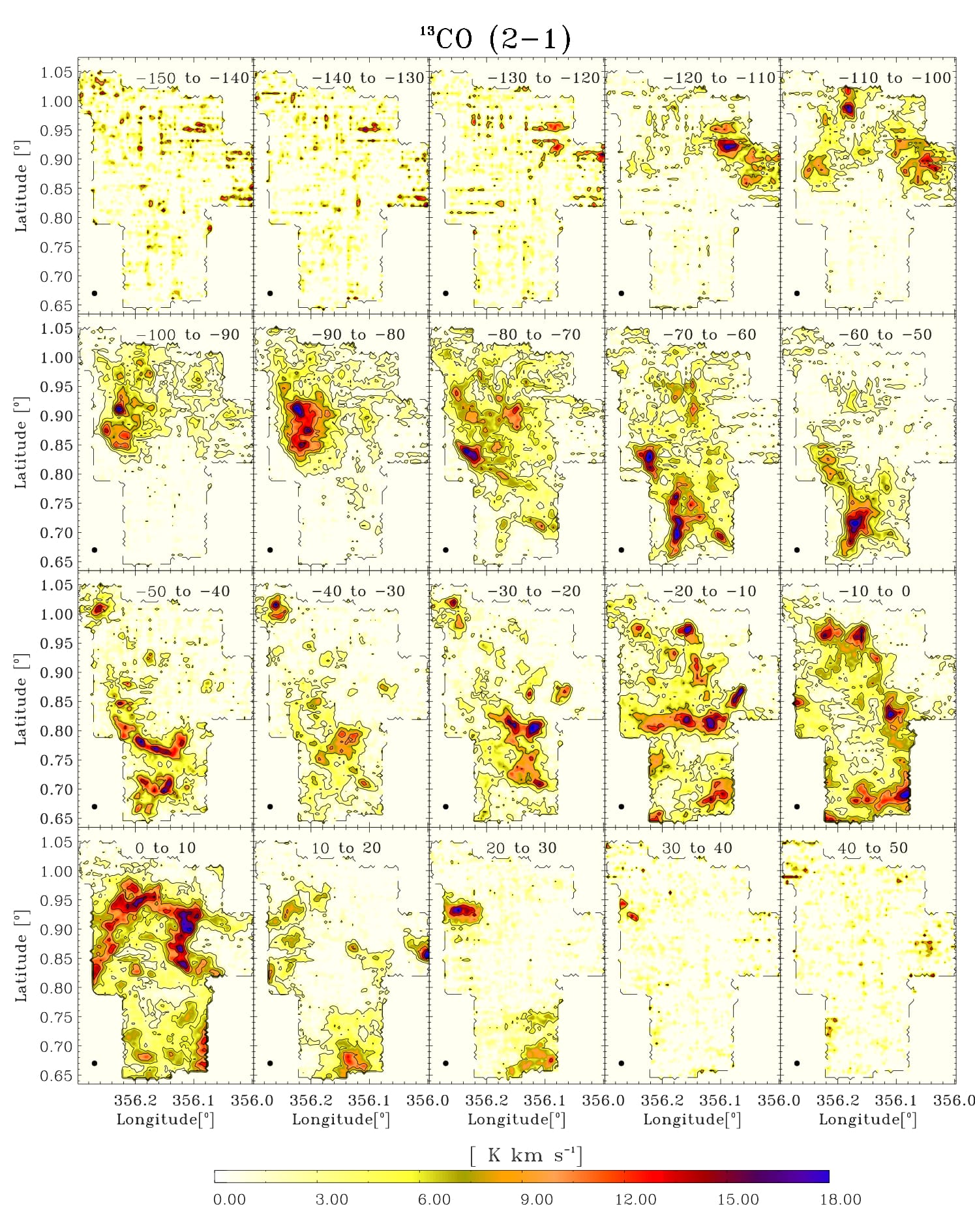}
\caption{Integrated brightness temperature of the M$-3.8+0.9$ molecular cloud in $^{13}$CO $(2-1)$
in velocity intervals of $10$ \kms.}
\label{clumpC_vel13CO}
\end{figure*}

\begin{figure*}
\centering
\includegraphics[width= 1.0 \textwidth]{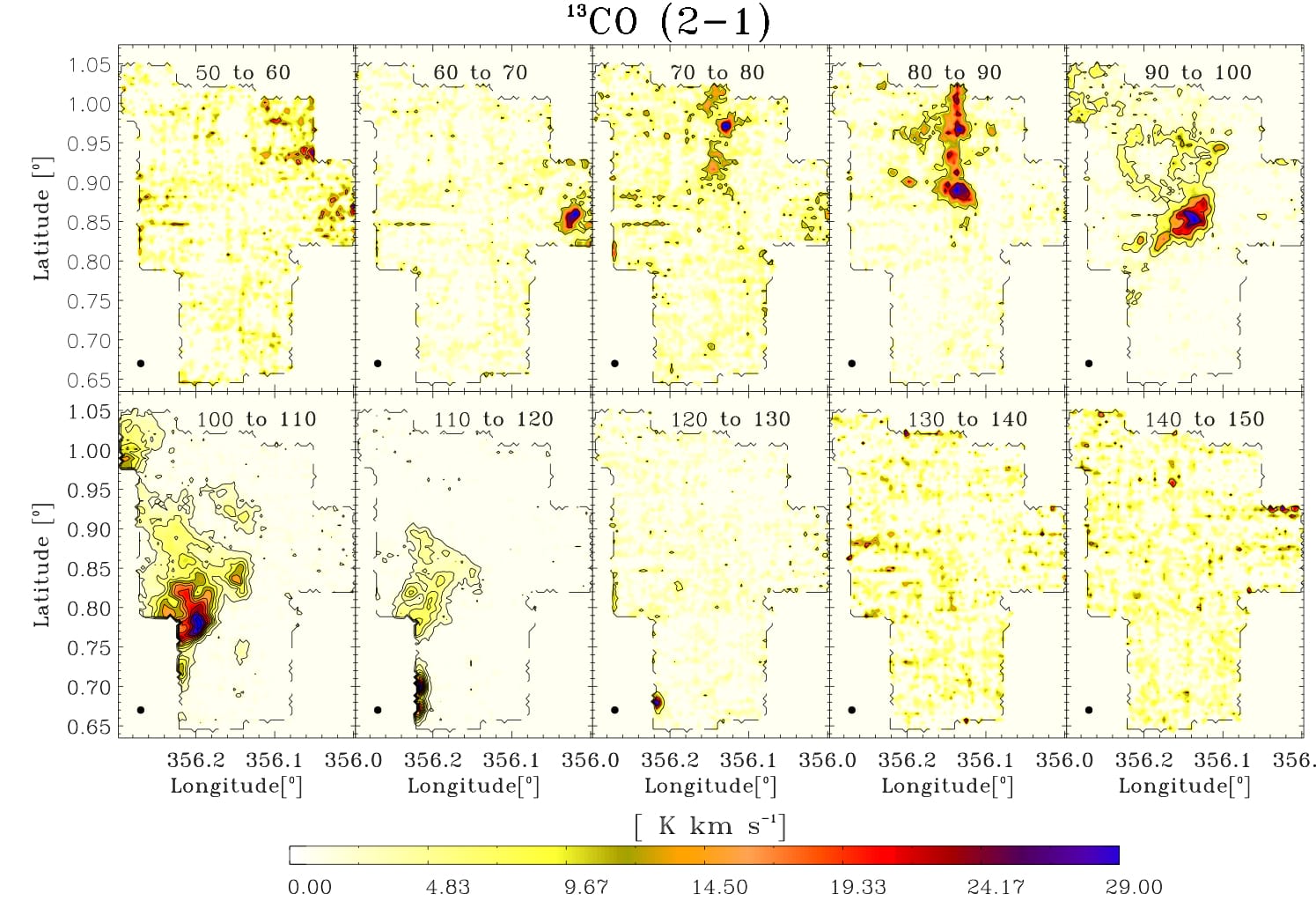}
\caption{Integrated brightness temperature of the M$-3.8+0.9$ molecular cloud in $^{13}$CO $(2-1)$
in velocity intervals of $10$ \kms, continuation.}
\label{clumpC_vel213CO}
\end{figure*}

\begin{table*}
\caption{Column density of the detected molecules. \label{columndensity_completerange}}
\begin{tabular}{ccccccccc}
\hline\hline
     & velocity range &H$^{13}$CN & SiO & HNCO & HCN & HCO$^+$ & HNC & HC$_3$N \\
     & [km\,s$^{-1}$] &[x 10$^{13}$ cm$^{-2}$] &[x 10$^{13}$ cm$^{-2}$] &[x 10$^{13}$ cm$^{-2}$] & [x 10$^{13}$ cm$^{-2}$]&[x 10$^{13}$ cm$^{-2}$] &[x 10$^{13}$ cm$^{-2}$] & [x 10$^{13}$ cm$^{-2}$]\\\hline
1.a  & -90  -20   & 0.29 $\pm$ 0.06 & 0.75 $\pm$ 0.05 & 1.33 $\pm$ 0.33 & 4.71 $\pm$ 0.05 & 0.78 $\pm$ 0.02 & 0.45 $\pm$ 0.04 & 0.48          \\        
2.a  & -120 -20   & 0.61 $\pm$ 0.06 & 0.96 $\pm$ 0.06 & 2.70 $\pm$ 0.54 & 8.96 $\pm$ 0.07 & 2.32 $\pm$ 0.03 & 1.05 $\pm$ 0.05 & 1.78 $\pm$ 0.22\\ 
2.b  & -120 -20   & 0.66 $\pm$ 0.07 & 0.92 $\pm$ 0.07 & 3.34 $\pm$ 0.49 &10.39 $\pm$ 0.06 & 3.25 $\pm$ 0.03 & 1.30 $\pm$ 0.05 & 2.19 $\pm$ 0.26\\
2.c  & -120 -20   & 0.44 $\pm$ 0.07 & 0.88 $\pm$ 0.07 & 1.85            & 6.12 $\pm$ 0.08 & 2.46 $\pm$ 0.03 & 0.79 $\pm$ 0.06 & 0.79           \\
3.a  & -120 -50   & 0.55 $\pm$ 0.07 & 1.29 $\pm$ 0.07 & 5.95 $\pm$ 0.72 & 5.17 $\pm$ 0.09 & 1.83 $\pm$ 0.03 & 1.52 $\pm$ 0.06 & 3.32 $\pm$ 0.27 \\
3.b  & -120 -20   & 0.65 $\pm$ 0.09 & 1.48 $\pm$ 0.08 & 2.96 $\pm$ 0.59 & 8.51 $\pm$ 0.08 & 3.01 $\pm$ 0.04 & 1.55 $\pm$ 0.07 & 3.16 $\pm$ 0.31 \\ 
4.a  & -100 -20   & 0.55 $\pm$ 0.07 & 0.92 $\pm$ 0.07 & 2.02 $\pm$ 0.50 & 4.15 $\pm$ 0.07 & 1.76 $\pm$ 0.03 & 0.76 $\pm$ 0.06 & 1.30 $\pm$ 0.24 \\ 
5.a  & -100 -20   & 0.21 $\pm$ 0.06 & 1.34 $\pm$ 0.06 & 1.59            & 3.87 $\pm$ 0.07 & 1.47 $\pm$ 0.03 & 0.91 $\pm$ 0.05 & 0.70 $\pm$ 0.22 \\
6.a  & -85  -20   & 0.25            & 0.88 $\pm$ 0.08 & 3.32 $\pm$ 0.78 & 3.69 $\pm$ 0.12 & 1.09 $\pm$ 0.04 & 0.70 $\pm$ 0.07 & 1.22 $\pm$ 0.30 \\ 
6.b  & -85  -30   & 0.39 $\pm$ 0.07 & 0.97 $\pm$ 0.07 & 2.35 $\pm$ 0.47 & 4.63 $\pm$ 0.06 & 1.70 $\pm$ 0.03 & 1.13 $\pm$ 0.05 & 1.40 $\pm$ 0.21 \\ 
6.c  & -80  -20   & 0.20            & 0.82 $\pm$ 0.07 & 2.20 $\pm$ 0.39 & 3.12 $\pm$ 0.05 & 1.42 $\pm$ 0.02 & 0.85 $\pm$ 0.04 & 0.61           \\ 
     &		      &               	&	          &	            &	              &	                &	           &	\\ \hline\hline
     & velocity range & N$_2$H$^+$ & CH$_3$OH & OCS & CS & SO & $^{13}$CO & H$_2$ \\
     & [km\,s$^{-1}$] &[x 10$^{13}$ cm$^{-2}$] &[x 10$^{13}$ cm$^{-2}$] &[x 10$^{13}$ cm$^{-2}$] & [x 10$^{13}$ cm$^{-2}$]&[x 10$^{13}$ cm$^{-2}$] &[x 10$^{16}$ cm$^{-2}$] & [x 10$^{21}$ cm$^{-2}$]\\ \hline
1.a  &  -90 -20   & 0.16 $\pm$ 0.03 & 29.94 $\pm$ 0.74 & 12.32           & 2.50 $\pm$ 0.11 & 2.49 $\pm$ 0.36 & 0.83 $\pm$ 0.02 & 4.37 $\pm$ 0.12 \\
2.a  & -120 -20   & 0.18 $\pm$ 0.03 & 54.68 $\pm$ 1.26 & 18.90           & 6.08 $\pm$ 0.17 & 4.11 $\pm$ 0.56 & 1.15 $\pm$ 0.03 & 6.06 $\pm$ 0.14 \\
2.b  & -120 -20   & 0.20 $\pm$ 0.04 & 64.17 $\pm$ 1.69 & 27.86 $\pm$ 8.55 & 8.90 $\pm$ 0.24 & 2.71 $\pm$ 0.74 & 1.57 $\pm$ 0.02 & 8.25 $\pm$ 0.12 \\
2.c  & -120 -20   & 0.28 $\pm$ 0.04 & 36.51 $\pm$ 1.73 & 26.30           & 4.81 $\pm$ 0.25 & 2.40           & 1.36 $\pm$ 0.04 & 7.16 $\pm$ 0.21 \\
3.a  & -120 -50   & 0.59 $\pm$ 0.05 & 81.52 $\pm$ 1.29 & 33.19 $\pm$ 6.67 & 7.46 $\pm$ 0.17 & 5.06 $\pm$ 0.61 & 1.71 $\pm$ 0.02 & 9.01 $\pm$ 0.10 \\
3.b  & -120 -20   & 0.32 $\pm$ 0.05 & 90.86 $\pm$ 1.29 & 20.16 $\pm$ 6.06 & 10.93 $\pm$ 0.18 & 5.15 $\pm$ 0.56 & 2.27 $\pm$ 0.03 & 11.96 $\pm$ 0.13 \\
4.a  & -100 -20   & 0.14 $\pm$ 0.04 & 49.31 $\pm$ 1.01 & 16.19           & 6.50 $\pm$ 0.15 & 1.82 $\pm$ 0.49 & 1.82 $\pm$ 0.02 & 9.58 $\pm$ 0.10 \\
5.a  & -100 -20   & 0.17 $\pm$ 0.04 & 49.19 $\pm$ 1.11 & 17.31           & 6.12 $\pm$ 0.16 & 2.45 $\pm$ 0.44 & 1.30 $\pm$ 0.02 & 6.83 $\pm$ 0.09 \\
6.a  & -85  -20   & 0.18 $\pm$ 0.05 & 39.86 $\pm$ 1.19 & 18.34           & 5.09 $\pm$ 0.17 & 1.71           & 1.46 $\pm$ 0.02 & 7.66 $\pm$ 0.11 \\
6.b  & -85  -30   & 0.32 $\pm$ 0.03 & 52.39 $\pm$ 0.87 & 13.97           & 6.76 $\pm$ 0.12 & 2.49 $\pm$ 0.42 & 1.47 $\pm$ 0.02 & 7.73 $\pm$ 0.09 \\
6.c  & -80  -20   & 0.30 $\pm$ 0.03 & 29.53 $\pm$ 0.90 & 28.26 $\pm$ 4.29 & 3.28 $\pm$ 0.12 & 1.19           & 1.02 $\pm$ 0.02 & 5.35 $\pm$ 0.09 \\
\end{tabular}
\tablefoot{$^a$ Column density obtained from HC$_3$N (10-9) because this transition has higher signal-to-noise ratio than the (11-10).\\
$^b$ N(H$_2$) derived from $^{13}$CO emission using a conversion factor of $1.9\times 10^{-6}$ (see text for details)}
\end{table*}    

\begin{figure*}
\centering
\includegraphics[angle=0,width= 1.0 \textwidth]{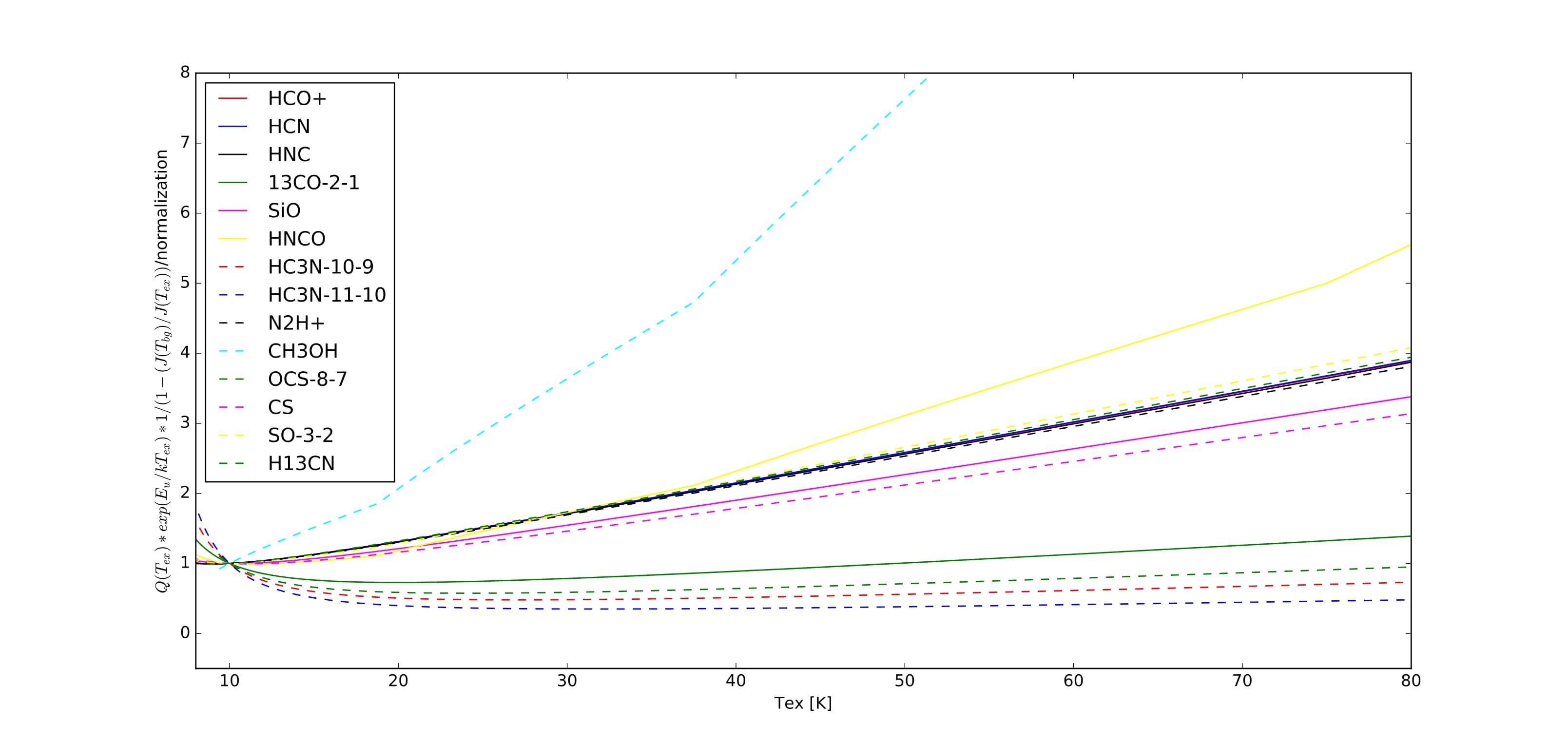}
\caption{Variation of column density with excitation temperature, T$_{\rm ex}$. The y-axis shows the T$_{\rm ex}$-dependent parameters in Eq. 1 normalized to T$_{\rm ex}=10$ K. For $^{13}$CO (2-1), we can see that the difference  between the assumption of T$_{\rm ex}=10$ K and 40 K (which is the temperature estimated by the detailed work of \citet{Torii_et_al_2010b} is  $\sim$ 10\%.  Most of the species in the 3-mm molecular transitions show a factor $< 2$ if we increase the T$_{\rm ex}$ at 40 K, with the exception of CH$_3$OH which is highly dependent of the T$_{\rm ex}$; and OCS and HC$_3$N. }
\label{variationTex}
\end{figure*}

\begin{figure*}
\centering
\includegraphics[angle=90,width= 0.35 \textwidth]{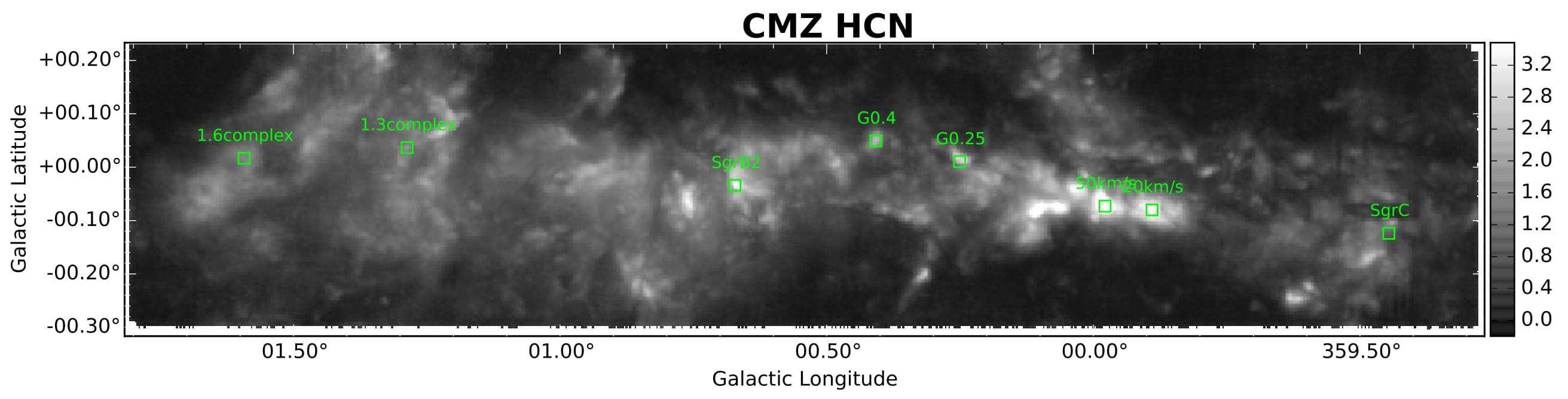}
\caption{Selected regions for deriving the column densities for the detected molecules of this work overlaid on the peak brightness images of HCN from \citet{Jones_et_al_2012}}
\label{coverturaCMZ}
\end{figure*}

\clearpage
\begin{figure*}
\vspace{0.2in}
\vbox{
\hbox{
\includegraphics[angle=0,width=0.3\textwidth]{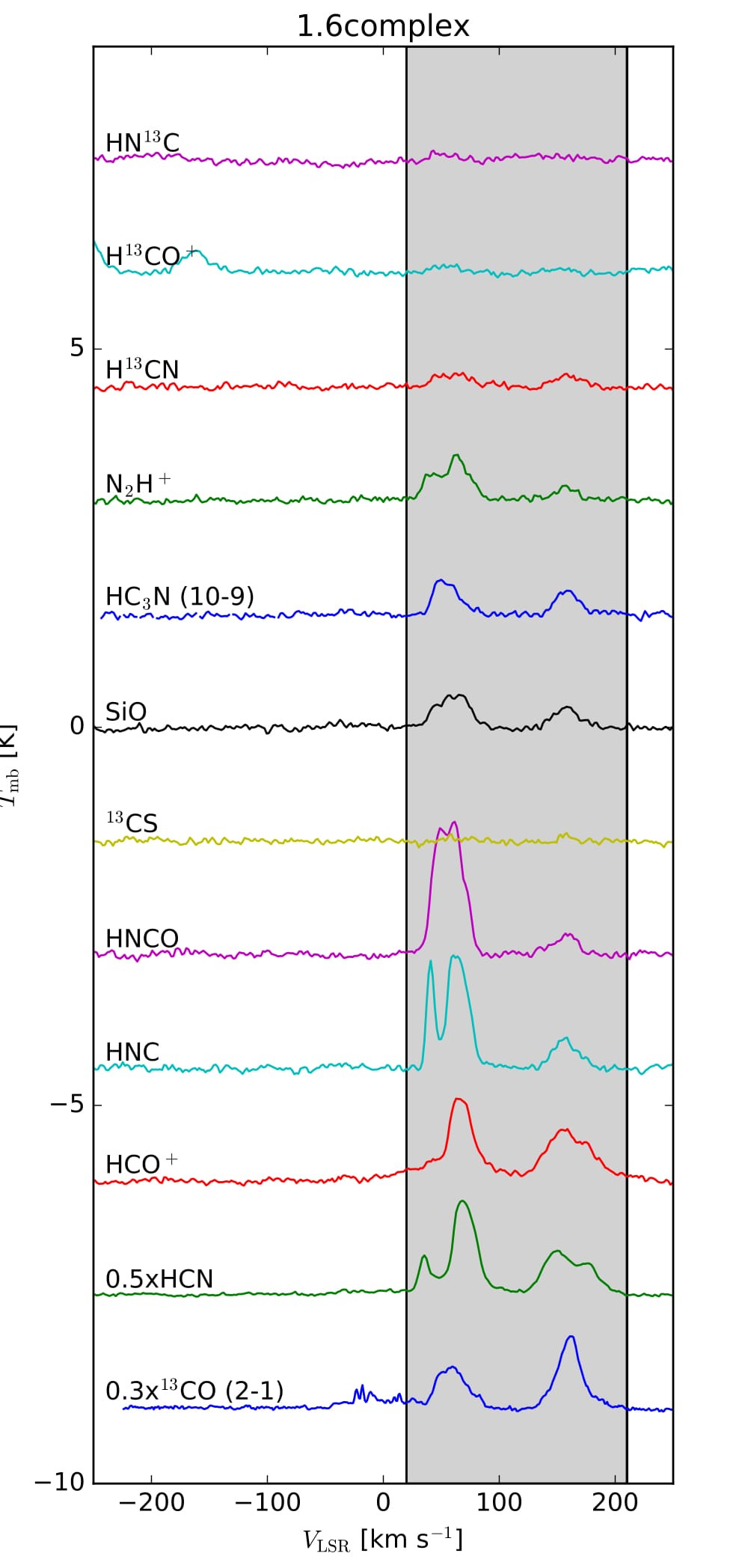}
\includegraphics[angle=0,width=0.3\textwidth]{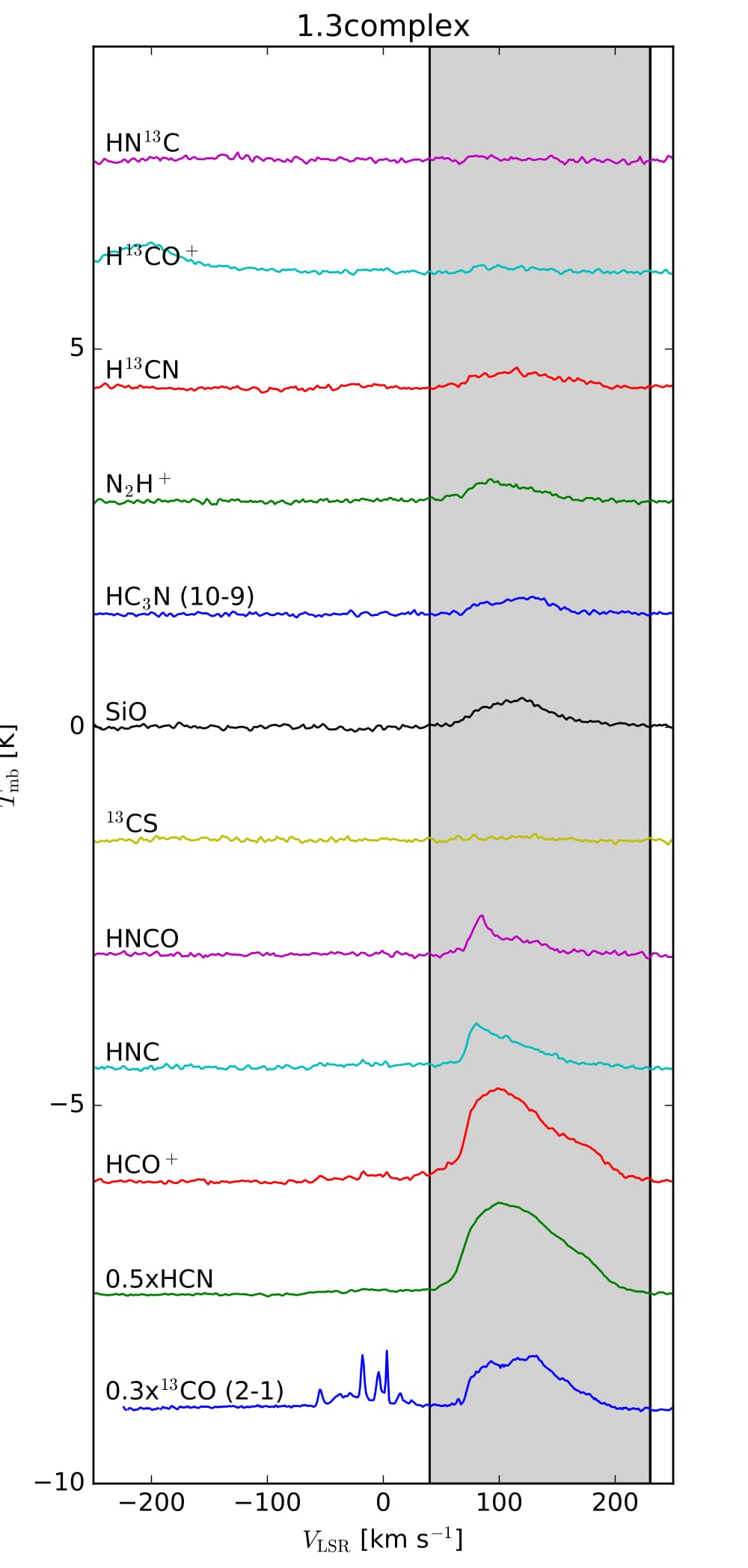}
\includegraphics[angle=0,width=0.3\textwidth]{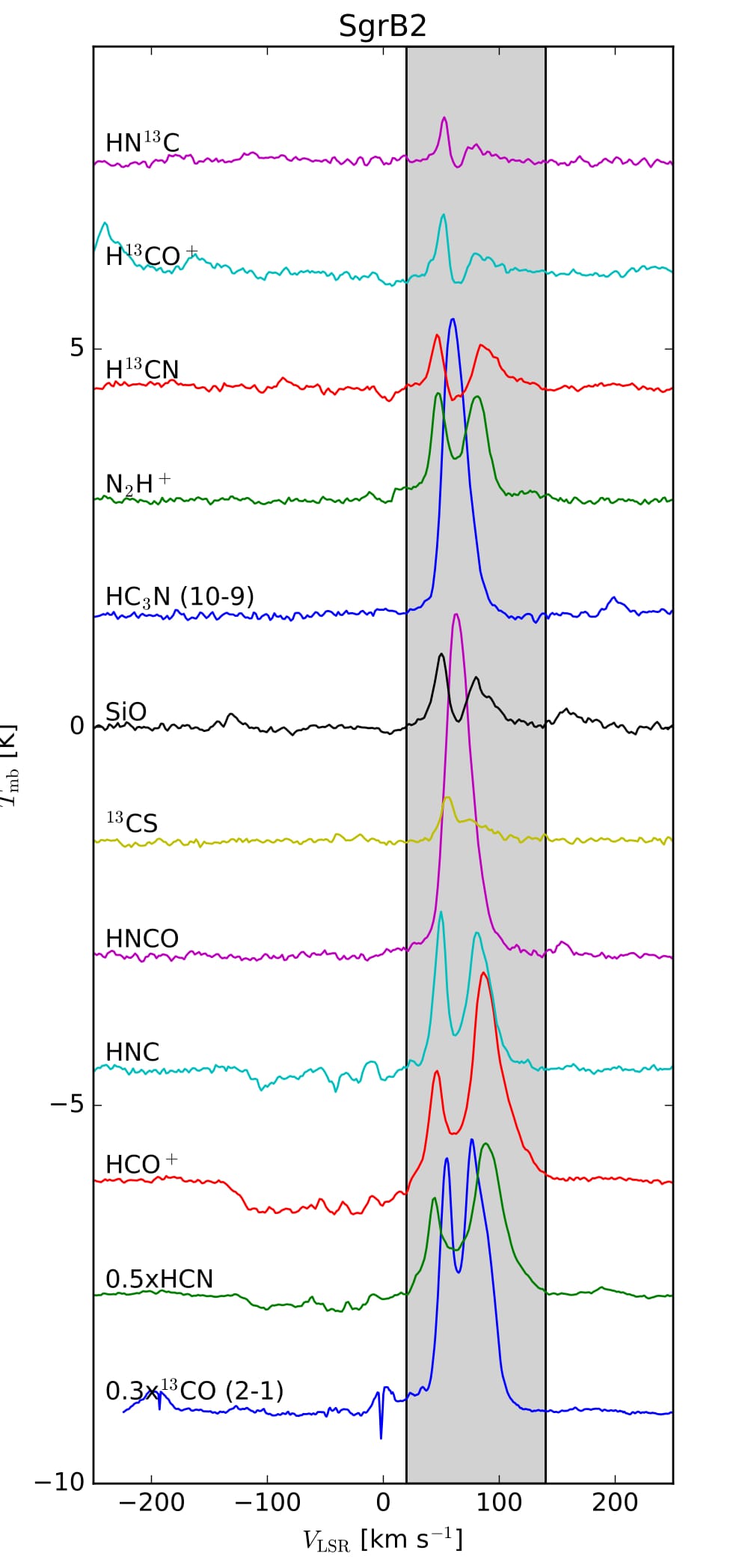}
}                                
\vspace{0.3in}                  
\hbox{                           
\includegraphics[angle=0,width=0.3\textwidth]{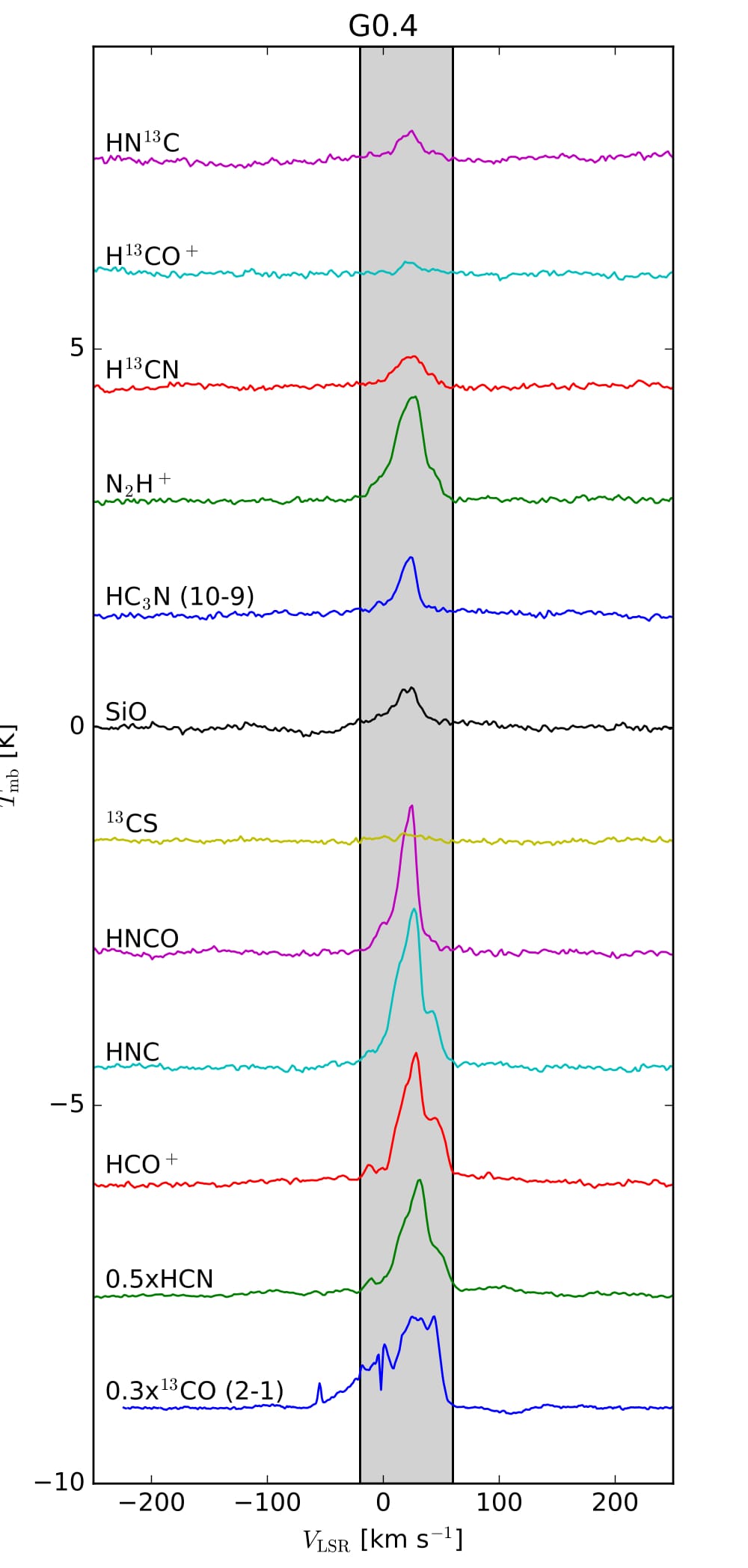}
\includegraphics[angle=0,width=0.3\textwidth]{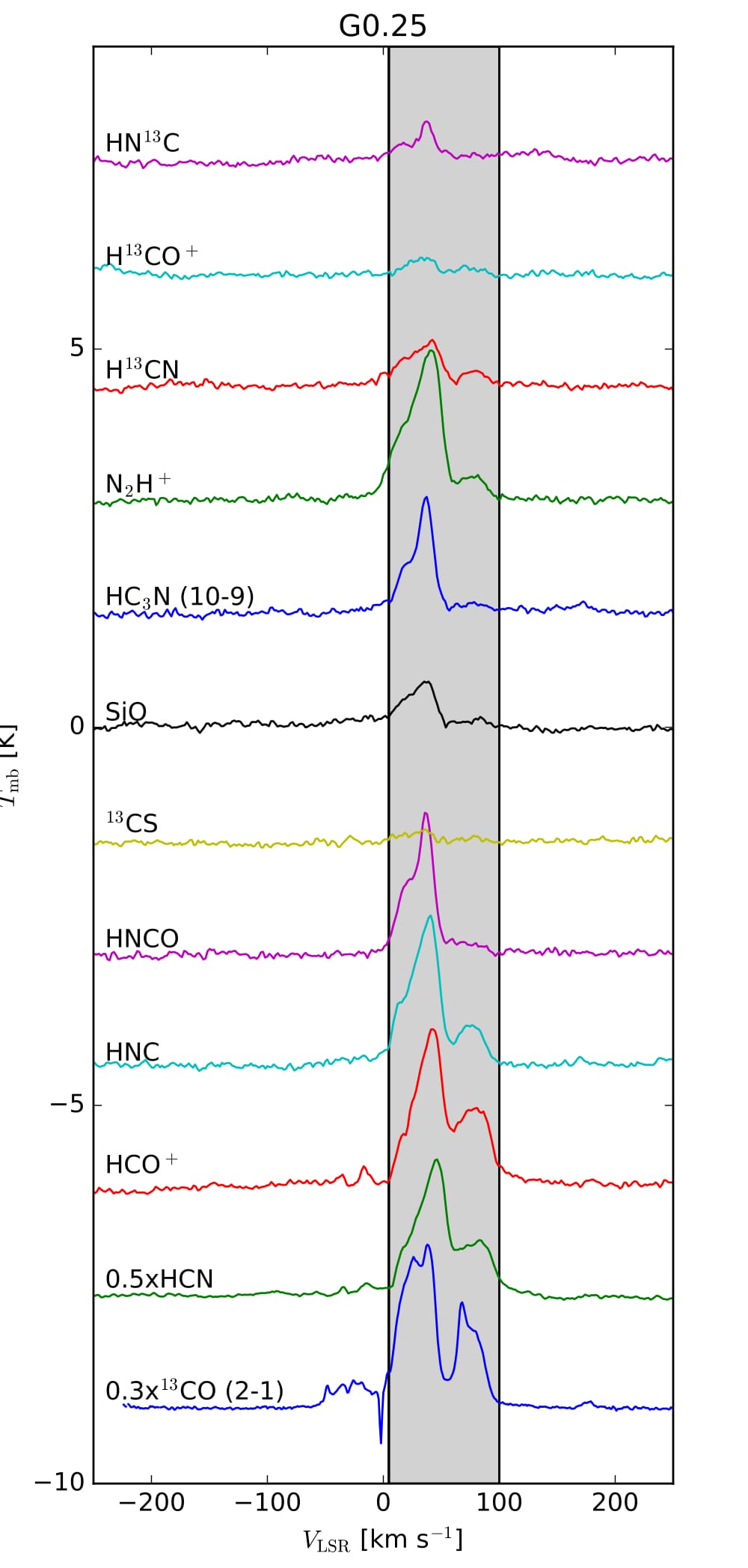}
\includegraphics[angle=0,width=0.3\textwidth]{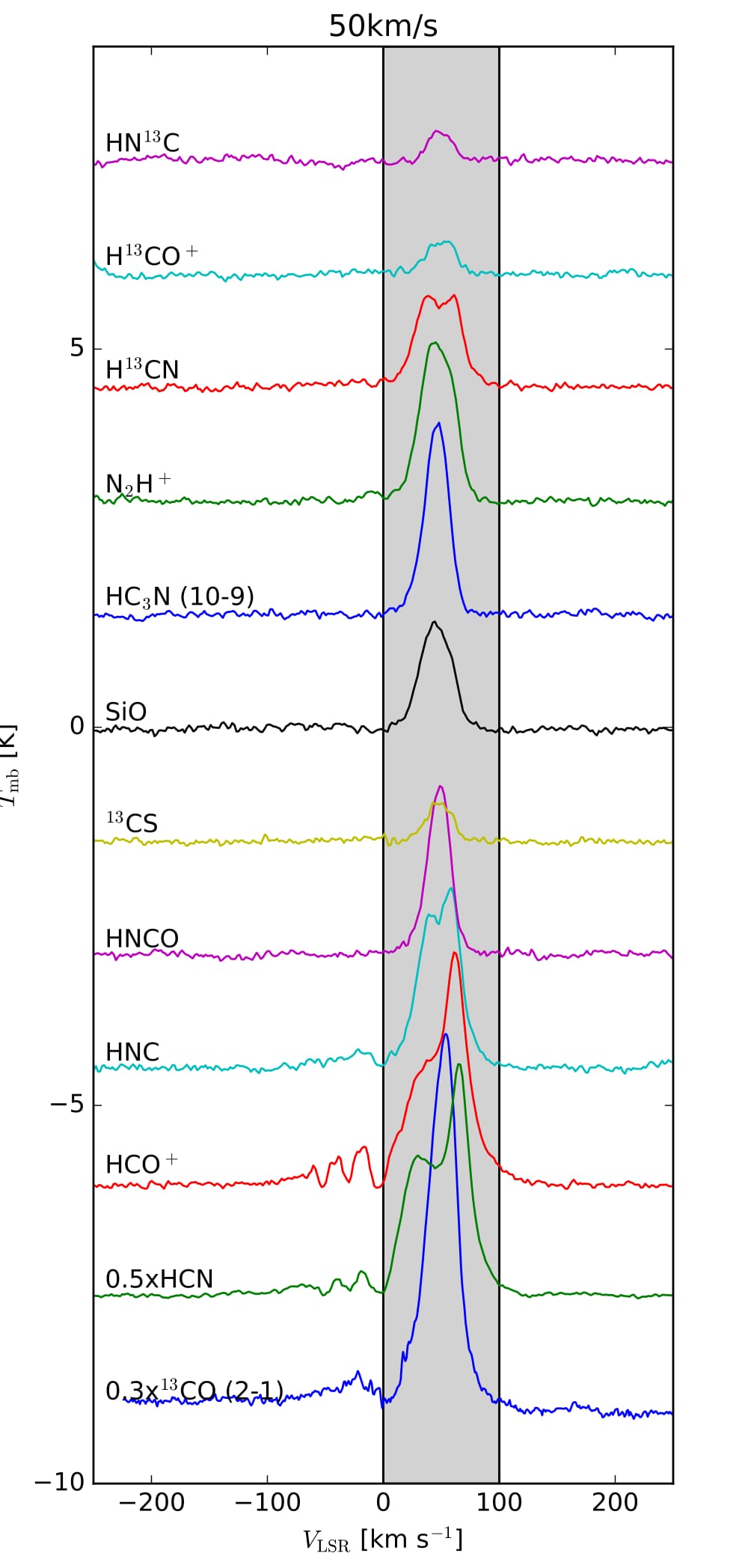}
}
}
\caption{Average spectra for each region shown in Fig. \ref{coverturaCMZ}. $^{13}$CO (2-1) spectra are scaled by a factor 0.3 and HCN spectra are scaled by a factor 0.5. }
\label{espectraregionesCMZ_jones}
\end{figure*}

\begin{figure*}
\vspace{0.2in}
\vbox{
\hbox{
\includegraphics[angle=0,width=0.3\textwidth]{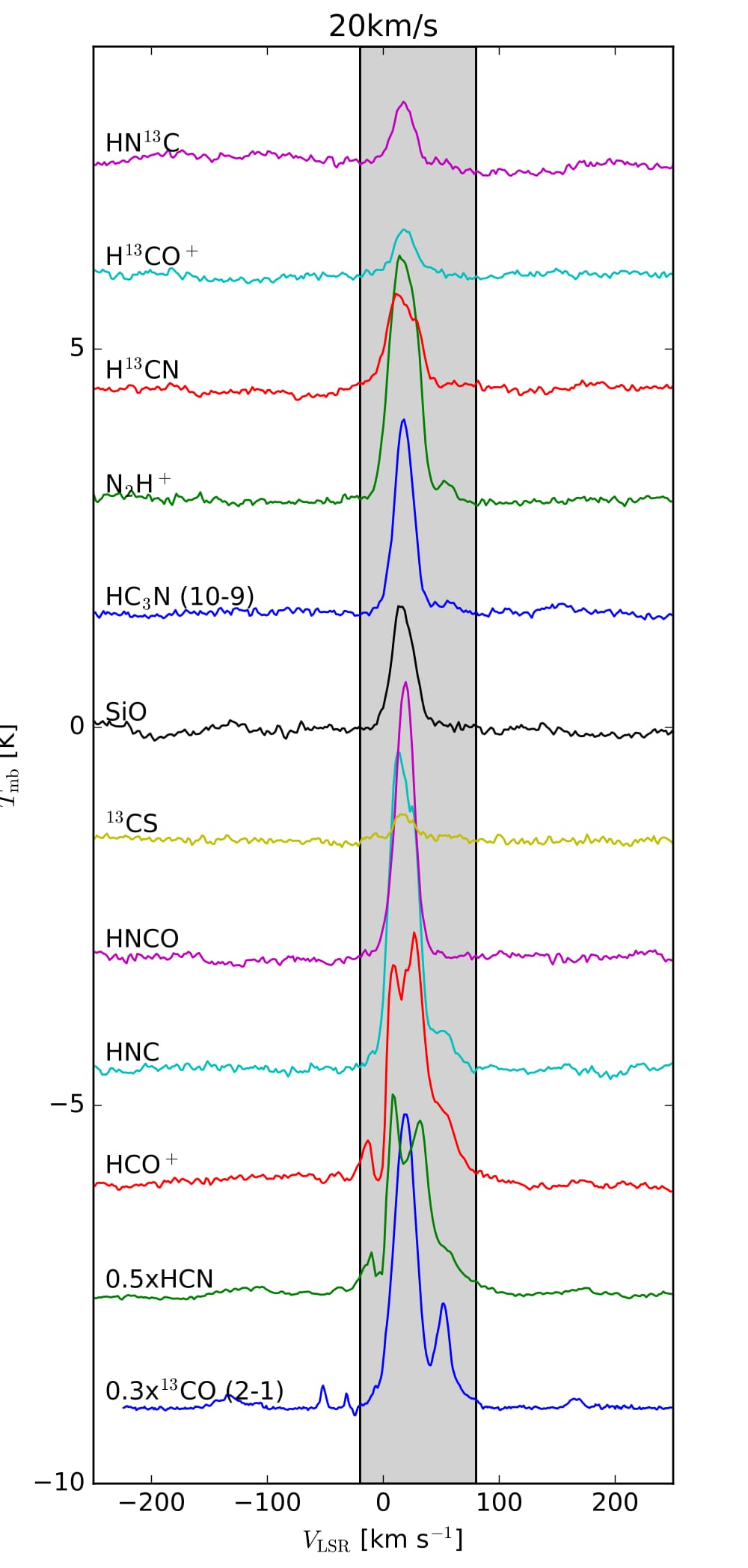}
\includegraphics[angle=0,width=0.3\textwidth]{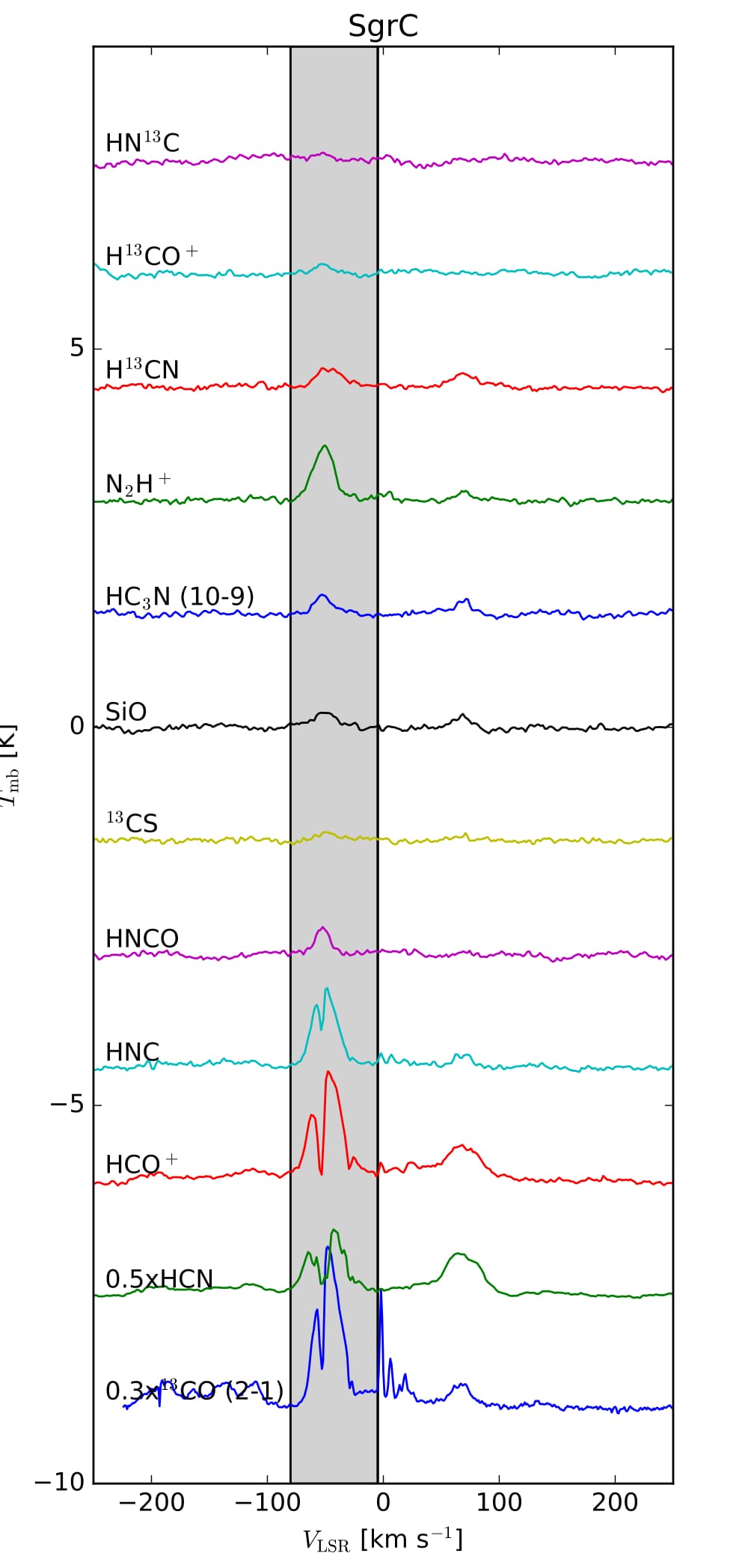}
}
}
\caption{Average spectra for each region shown in Fig. \ref{coverturaCMZ}. $^{13}$CO (2-1) spectra are scaled by a factor 0.3 and HCN spectra are scaled by a factor 0.5.}
\label{espectraregionesCMZ_jones}
\end{figure*}

\clearpage

\section{Beam efficiency for APEX observations \label{cal}}
Since the Galactic center molecular emission is extended in comparison with the beam size, we estimate the beam efficiency for the APEX telescope considering an extended source. 
Considering the main beam efficiencies measured in November 2012 using Jupiter\footnote{\url{http://www3.mpifr-bonn.mpg.de/div/submmtech/}} (see Table \ref{beameff}), we performed a least squares fit for the Ruze formula to get the values for B$_0$ and $\sigma$ (B$_0=0.69$, $\sigma=0.019$).
\begin{equation}
 B_{\rm{eff}}(\lambda) = B_0*\exp(-(4\times\pi \times \sigma/\lambda)^2)
\end{equation}

\begin{table*}
\caption{\label{beameff} Beam efficiencies for APEX telescope using Jupiter}
\centering
\begin{tabular}{cccccc}
\hline\hline
Frequency & Source & Source  &Main beam & Forward & Beam \\
	  &        &  size   & size     & efficiency & efficiency\\
$[$GHz$]$ &        &   ['']      &  ['']        &                   &      \\\hline
 349      & Jupiter&  44.8       & 17.6         & 0.95              & 0.67 \\
 466      & Jupiter&             & 13.2         & 0.95              & 0.58 \\
 691.9    & Jupiter&  47.4       & 8.9          & 0.95              & 0.49 \\
 807.1    & Jupiter&             & 7.7          & 0.95              & 0.48 \\\hline
\hline
\end{tabular}
\end{table*}

\begin{figure}
\centering
\includegraphics[width= 0.55\textwidth]{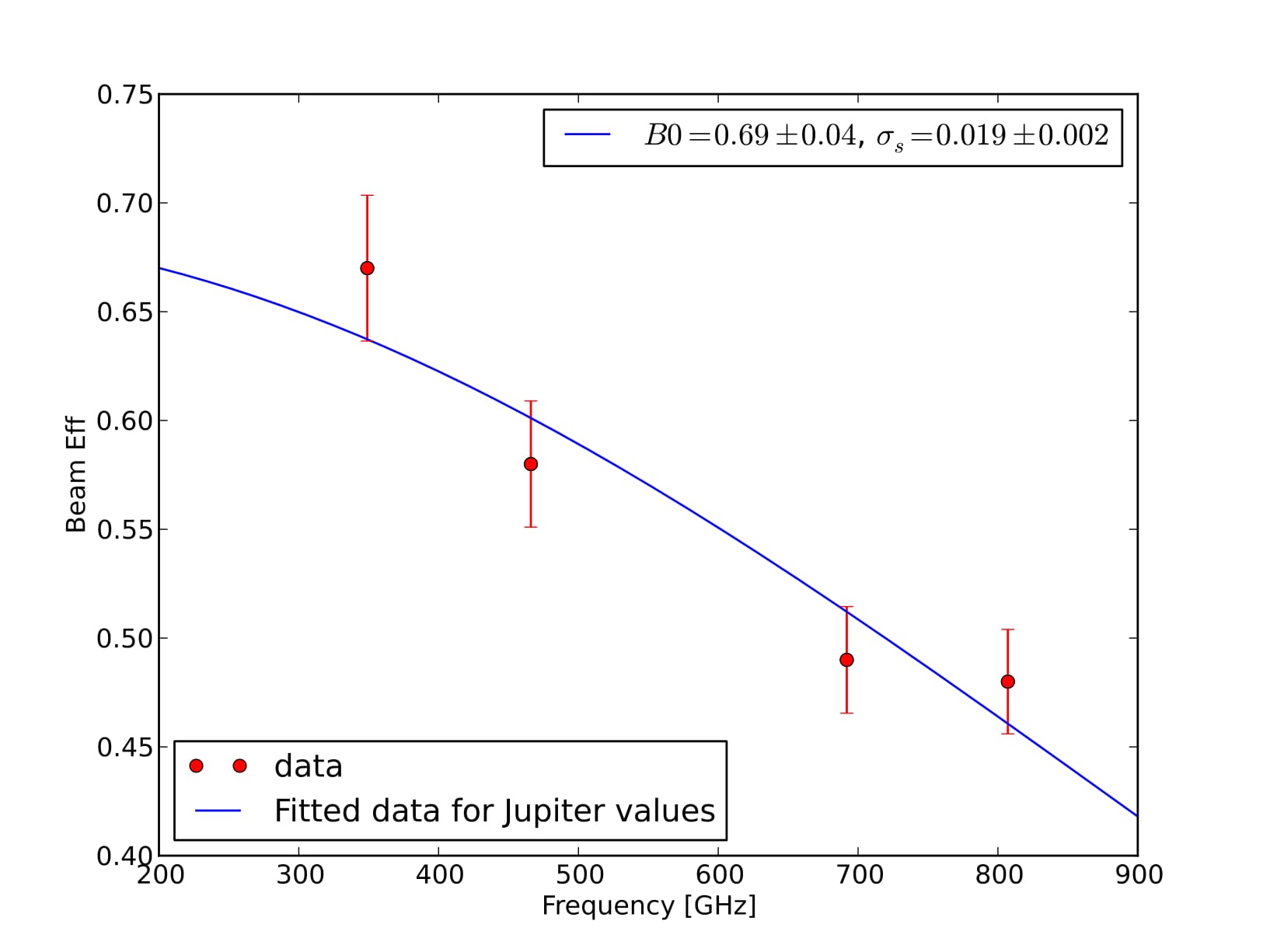}
\caption{Least squares fit for the values of the main beam efficiency measured on November 2012 at the APEX telescope. We fit the Ruze formula to get the values for B$_0$ and $\sigma$.}
\label{BeamAPEX}
\end{figure}

\clearpage

\section{Velocity integrated emission \label{velocityintegratedmolecules}}

\begin{figure*}
\includegraphics[angle=0,width=1.0 \textwidth]{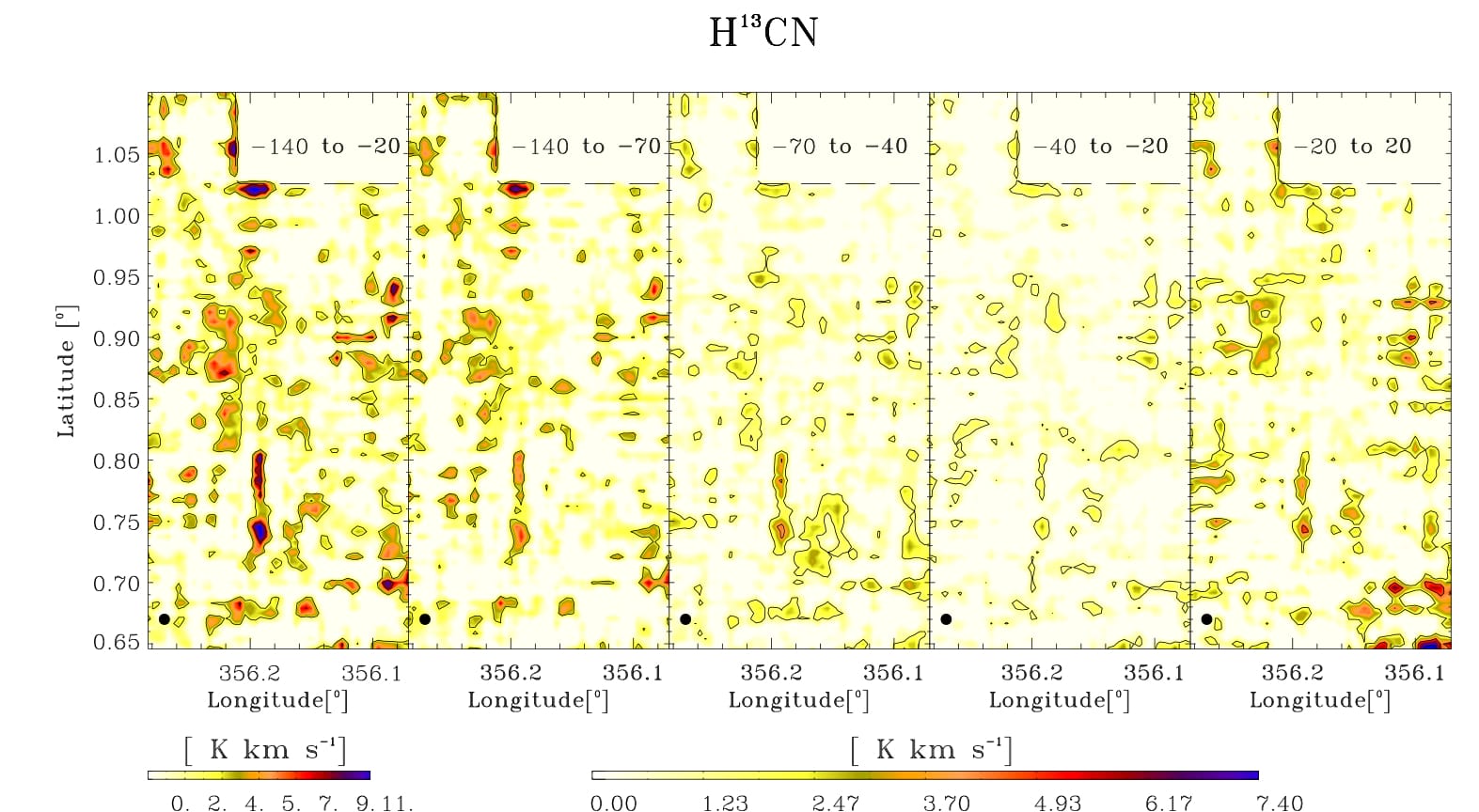}
\caption{Velocity integrated emission in H$^{13}$CN. From left to right: velocity range from -140 to 20 \kms (the complete velocity range covered by the GMLs); velocity range from -140 to -70 \kms; -70 to -40 \kms, -40 to -20\kms and -20 to 20 \kms \label{inicio}}
\end{figure*}

\begin{figure*}
\includegraphics[angle=0,width=1.0 \textwidth]{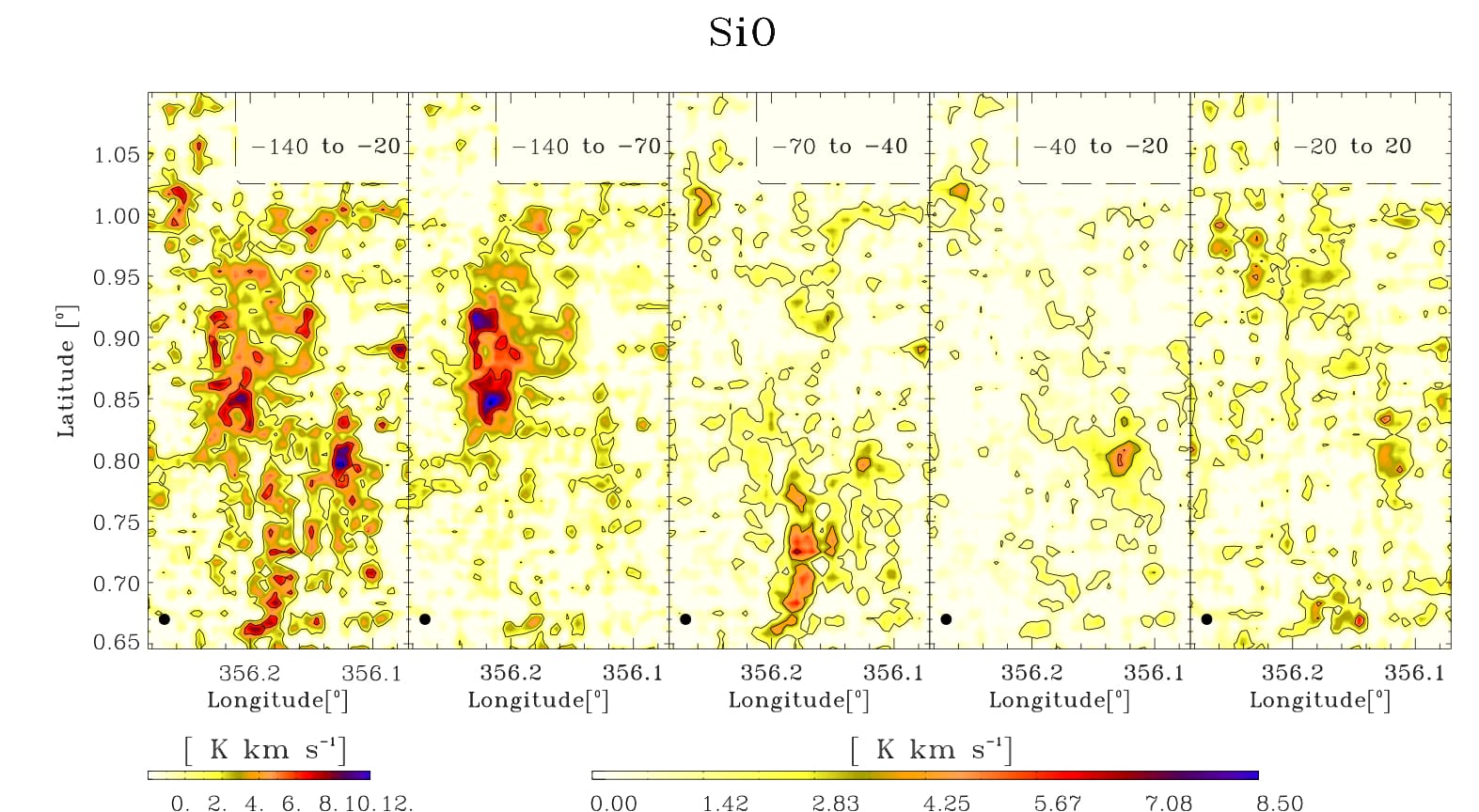}
\caption{Velocity integrated emission in SiO. From left to right: velocity range from -140 to 20 \kms (the complete velocity range covered by the GMLs); velocity range from -140 to -70 \kms; -70 to
-40 \kms, -40 to -20\kms and -20 to 20 \kms}
\end{figure*}

\begin{figure*}
\includegraphics[angle=0,width=1.0 \textwidth]{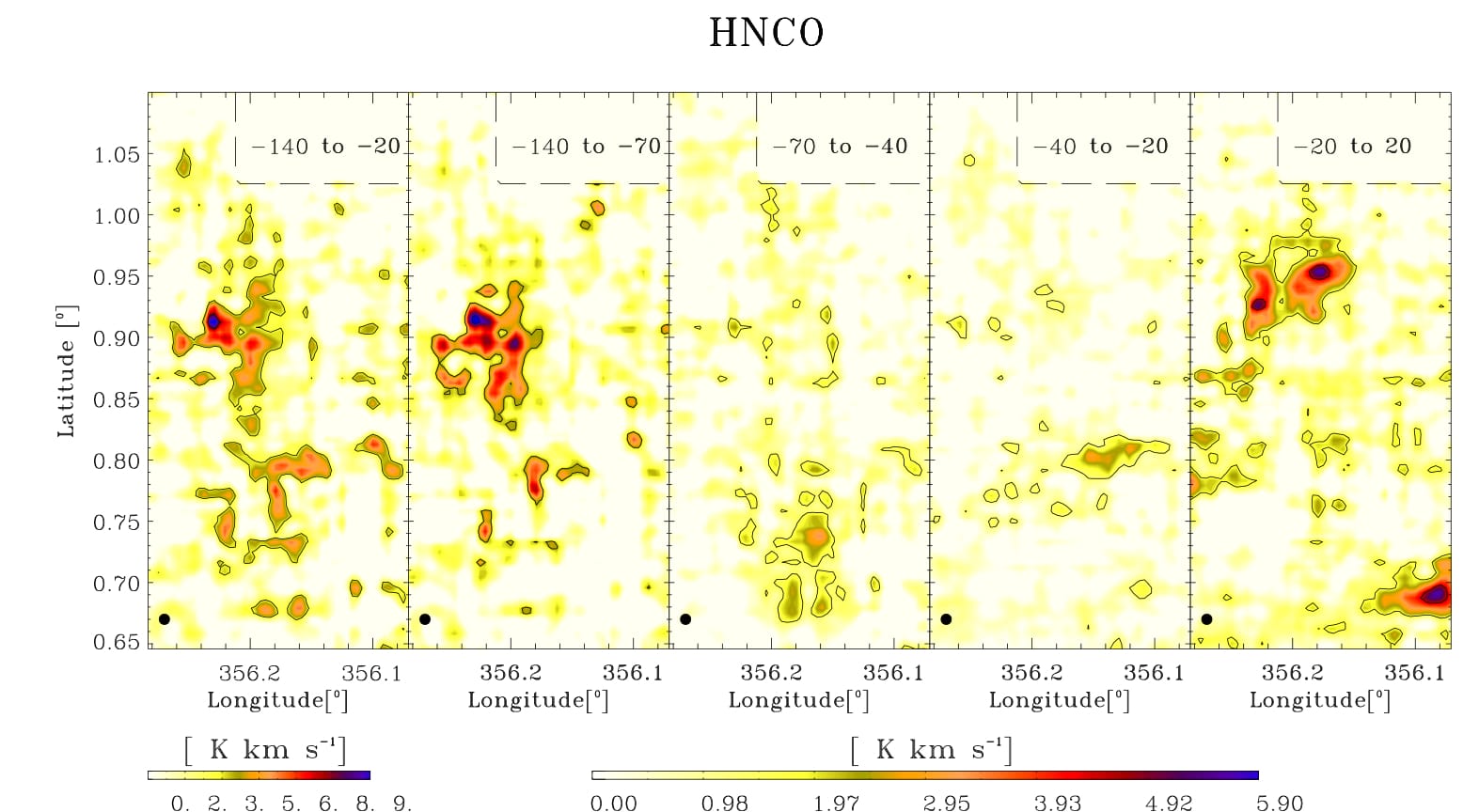}

\caption{Velocity integrated emission in HNCO. From left to right: velocity range from -140 to 20 \kms (the complete velocity range covered by the GMLs); velocity range from -140 to -70 \kms; -70 to
-40 \kms, -40 to -20\kms and -20 to 20 \kms}
\label{HNCO}
\end{figure*}

\begin{figure*}
\includegraphics[angle=0,width=1.0 \textwidth]{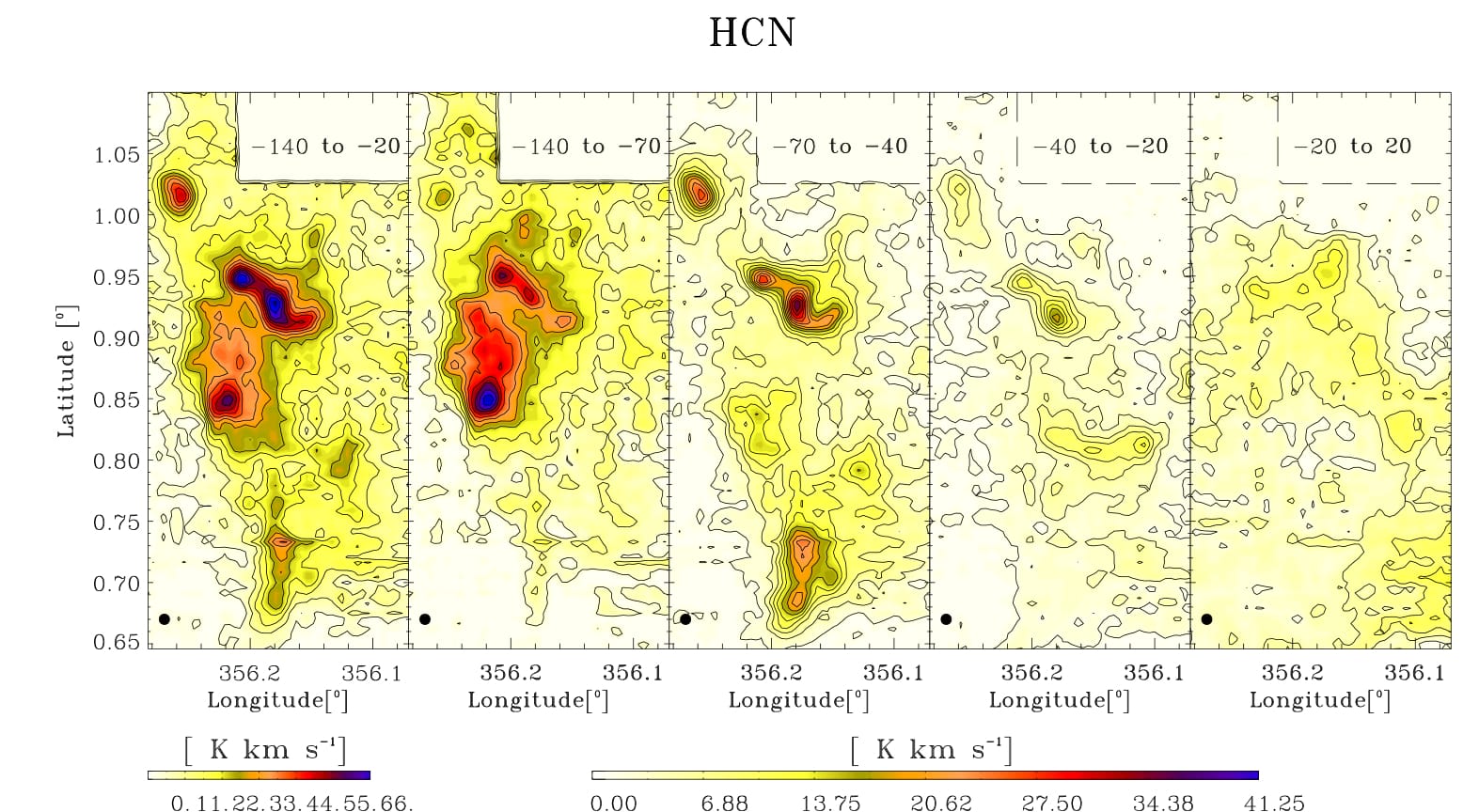}
\caption{Velocity integrated emission in HCN. From left to right: velocity range from -140 to 20 \kms (the complete velocity range covered by the GMLs); velocity range from -140 to -70 \kms; -70 to
-40 \kms, -40 to -20\kms and -20 to 20 \kms \label{inicio}}
\end{figure*}

\begin{figure*}
\includegraphics[angle=0,width=1.0 \textwidth]{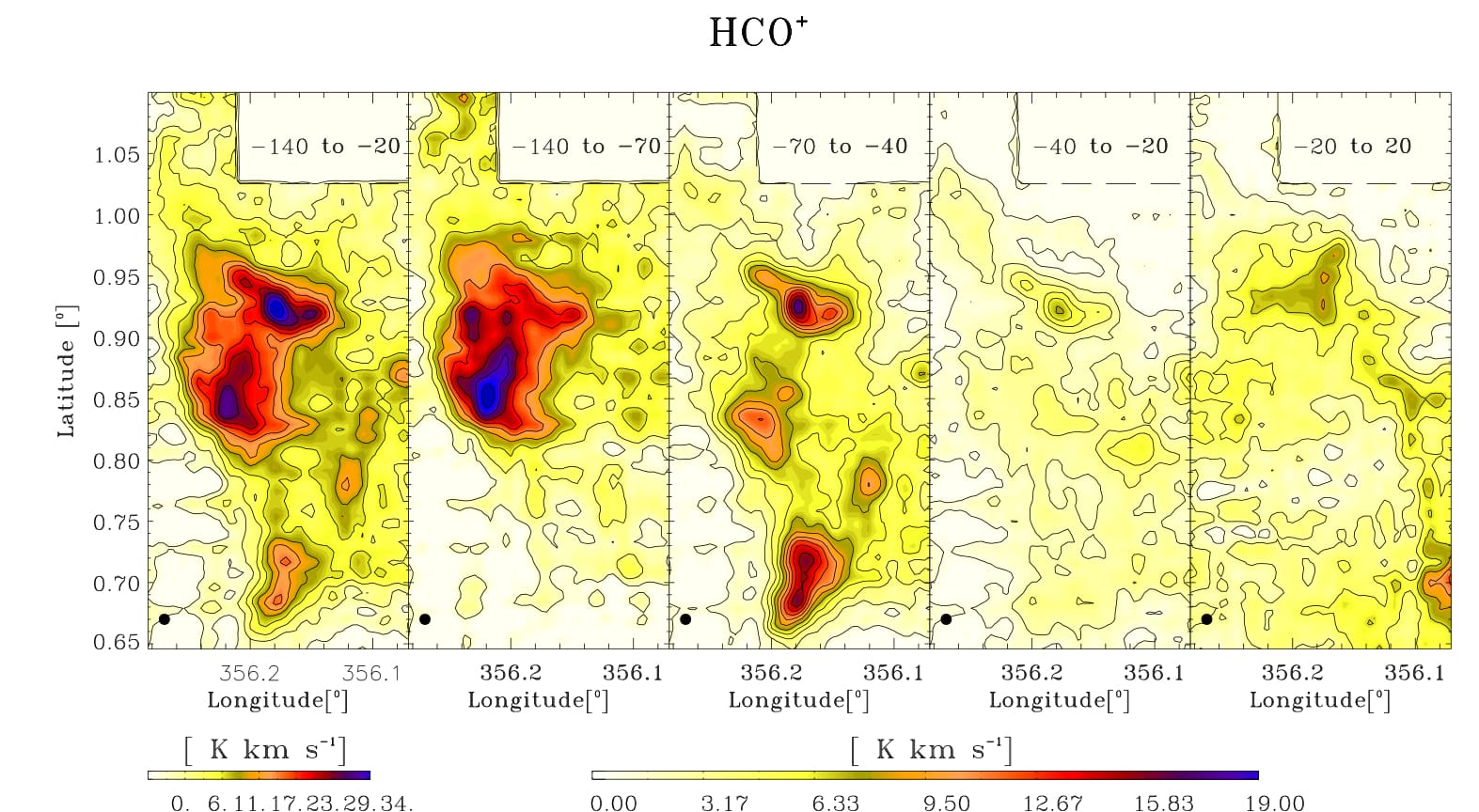}
\caption{Velocity integrated emission in HCO$^+$. From left to right: velocity range from -140 to 20 \kms (the complete velocity range covered by the GMLs); velocity range from -110 to -70 \kms; -70
to -40 \kms, -40 to -20\kms and -20 to 20 \kms \label{HCO+_intervals}}
\end{figure*}

\begin{figure*}
\includegraphics[angle=0,width=1.0 \textwidth]{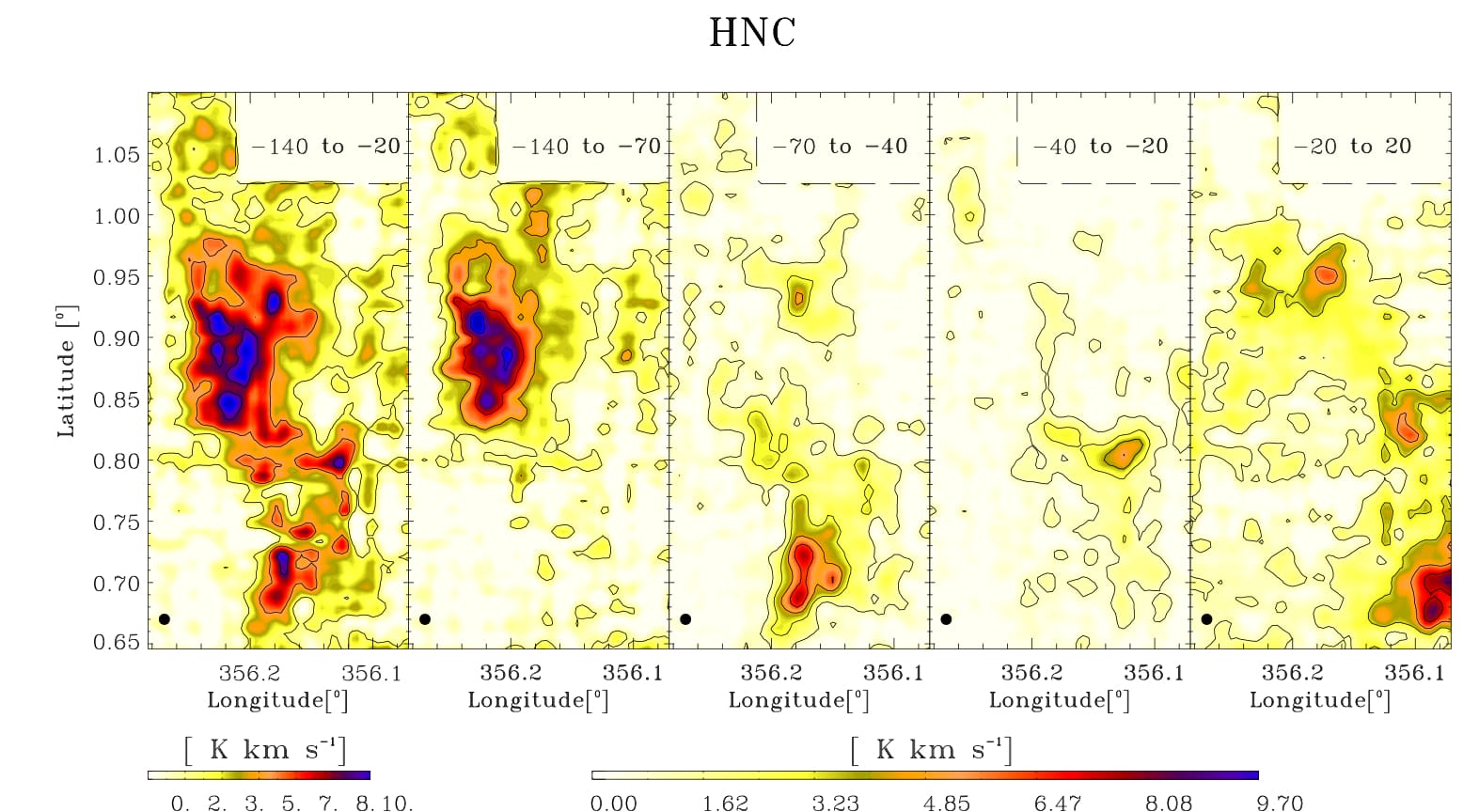}

\caption{Velocity integrated emission in HNC. From left to right: velocity range from -140 to 20 \kms (the complete velocity range covered by the GMLs); velocity range from -140 to -70 \kms; -70 to
-40 \kms, -40 to -20\kms and -20 to 20 \kms \label{HNC_intervals}}
\end{figure*}

\begin{figure*}
\includegraphics[angle=0,width=1.0 \textwidth]{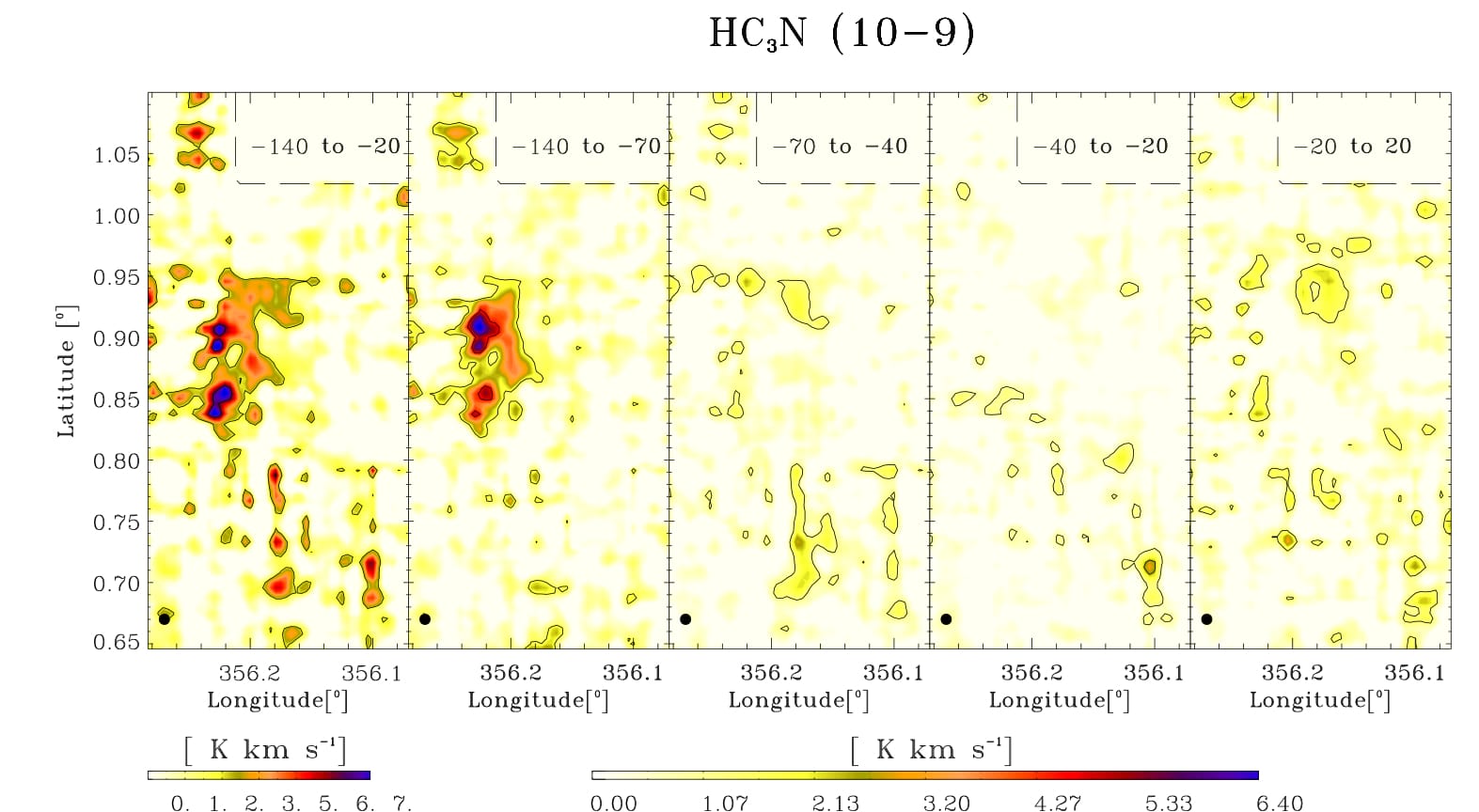}
\caption{Velocity integrated emission in HC$_3$N (10-9). From left to right: velocity range from -140 to 20 \kms (the complete velocity range covered by the GMLs); velocity range from -140 to -70
\kms; -70 to -40 \kms, -40 to -20\kms and -20 to 20 \kms}
\end{figure*}

\begin{figure*}
\includegraphics[angle=0,width=1.0 \textwidth]{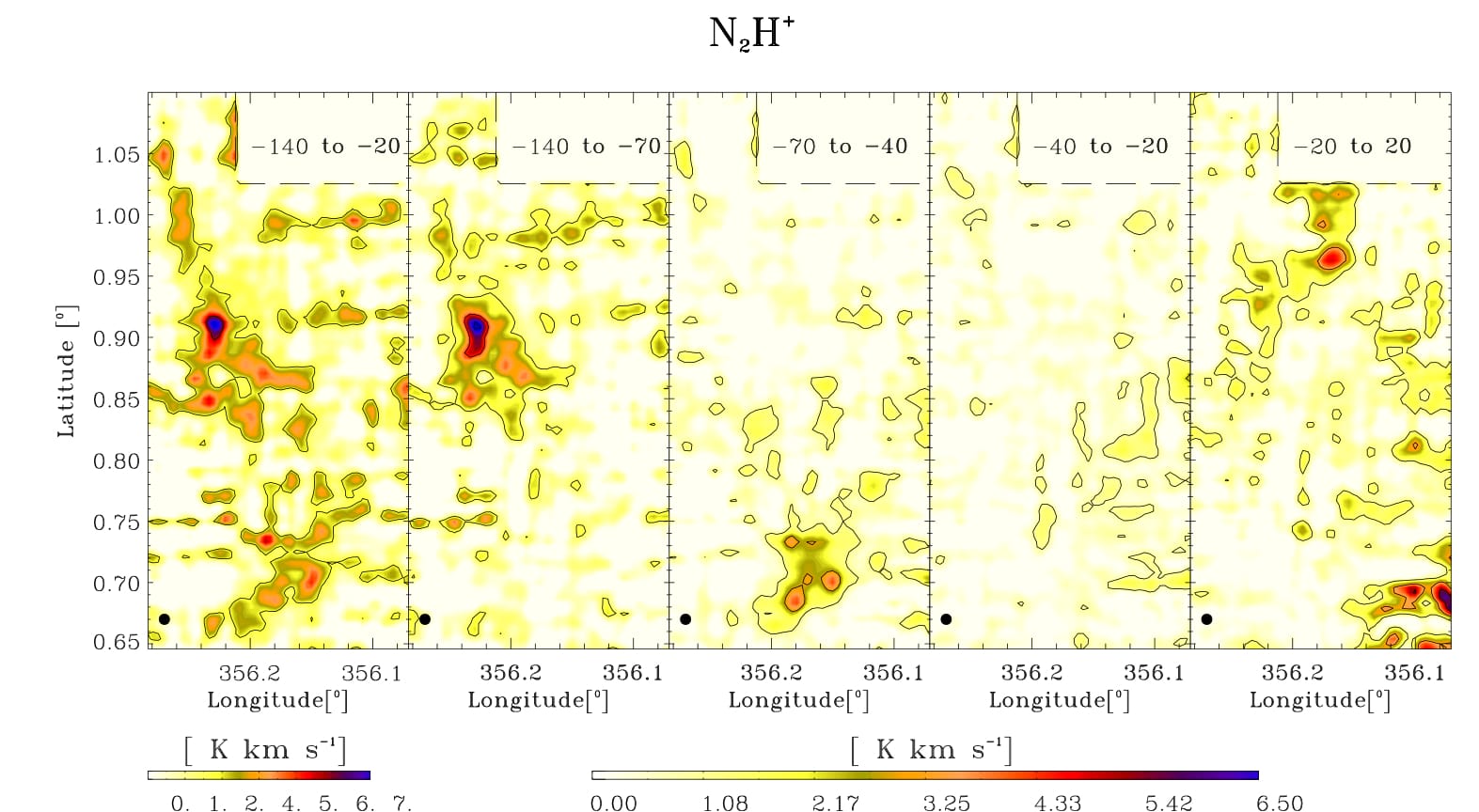}
\caption{Velocity integrated emission in N$_2$H$^+$. From left to right: velocity range from -140 to 20 \kms (the complete velocity range covered by the GMLs); velocity range from -140 to -70 \kms;
-70 to -40 \kms, -40 to -20\kms and -20 to 20 \kms}
\end{figure*}

\begin{figure*}
\includegraphics[angle=0,width=1.0 \textwidth]{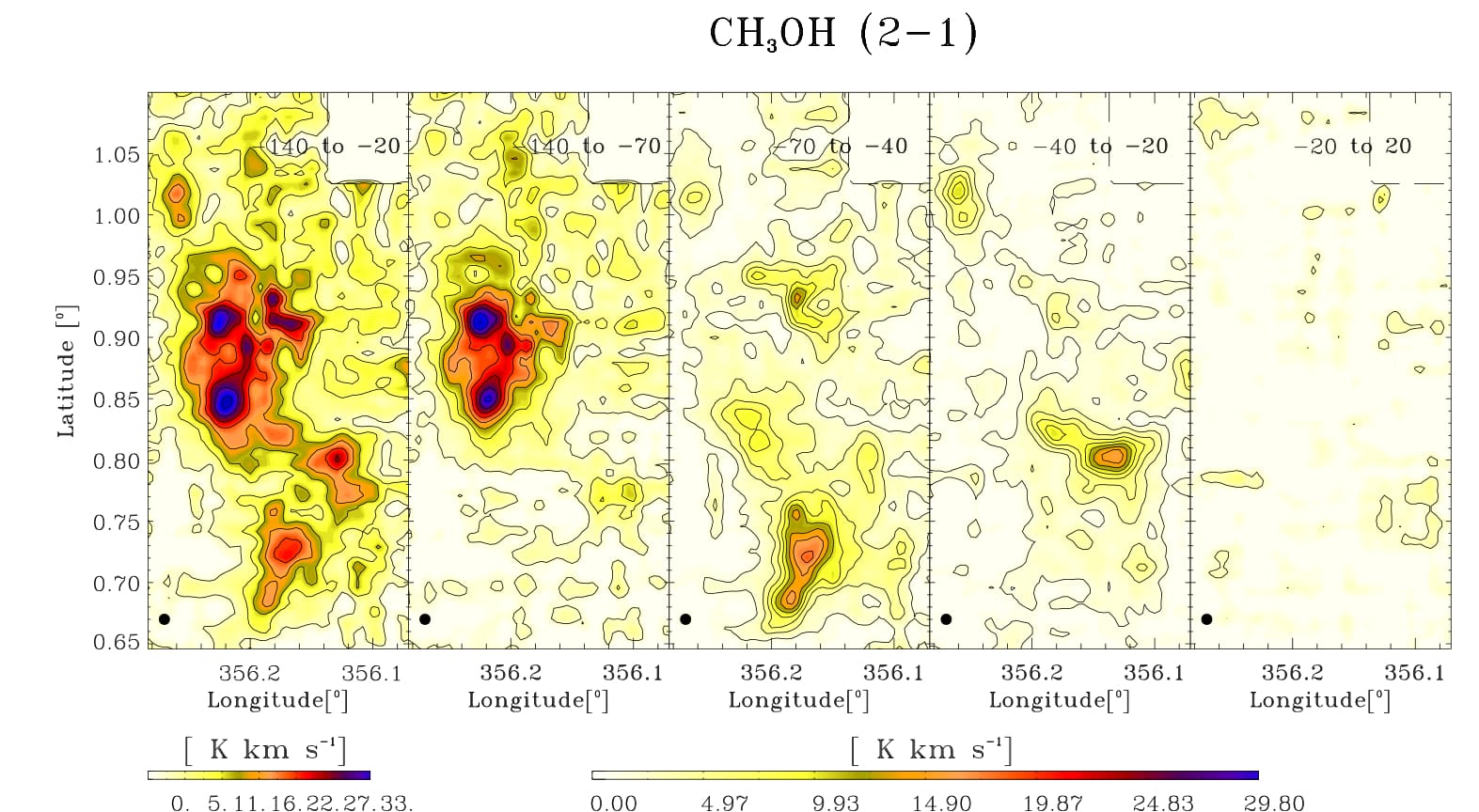}
\caption{Velocity integrated emission in CH$_3$OH. From left to right: velocity range from -140 to 20 \kms (the complete velocity range covered by the GMLs); velocity range from -140 to -70 \kms; -70 to -40 \kms, -40 to -20\kms and -20 to 20 \kms }
\end{figure*}

\begin{figure*}
\includegraphics[angle=0,width=1.0 \textwidth]{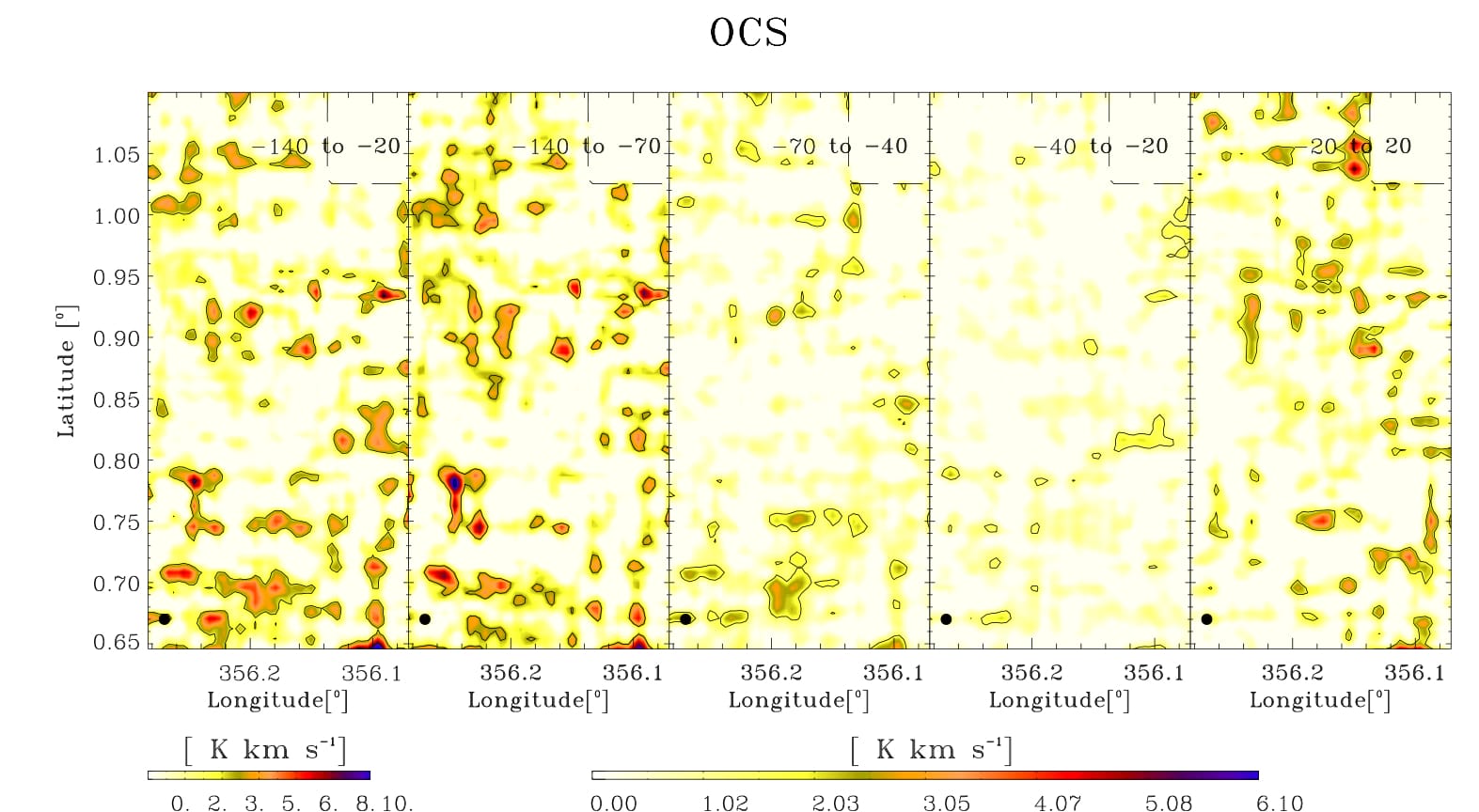}
\caption{Velocity integrated emission in OCS. From left to right: velocity range from -140 to 20 \kms (the complete velocity range covered by the GMLs); velocity range from -140 to -70 \kms; -70 to
-40 \kms, -40 to -20\kms and -20 to 20 \kms}
\end{figure*}

\begin{figure*}\label{cs}
\includegraphics[angle=0,width=1.0 \textwidth]{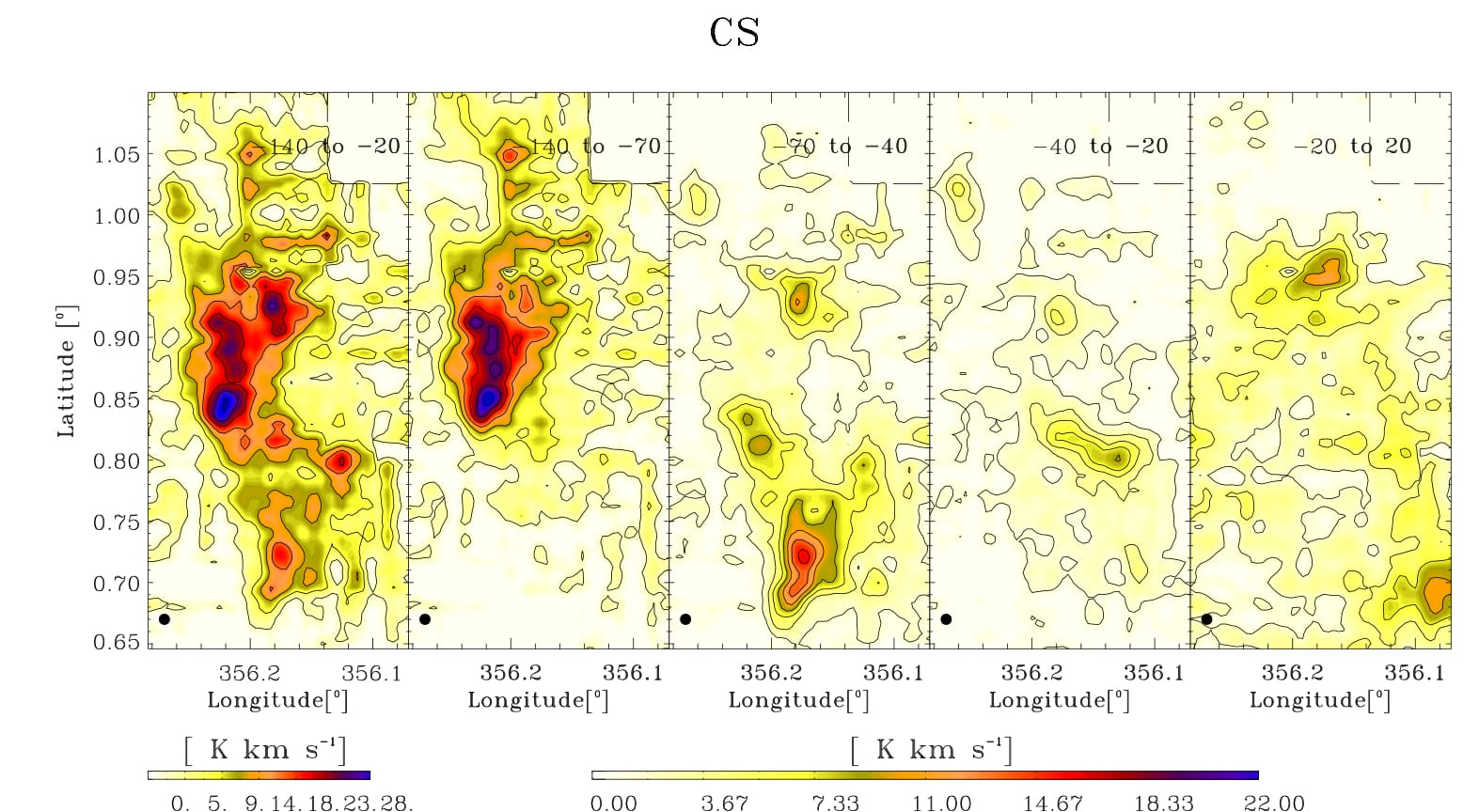}
\caption{Velocity integrated emission in CS. From left to right: velocity range from -140 to 20 \kms (the complete velocity range covered by the GMLs); velocity range from -140 to -70 \kms; -70 to
-40 \kms, -40 to -20\kms and -20 to 20 \kms}
\end{figure*}

\begin{figure*}
\includegraphics[angle=0,width=1.0 \textwidth]{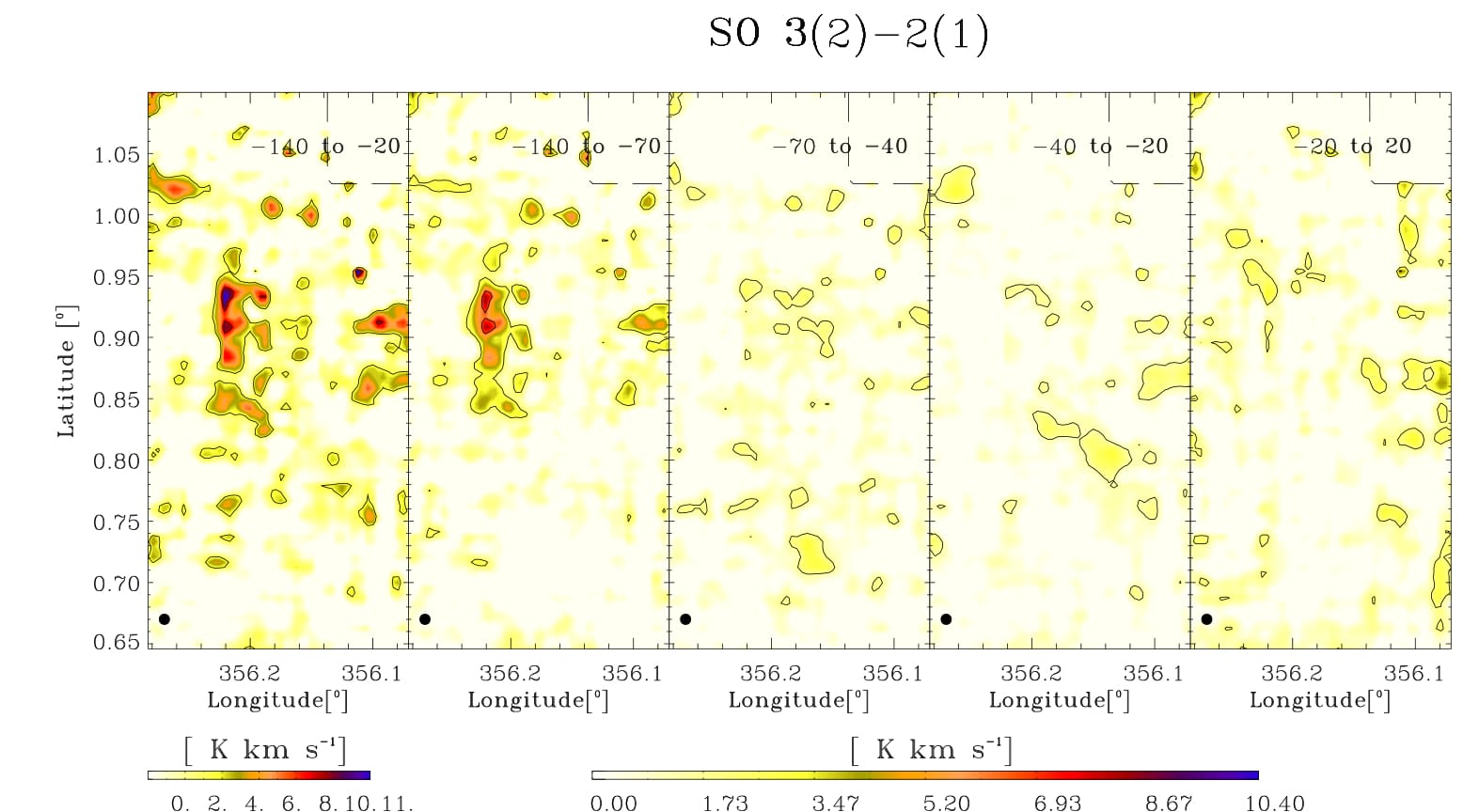}
\caption{Velocity integrated emission in SO ($3_2-2_1$). From left to right: velocity range from -140 to 20 \kms (the complete velocity range covered by the GMLs); velocity range from -140 to -70 \kms; -70
to -40 \kms, -40 to -20\kms and -20 to 20 \kms}
\end{figure*}

\begin{figure*}
\includegraphics[angle=0,width=1.0 \textwidth]{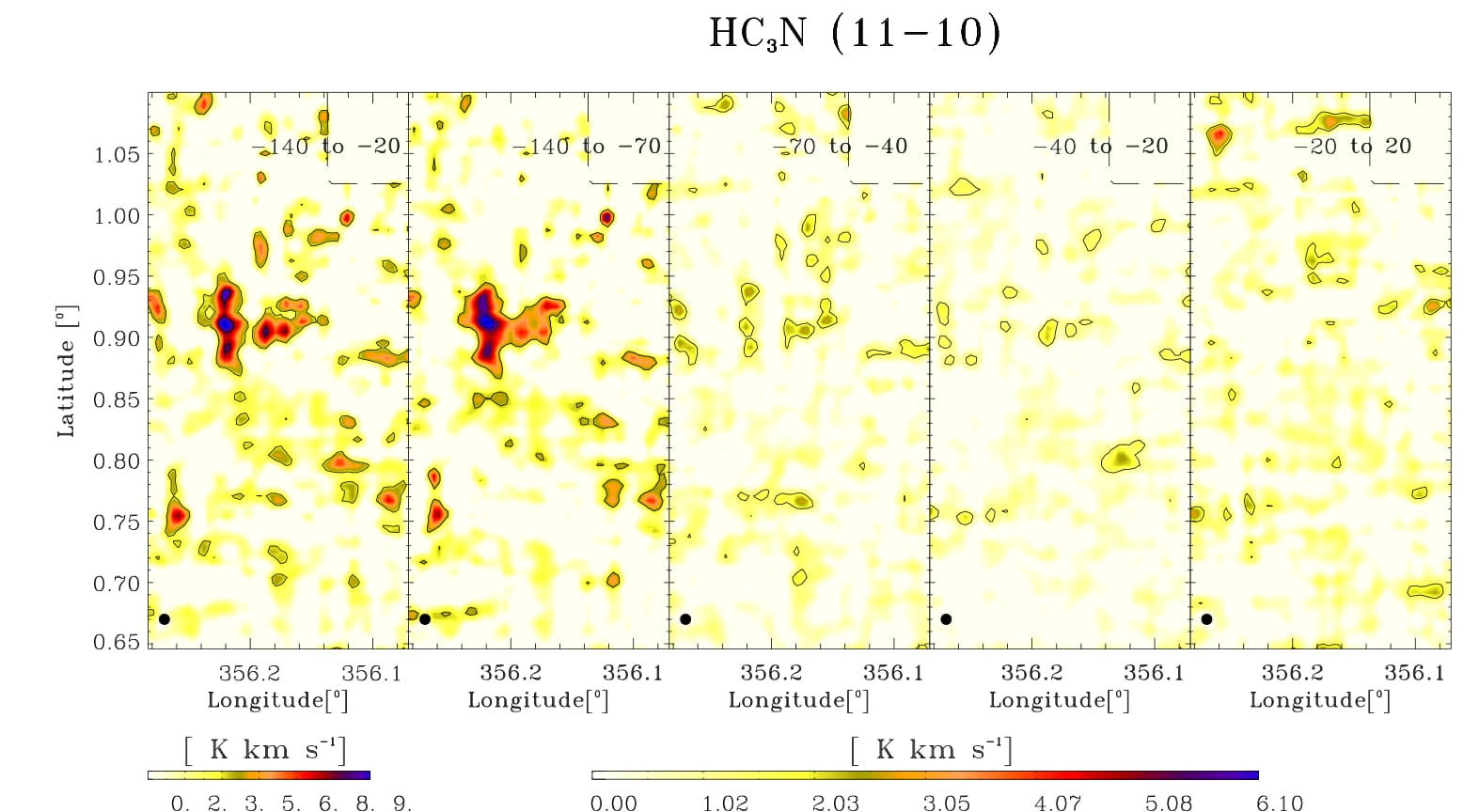}

\caption{Velocity integrated emission in HC$_3$N (11-10). From left to right: velocity range from -140 to 20 \kms (the complete velocity range covered by the GMLs); velocity range from -140 to -70
\kms; -70 to -40 \kms, -40 to -20\kms and -20 to 20 \kms\label{final}}
\end{figure*}

\clearpage

\section{Latitude-velocity maps}

\begin{figure*}
\includegraphics[angle=0,width=1.0 \textwidth]{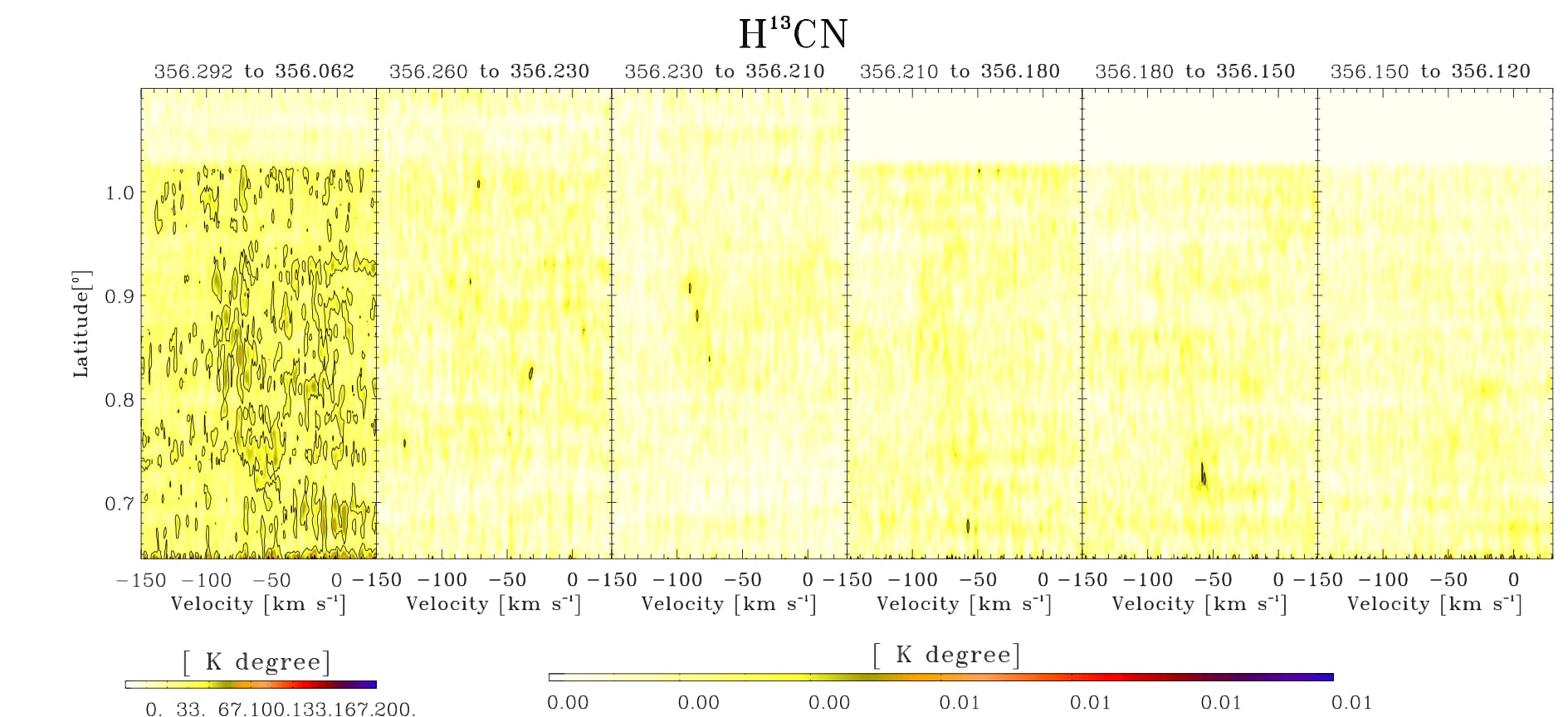}
\caption{Latitude-velocity maps of the H$^{13}$CN emission integrated in the complete longitude range from 356\deg.29165 to 356\deg.06249 (Left), and in longitude steps of 108'' (second panel to the right).}
\label{bvini}
\end{figure*}

\begin{figure*}
\includegraphics[angle=0,width=1.0 \textwidth]{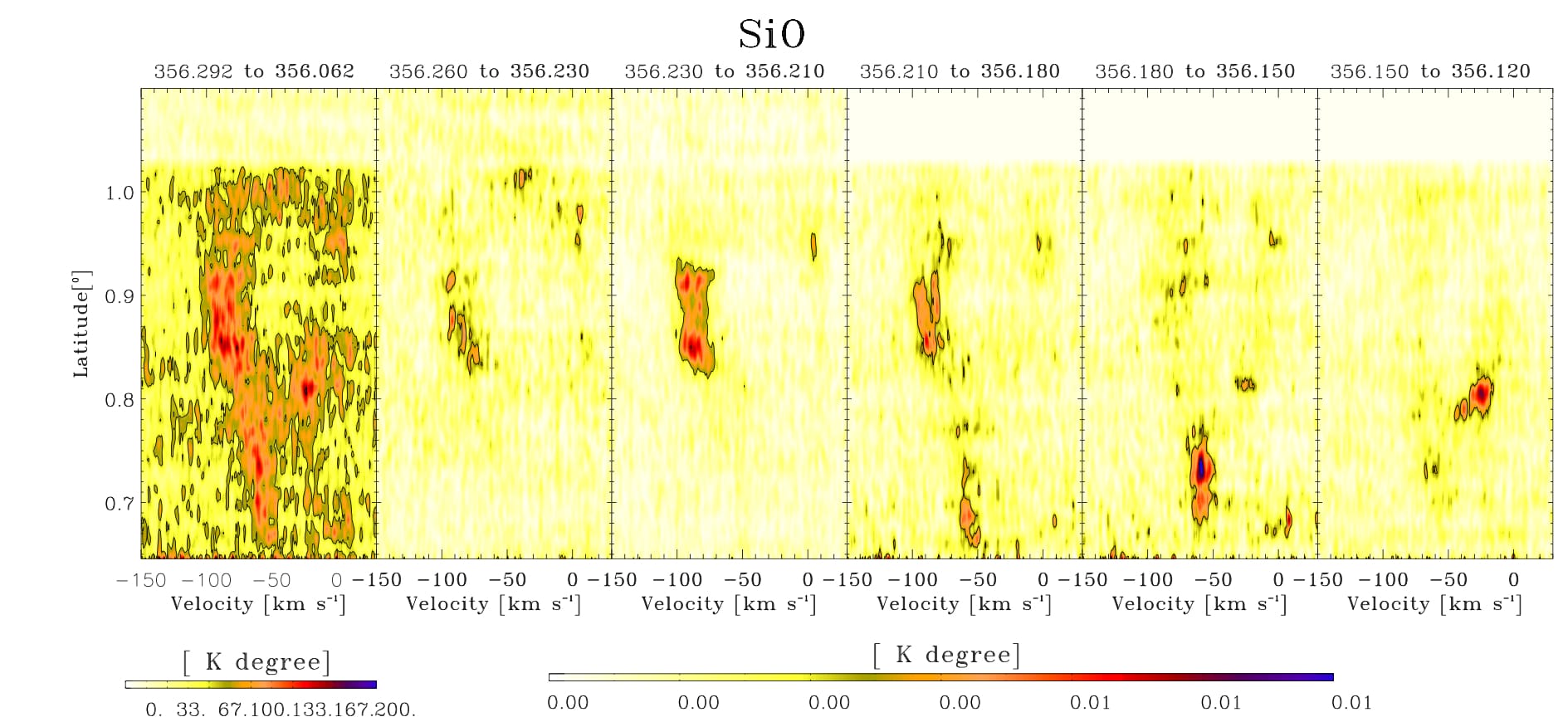}
\caption{Latitude-velocity maps of the SiO emission integrated in the complete longitude range from 356\deg.29165 to 356\deg.06249 (Left), and in longitude steps of 108'' (second panel to the right).}
\label{bv_cut}
\end{figure*}

\begin{figure*}
\includegraphics[angle=0,width=1.0 \textwidth]{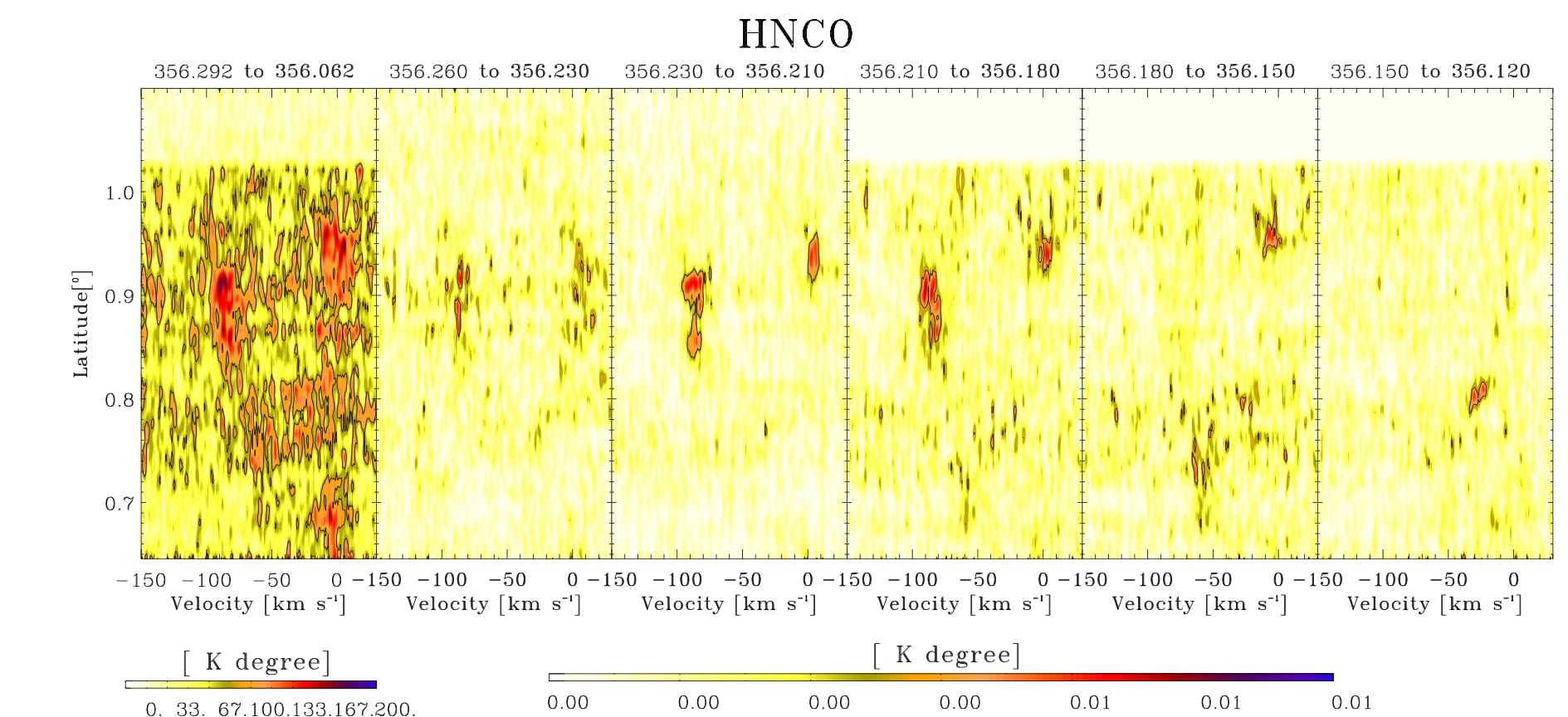}
\caption{Latitude-velocity maps of the HNCO emission integrated in the complete longitude range from 356\deg.29165 to 356\deg.06249 (Left), and in longitude steps of 108'' (second panel to the right).}
\label{bv_cut}
\end{figure*}

\begin{figure*}
\includegraphics[angle=0,width=1.0 \textwidth]{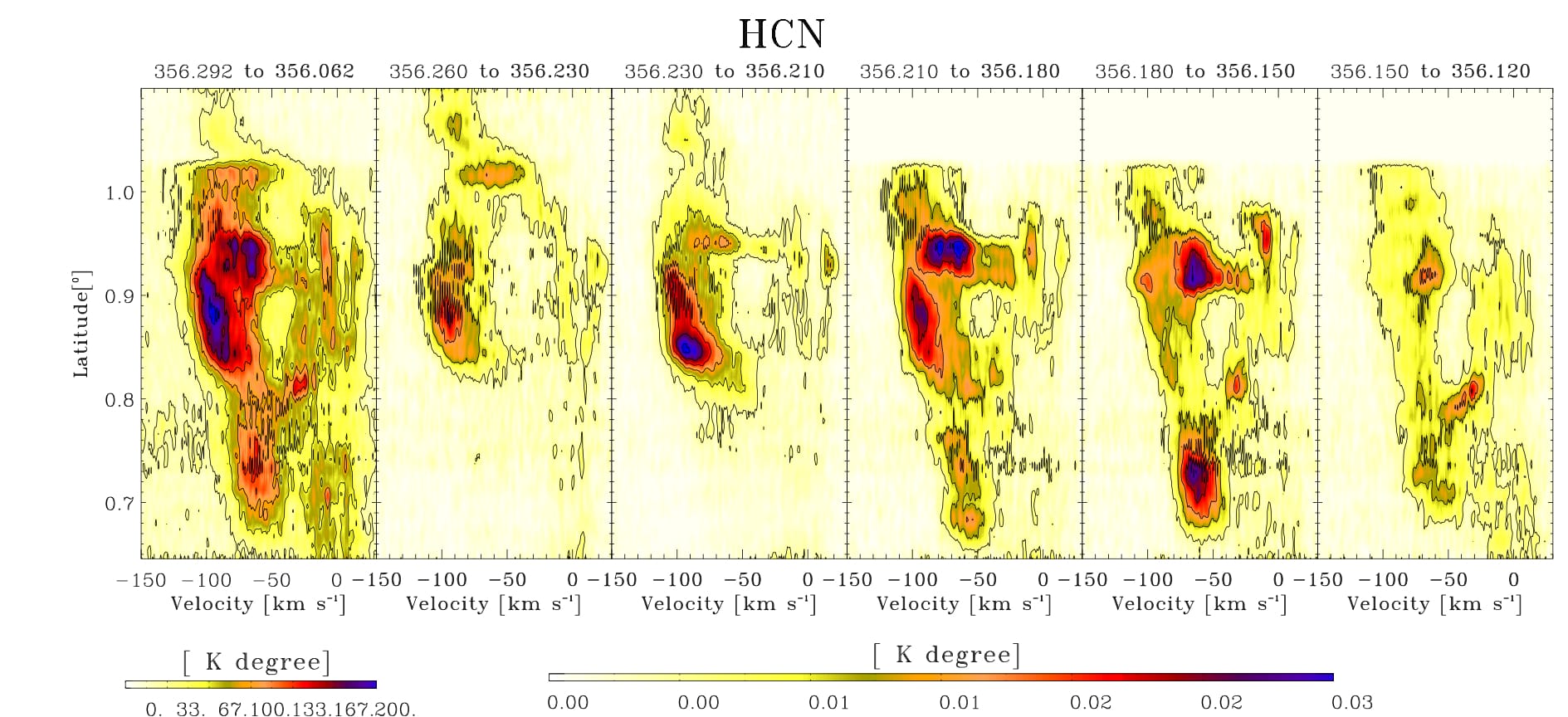}
\caption{Latitude-velocity maps of the HCN emission integrated in the complete longitude range from 356\deg.29165 to 356\deg.06249 (Left), and in longitude steps of 108'' (second panel to the right).}
\label{bv_cut}
\end{figure*}

\begin{figure*}
\includegraphics[angle=0,width=1.0 \textwidth]{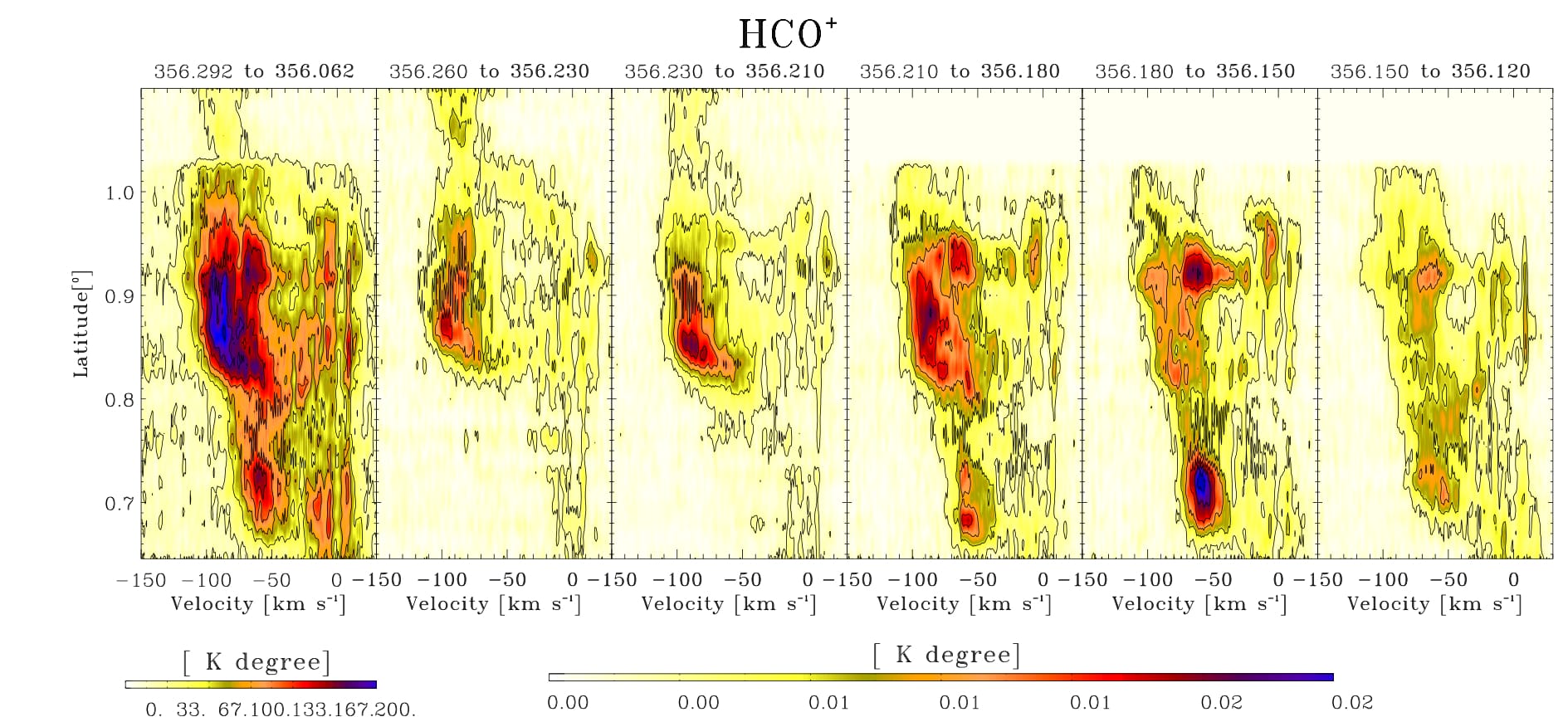}
\caption{Latitude-velocity maps of the HCO$^+$ emission integrated in the complete longitude range from 356\deg.29165 to 356\deg.06249 (Left), and in longitude steps of 108'' (second panel to the right).}
\label{bv_cut}
\end{figure*}

\begin{figure*}
\includegraphics[angle=0,width=1.0 \textwidth]{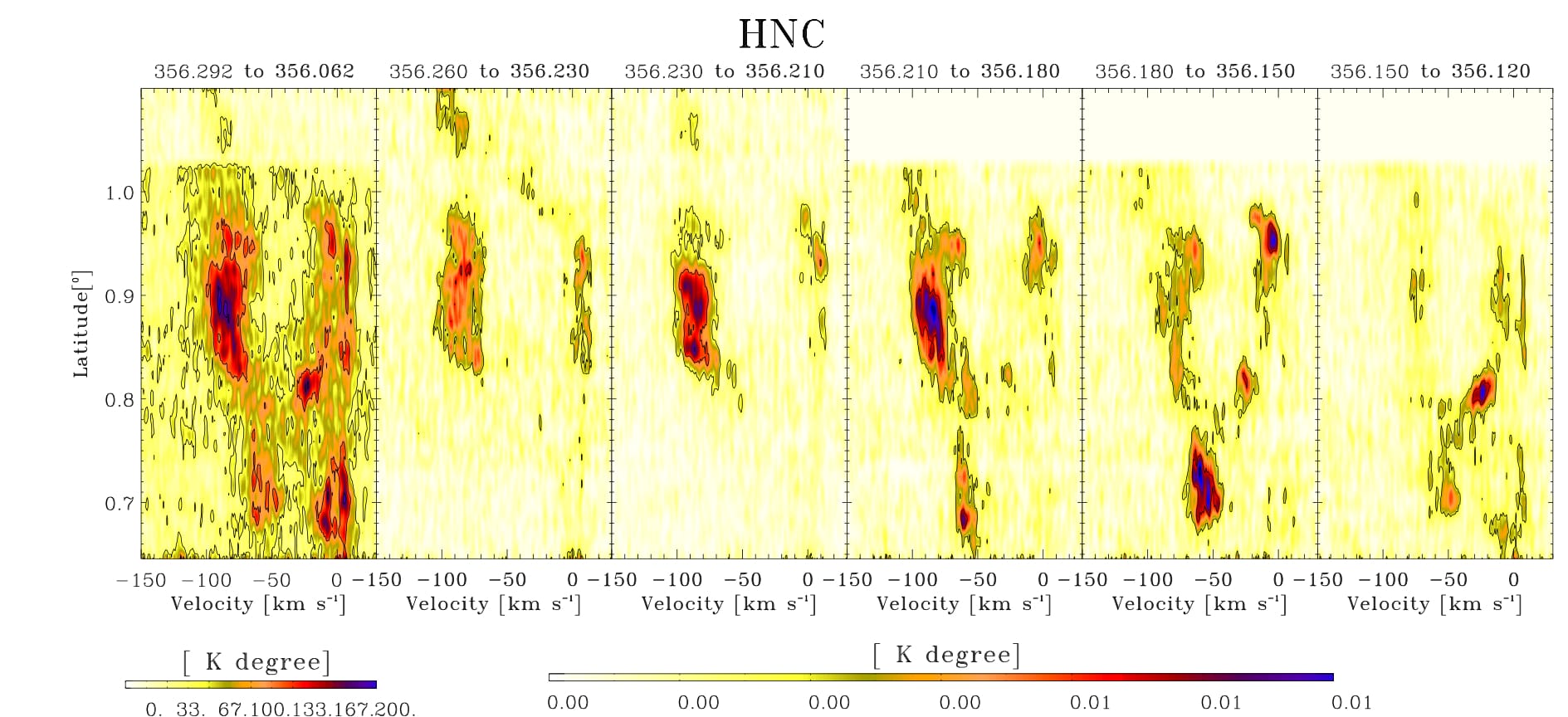}
\caption{Latitude-velocity maps of the HNC emission integrated in the complete longitude range from 356\deg.29165 to 356\deg.06249 (Left), and in longitude steps of 108'' (second panel to the right).}
\label{bv_cut}
\end{figure*}

\begin{figure*}
\includegraphics[angle=0,width=1.0 \textwidth]{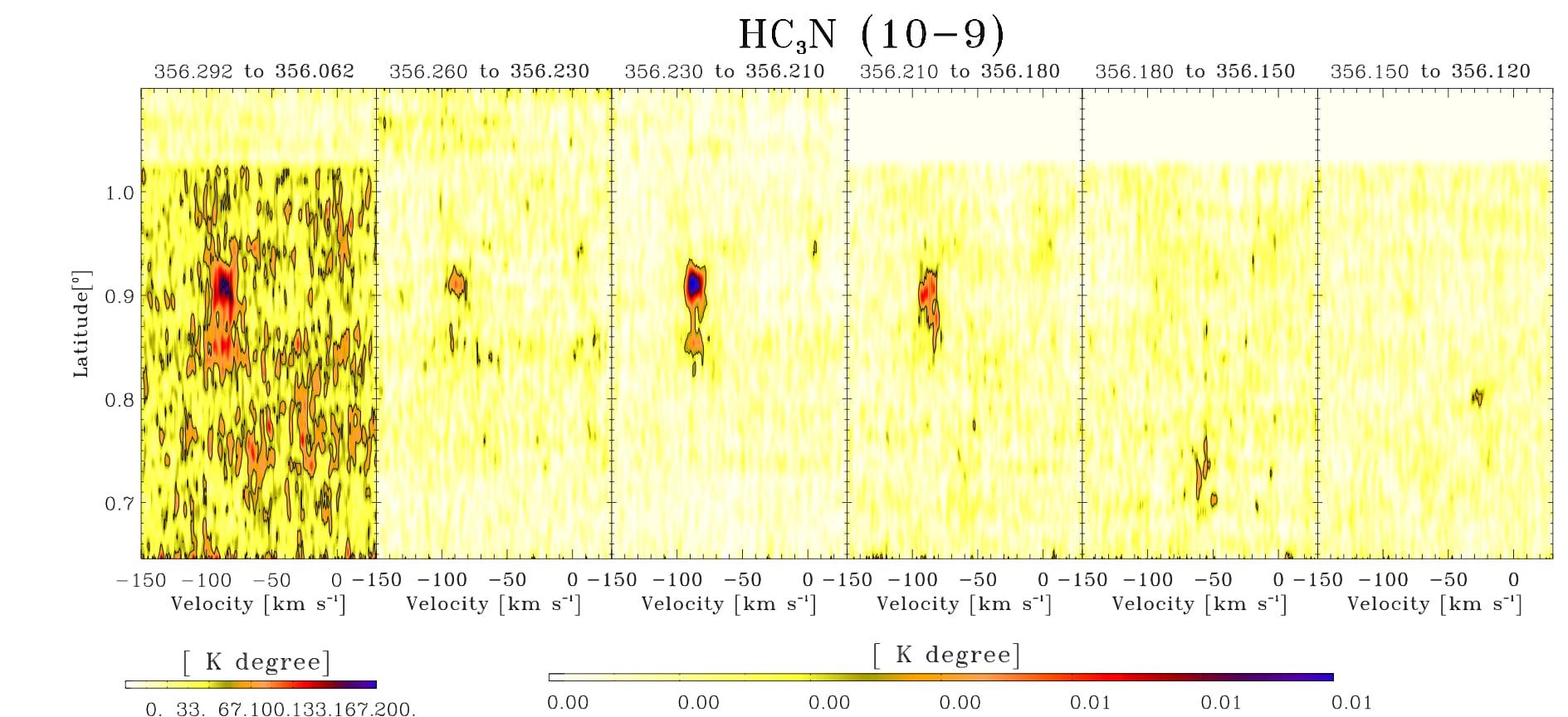}
\caption{Latitude-velocity maps of the HC$_3$N (10-9) emission integrated in the complete longitude range from 356\deg.29165 to 356\deg.06249 (Left), and in longitude steps of 108'' (second panel to the right).}
\label{bv_cut}
\end{figure*}

\begin{figure*}
\includegraphics[angle=0,width=1.0 \textwidth]{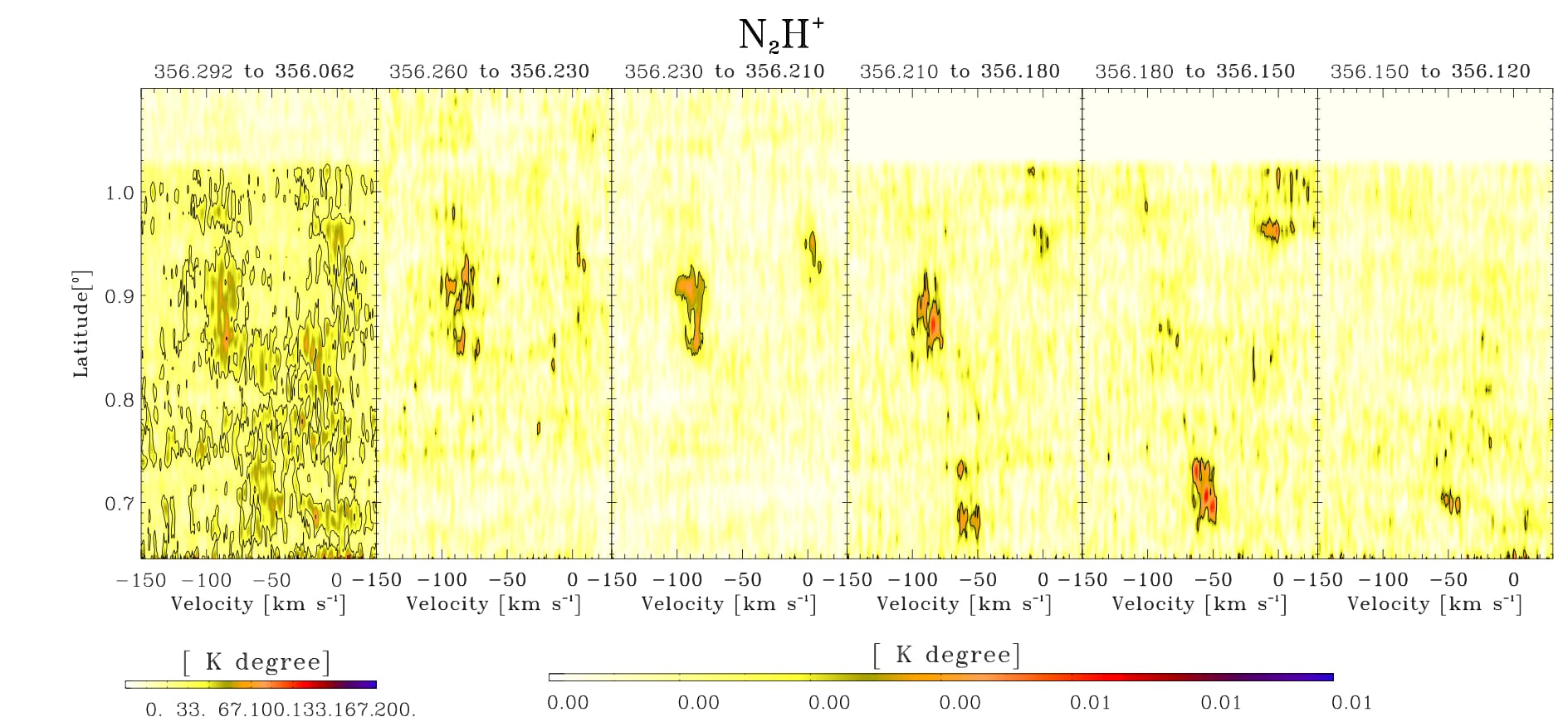}
\caption{Latitude-velocity maps of the N$_2$H$^+$ emission integrated in the complete longitude range from 356\deg.29165 to 356\deg.06249 (Left), and in longitude steps of 108'' (second panel to the right).}
\label{bv_cut}
\end{figure*}

\begin{figure*}
\includegraphics[angle=0,width=1.0 \textwidth]{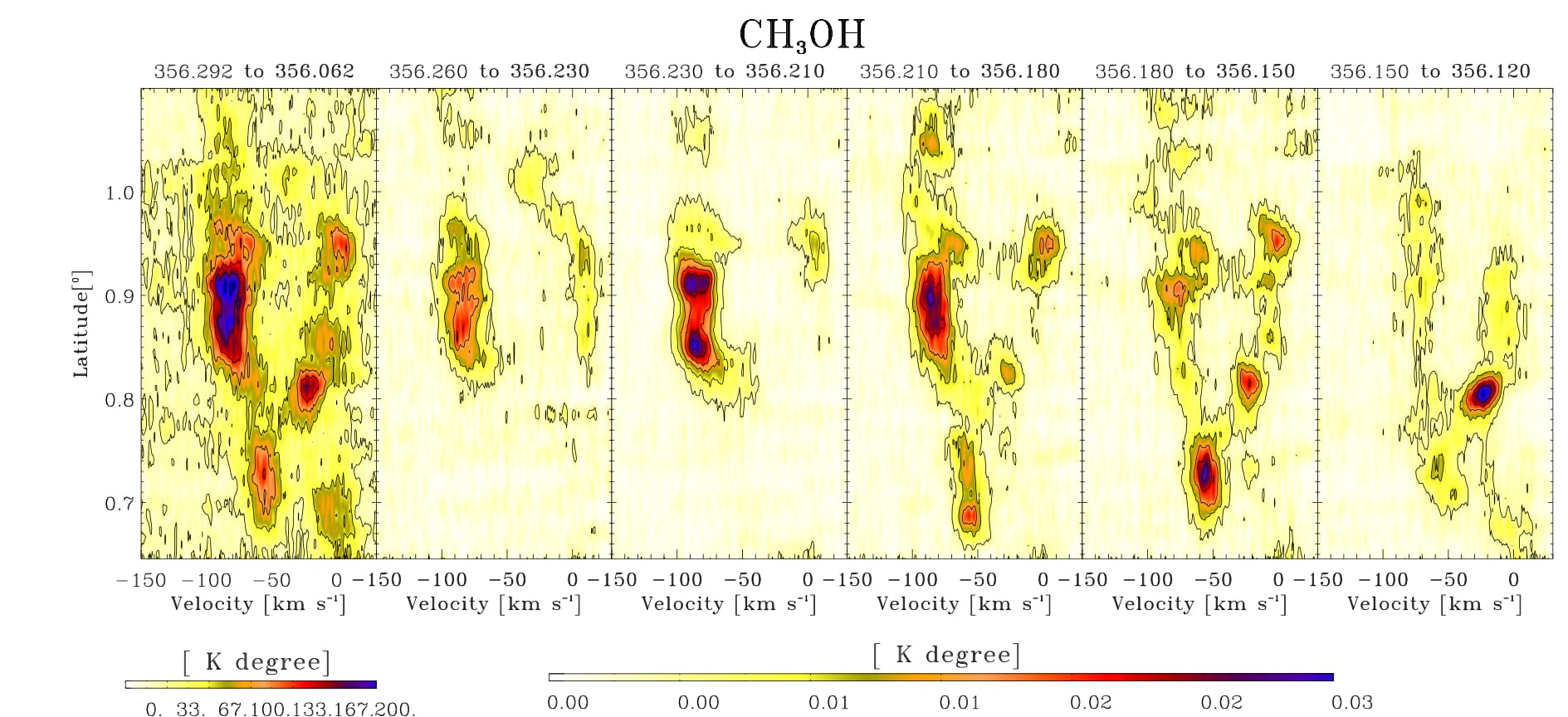}
\caption{Latitude-velocity maps of the CH$_3$OH emission integrated in the complete longitude range from 356\deg.29165 to 356\deg.06249 (Left), and in longitude steps of 108'' (second panel to the right).}
\label{bv_cut}
\end{figure*}

\begin{figure*}
\includegraphics[angle=0,width=1.0 \textwidth]{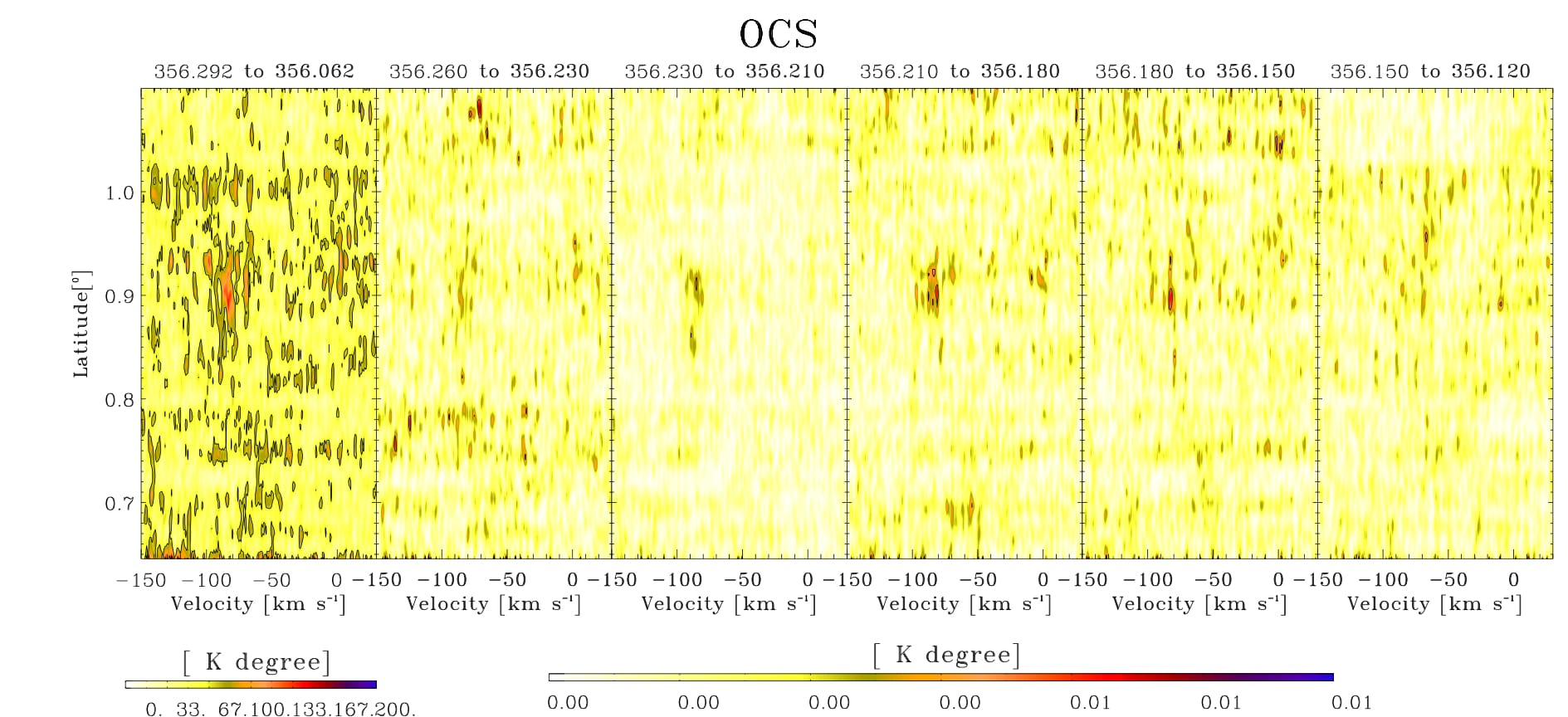}
\caption{Latitude-velocity maps of the OCS emission integrated in the complete longitude range from 356\deg.29165 to 356\deg.06249 (Left), and in longitude steps of 108'' (second panel to the right).}
\label{bv_cut}
\end{figure*}

\begin{figure*}
\includegraphics[angle=0,width=1.0 \textwidth]{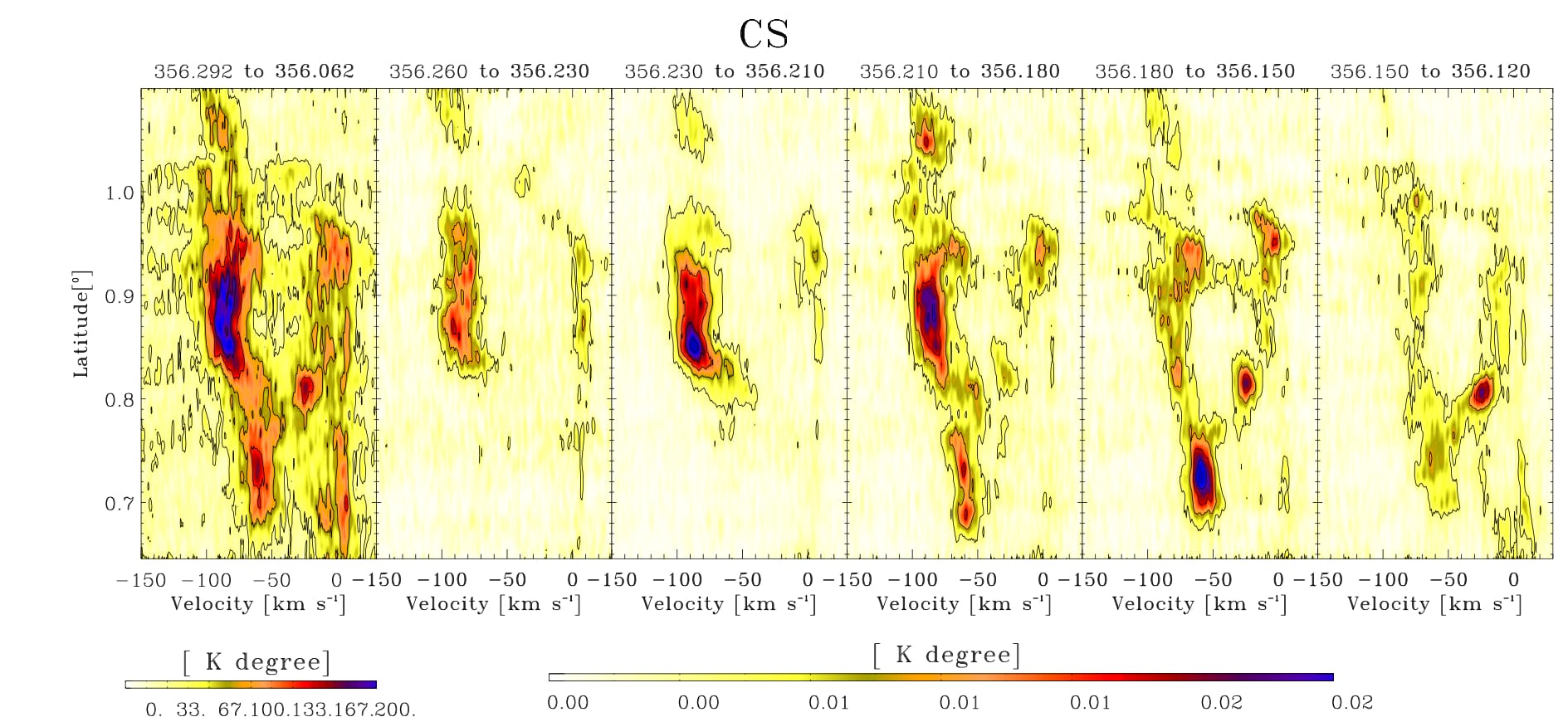}
\caption{Latitude-velocity maps of the CS emission integrated in the complete longitude range from 356\deg.29165 to 356\deg.06249 (Left), and in longitude steps of 108'' (second panel to the right).}
\label{bv_cut}
\end{figure*}

\begin{figure*}
\includegraphics[angle=0,width=1.0 \textwidth]{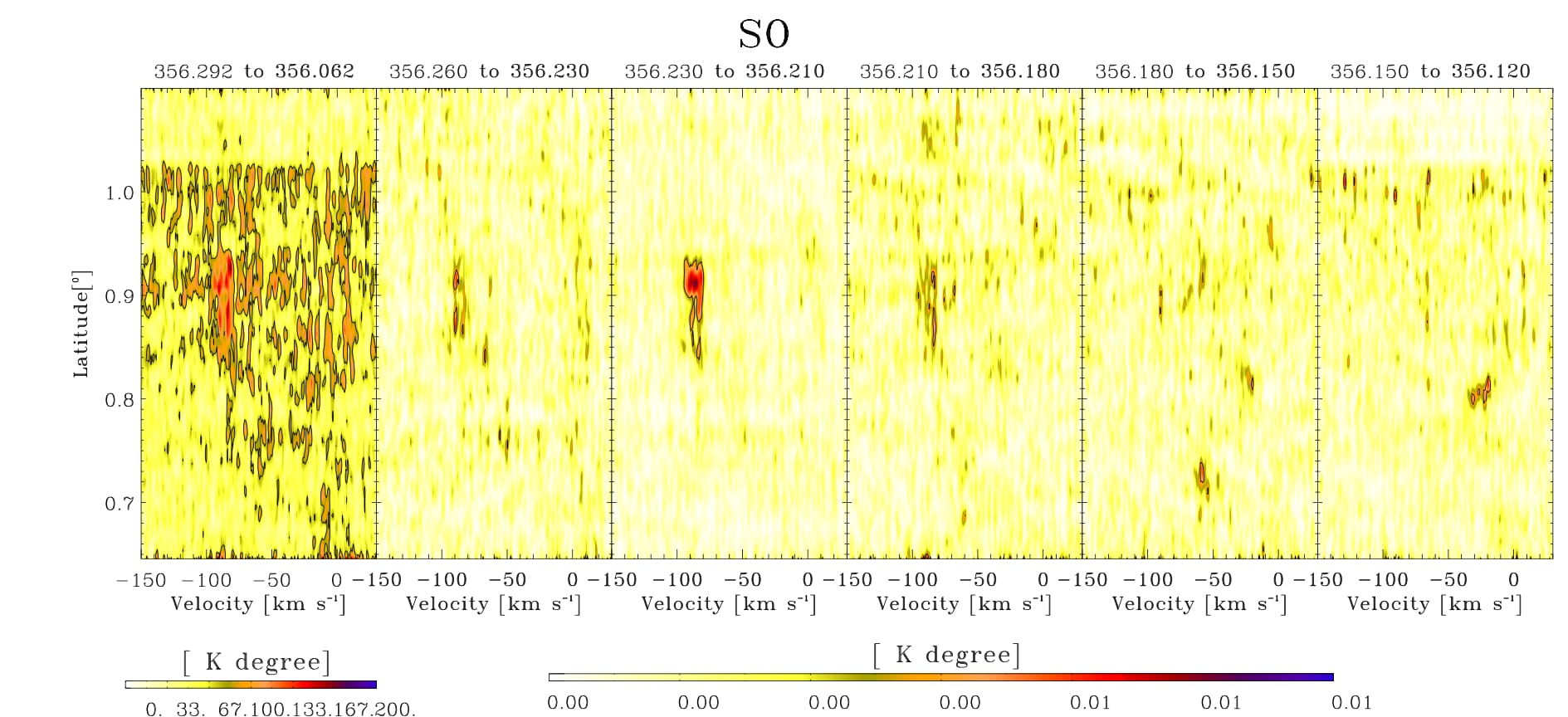}
\caption{Latitude-velocity maps of the SO emission integrated in the complete longitude range from 356\deg.29165 to 356\deg.06249 (Left), and in longitude steps of 108'' (second panel to the right).}
\label{bv_cut}
\end{figure*}

\begin{figure*}
\includegraphics[angle=0,width=1.0 \textwidth]{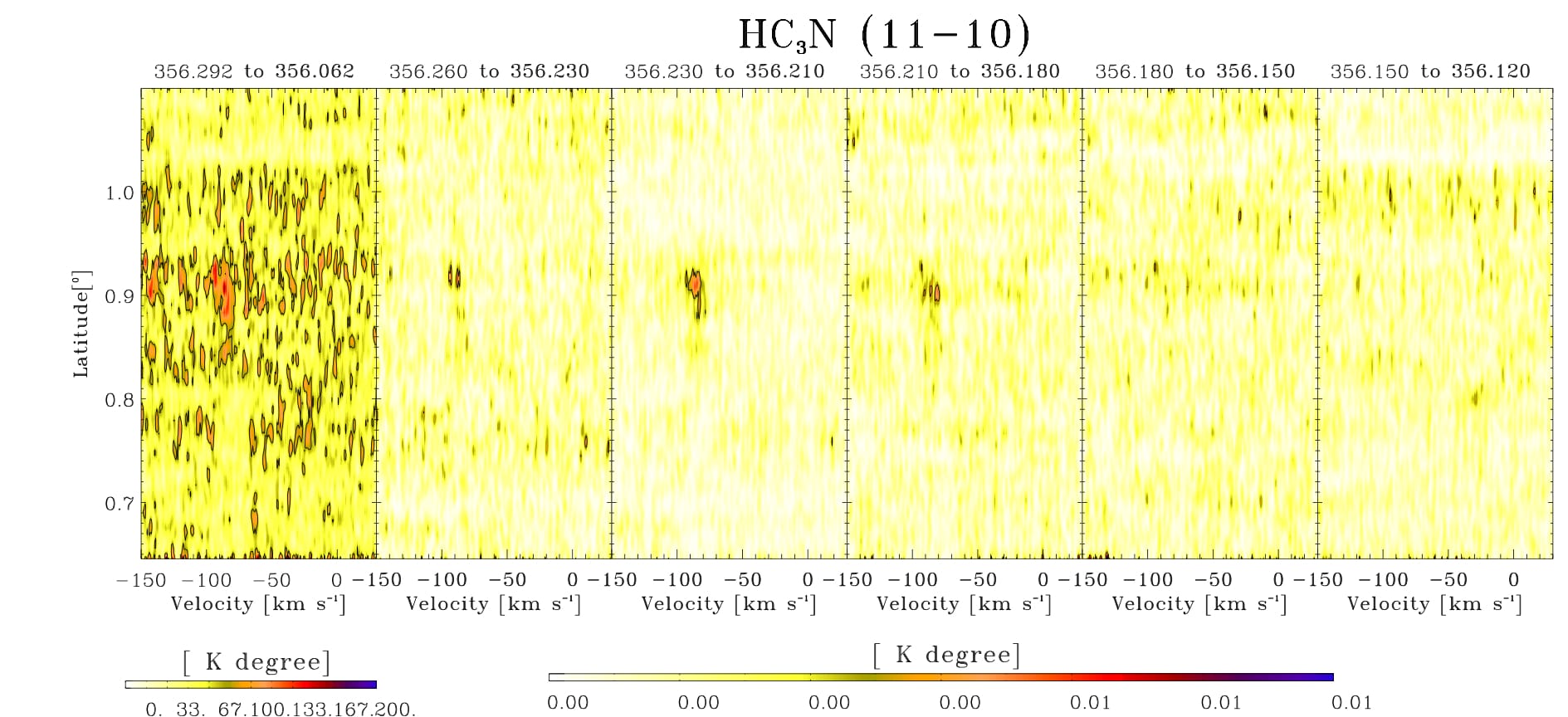}
\caption{Latitude-velocity maps of the HC$_3$N (11-10) emission integrated in the complete longitude range from 356\deg.29165 to 356\deg.06249 (Left), and in longitude steps of 108'' (second panel to the right).}
\label{bvfinal}
\end{figure*}

\clearpage
\end{appendix}
\end{document}